\documentclass{article}

\usepackage[english]{babel}

\usepackage[letterpaper,top=2cm,bottom=2cm,left=3cm,right=3cm,marginparwidth=1.75cm]{geometry}

\usepackage{cite}

\usepackage{amsmath}
\usepackage{graphicx}
\usepackage[normalem]{ulem}
\usepackage[colorlinks=true, allcolors=blue]{hyperref}
\usepackage{amsmath,amssymb,epsfig,amsfonts,mathrsfs}
\DeclareSymbolFontAlphabet{\mathrsfs}{rsfs}

\newcommand{\be}{\begin{equation}}
\newcommand{\ee}{\end{equation}}
\newcommand{\ba}{\begin{aligned}}
\newcommand{\ea}{\end{aligned}}

\newcommand\BR{\mathbb{R}}

\newcommand\BC{\mathbb{C}}

\newcommand\CalD{\mathcal{D}}

\newcommand\CalF{\mathcal{F}}
\newcommand\CalP{\mathcal{P}}

\newcommand\CalN{\mathcal{N}}
\newcommand\CalS{\mathcal{S}}
\newcommand\CalM{\mathcal{M}}

\newcommand\CalV{\mathcal{V}}

\title{A Panorama Of Physical Mathematics c. 2022
}
\author{
Ibrahima Bah$^1$, 
Daniel S. Freed $^2$,
Gregory W. Moore$^3$, \\
 Nikita Nekrasov$^4$, 
Shlomo S. Razamat$^5$, 
Sakura Sch\"afer-Nameki$^6$}

\begin{document}
\maketitle

\begin{center}
{\it $^1$  Department of Physics and Astronomy, Johns Hopkins University, Baltimore, MD 21218, USA }\\
{\it $^2$ Department of Mathematics, University of Texas at Austin }\\
{\it $^3$ NHETC and Department of Physics and Astronomy, Rutgers University }\\
{\it $^4$ Simons Center for Geometry and Physics, \\
Stony Brook University, Stony Brook, NY 11794-3636, USA}\\
{\it $^5$ Department of Physics, Technion, Haifa, 32000, Israel}\\
{\it $^6$ Mathematical Institute,University of Oxford, Oxford OX2 6GG, United Kingdom }
\end{center}
\begin{abstract}
 \noindent
 What follows is a broad-brush overview of the 
 recent synergistic interactions 
between mathematics and theoretical physics of quantum 
field theory and string theory. The discussion is forward-looking, 
suggesting potentially useful and fruitful directions and problems, 
some old, some new, for further development of the subject. 
This paper is a much extended version of the Snowmass whitepaper on physical mathematics 
\cite{Bah:2022xfv}.  Version:    \today .
\end{abstract}

\tableofcontents

\section{General Remarks }

\label{sec:genrem}

At the dawn of the scientific revolution Galileo wrote that the
book of nature is written in the language of mathematics \cite{Galileo,drake2016controversy}. That
insight continues to illuminate the pursuit of science to this day.
As physics progresses in its quest to understand the most fundamental
laws of nature the mathematics in use does not remain static. It is,
rather, cumulative. Moreover, as stressed by both Einstein \cite{Einstein_MTP} and Dirac \cite{dirac_1940}, it becomes more
sophisticated and more abstract. The history of  the relation of physics
and mathematics teaches us that, very often, great advances in physics are
accompanied by profound innovations in mathematics, and vice versa. The past fifty years
have been a watershed moment in this venerable dialogue between physics
and mathematics. Astonishing advances in physics have led to profound
mathematical discoveries and conversely new areas of mathematics
have been brought to bear on attempts to understand some of the most
basic questions about nature. There is a community of scientists
that places equal stress on \underline{both} the applications to advances in mathematics
along with the drive to a deeper understanding of nature. A good name for
the enterprise pursued by this community is ``physical mathematics.''
The subject continues to be very active and vibrant. Foundations are
becoming firmer and some participants are thrusting into exciting
new territory.

This broad overview is meant to accompany the Snowmass whitepaper 
on physical mathematics \cite{Bah:2022xfv} We  will attempt to indicate some of the
promising areas for future research along these lines, while also
recalling some older promising but recently untrodden paths.
While many topics of interest are covered here the authors are 
keenly aware that there are a large number of important topics, 
quite relevant to current research in physical mathematics, which have 
been omitted. Some, but not all, of these are addressed in other 
Snowmass documents.  And then, there are the unknown unknowns: 
It must 
be borne in mind that in research the greatest treasures are often
unforeseen, and we expect that some of the future paths indicated here
will lead to unexpected and unanticipated new directions.

There are several other overviews of physical mathematics, or physmatics, 
as it is sometimes called to which we draw the reader's attention. 
These include \cite{DysonMissedOpportunities, JaffeNotices, ConceptualFoundationsQFT,Jaffe:1993qt,
Atiyah:1994qu, Zaslow:2005jw, Moore:PhysMathFuture, Aganagic:2015tka, Morrison:2018zvt,YujiReview}. 
In addition two ongoing world-wide 
seminar series \cite{QFTandGeo,WHCGP}, as well as the annual 
international String-Math conferences, cover many topics discussed herein and give an excellent 
snapshot of some of the most interesting current research topics. 

The plan of this paper  is as follows: 
Section \ref{sec:QFT} is a discussion of Quantum Field Theory (QFT). It begins with a summary of attempts to give a general definition of what QFT is and covers some topics in topological QFT. It also covers new algebraic structures
and new notions of symmetries in QFT, along with  updates on resurgence. 
Complementing this, section \ref{sec:STMT} concerns some of the fundamental questions relating to string theory and M-theory, and their mathematical foundation. Much recent progress has been made in the study of anomalies, which are the topic of section \ref{sec:Anomalies}, which  addresses both anomalies in a general QFT setting, as well as anomaly cancellation in string theory. The quantum consistency of string theory vacua plays another central role in section \ref{sec:HoloQG}, where we 
discuss some specific and concrete mathematical problems connected to general issues in quantum gravity 
such as holography as well as certain aspects of the so-called ``swampland program''. Section 
 \ref{sec:NumberTheory} touches on some of the many interactions between string theory, QFT, and certain 
 aspects of number theory. Section   \ref{sec:CondMat} is an all-too-brief hint of some of the many and 
 profound connections to condensed matter physics.
Geometry and topology provide a large set of connections to QFT and string theory and are the topic of sections \ref{sec:InteractionsLowDimensional} and  \ref{sec:GeoQFT}, where we discuss the relation between geometric invariants and QFTs, as well as the geometrization program of QFTs, and superconformal field theories (SCFTs), respectively. A former section of this panorama has evolved into a separate publication \cite{Commando}. A partial list of topics to be covered in \cite{Commando} is given in section \ref{sec:Omissions}.

\section{Quantum Field Theory}
\label{sec:QFT}

The \emph{lingua franca} of physical mathematics is quantum field theory.
There are many dialects of quantum field theory, and altogether, 
quantum field theory is perhaps one of the most spectacularly successful frameworks for describing nature
in the history of science. Yet many physicists---and
most (all?) mathematicians---feel that we still lack a truly satisfactory and
universally applicable definition.  Therefore, it is a major open question to
find an axiom system for quantum field theory that adequately captures the
astonishing range of phenomena, both physical and mathematical, that it
describes.

\subsection{Approaches To Defining Quantum Field Theory}

Axiom systems for quantum mechanics date from the late 1920s and 1930s.  The
\emph{Dirac-von Neumann} axioms~\cite{Dirac,vNeumann,Mackey} start with a
complex separable Hilbert space~$\mathscr{H}$ and define \emph{states} and
\emph{observables} in terms of~$\mathscr{H}$.  Time evolution is a unitary
1-parameter group acting on~$\mathscr{H}$ generated by a self-adjoint operator, the
Hamiltonian.  An alternative set of axioms for \emph{algebraic quantum
mechanics} was put forth by von Neumann and Irving Segal~\cite{ISegal}.  Here
one focuses on observables, which form a $C^*$-algebra~$\mathfrak{A}$.
States are then positive unital functionals $\mathfrak{A}\to \mathbb{C}$.
These frameworks for quantum mechanics are mathematically rigorous, and in
turn quantum mechanics was a main impetus for the development of much of this
mathematics, such as the theory of operators.
 
Quantum field theory (QFT) has long resisted such a clean mathematical
treatment. There are, of course, several traditional approaches in the
physics literature.  These include: perturbing away from free field theories,
which can be defined via an action principle; and lattice theories, which are 
particularly well-suited to computations at strong coupling.  Renormalization
theory is a key ingredient in all of these frameworks. In this
section we focus on mathematical axiom systems for QFT.  We remark that each 
known approach to QFT has important limitations.  
For essays on general issues at the foundations of QFT, see \cite{conceptual}.

The first attempts at an axiomatic formulation of QFT take place in flat
Minkowski spacetime~$\mathbb{M}$.  The \emph{Wightman axioms}~\cite{PCT}
postulate a unitary representation of the Poincar{\'e} group on a Hilbert
space~$\mathscr{H}$ as well as a field map which assigns (possibly unbounded)
operators on~$\mathscr{H}$ to test functions on~$\mathbb{M}$.  Roughly, these
axioms specialize to the Dirac-von Neumann axioms in case $\dim\mathbb{M}=1$.
The alternative \emph{Haag-Kastler} axioms~\cite{HaagKastler} for QFT
on~$\mathbb{M}$ are in the spirit of algebraic quantum mechanics.  They
assign a $C^*$-algebra~$\mathfrak{A}(U)$ (of observables) to each open subset
$U\subset \mathbb{M}$ with compact closure, compatibly with the Poincar{\'e}
group and causality.  These axiom systems---though incomplete and not universally applicable---are the foundation and inspiration for much
mathematical work in QFT.  
 
Before we turn to modern axiom systems for QFT, we list some of the key
issues one confronts in formulating axioms:

\begin{enumerate}

\item Some QFT's have no action principle and no obvious
``fundamental'' degrees of freedom.

\item Some QFT's have many action principles, with completely
different ``fundamental'' degrees of freedom.

\item Even when there is an action principle, interacting QFT's with
running coupling constants are not clearly and rigorously mathematically
defined.

\item S-matrix amplitudes exhibit remarkable properties making startling connections
to mathematical fields not traditionally related to physics. Moreover, they  encode
physical properties such as locality in highly nontrivial ways.

\item The universe of  ``theories'' is \emph{not} a disjoint union of string
theory and QFT. From AdS/CFT we believe that some field theories are equivalent
to string theories. On the other hand, there are other ``theories,'' which are
neither local quantum field theories nor traditional string theories (with gravity).

\item Many field theoretic phenomena have \emph{geometrical reformulations},
reducing highly nontrivial facts of field theory to simple geometrical
constructions.  See section \ref{sec:GeoQFT} for more discussion. 

\item A fully local theory should associate physical quantities to subspaces
of all codimension and have coherent gluing rules for gluing of such
subspaces.

\item Even when there is an action principle, a QFT is not completely defined
by an action and a list of local operators. One must include (higher
categories of) defects of all dimensions, in accord with the previous point. 

\end{enumerate}

In the mid 1980's Graeme Segal introduced a novel axiom system for
two-dimensional conformal field theory~\cite{SegalCFT}.  In this telling a field theory
is a functorial assignment of a topological vector space to a circle and of a
trace class (nuclear) linear map to a 2-dimensional bordism endowed with a
conformal structure.  In brief: a 2d conformal field theory is a linear
representation of the bordism category of Riemann surfaces.  As opposed to
the Wightman axioms, which model QFT on flat Minkowski spacetime, the Segal
axioms model \emph{Wick-rotated} QFT on compact Riemannian manifolds.
Significantly, while they were originally developed for two-dimensional
conformal field theories, the Segal axioms are now understood to apply to quantum field
theories in all dimensions.  For example, they apply to topological field theory,
where the explicit relationship to classical bordism was flagged by Michael Atiyah~\cite{Atiyah}.
Within the past year Segal, together
with Maxim Kontsevich, posted a preprint \cite{Kontsevich:2021dmb} which takes up
the axiom system in the general QFT framework.  Furthermore, they prove an
analog of the Osterwalder-Schrader theorem: they construct a QFT on globally
hyperbolic Lorentz manifolds from a QFT on Riemannian manifolds in which
positivity of energy is encoded by having the theory defined on manifolds
with certain complex metrics.  The Segal axioms have provided an inroad into
QFT for mathematicians of all stripes.  Much remains to investigate,
particularly analytic aspects, and this is a fertile ground for future
investment. An interesting application of some observations of \cite{Kontsevich:2021dmb} to   
quantum gravity appeared in \cite{Witten:2021nzp}.

In the past decade Kevin Costello and Owen Gwilliam developed an axiom system
for Wick-rotated QFT which is an analog of the Haag-Kastler axioms in flat
Minkowski spacetime~\cite{Costello:2016vjw,CostelloGwilliam2}.  They define
\emph{factorization algebras} on smooth manifolds, modeled after the
\emph{chiral algebras} of Beilinson-Drinfeld~\cite{BeilinsonDrinfeld}.  The
latter were motivated by two-dimensional conformal field theory, whereas the
former are meant to apply to QFT in all dimensions.  A factorization
algebra~$\mathcal{F}$ on a smooth manifold~$M$ assigns a chain
complex~$\mathcal{F}(U)$ to each open subset~$U\subset M$, and the assignment
satisfies a multiplicative version of duals to the usual sheaf axioms (i.e.,
it satisfies a multiplicative cosheaf axiom).  Many examples have been
constructed using Costello's previous work on perturbation theory.  
Here is a very small quirky
sample of applications of these ideas to physical theories: 
\cite{CoLi,MR3869646,MR3869645,Costello:2017dso,MR4310891, Koszul2021,Costello:2022wso}.  This
framework has opened new mathematical avenues into modern aspects of QFT,
such as the AdS/CFT correspondence.  Again there is much for the future that
is worthy of robust support.

For further discussion of axiomatic approaches to defining QFT, together 
with an ample set of references to the older literature, see the recent 
Snowmass whitepaper \cite{Dedushenko:2022zwd}. 

\bigskip\bigskip

\subsection{Topological Quantum Field Theory}\label{sec:TQFT}

From the mathematical point of view, the best established QFT's are the
topological quantum field theories (TQFT). These have been put on a firm
footing and capture some of the essential aspects of locality. One way
of taking the idea  of locality  to its logical conclusion  is the
notion of a fully extended (a.k.a.~ ``fully local'')
 TQFT based on symmetric monoidal $n$-categories, where $n$ is the dimension
of spacetime in a physical theory
\cite{FreedHigher,Lawrence,BaezDolan,Lurie,Freed:2012hx,Teleman-Berkeley}.
In recent years there has been a vigorous development of this topic and
mathematicians have extended these ideas in very sophisticated ways; a small
sample is~\cite{AyFr,CaSch}. (There is a recent discussion of fully local non-topological field theories in~\cite{GP}.)

Although the subject has matured, and the language of TQFT is a commonly
used and powerful tool in mathematics, there remain many important open problems in the subject.
Many bordism categories of physical interest (e.g. including various kinds of tangential structures)
remain relatively unexplored. Other extensions under exploration include  equivariant, nonunitary, and nonsemisimple TQFT's.  The study of defects and boundary conditions has much future potential.    TQFT's have
 also proven to be quite powerful in addressing questions about anomalies in field theory
 and topological phases of matter, and we will comment further on that below.

Many important classification questions remain unanswered. The
 classification of deformation types of invertible TQFT's \cite{Freed:2016rqq} gives some hope that
further and more general classification questions are not completely
out of reach.  An example of a long-standing open question is the classification of 3d unitary
TQFT's. An old conjecture \cite{Moore:1989vd} posits that they are all associated with
3d Chern-Simons-Witten theory, once one has chosen a suitable level and compact gauge group.
(Closely related to this, the modular tensor categories (MTC's)  of rational conformal field theories 
(RCFT's) should all be built by using the
Wess-Zumino-Novikov-Witten model \cite{Witten:1983ar} using a small set of standard
constructions.) There are recent interesting challenges to this conjecture
\cite{Evans:2010yr,EvansGannon-Tambara,Huang:2021nvb,Vanhove:2021zop} which should be further investigated.
In particular, Teleman has shown that the Haagerup MTC cannot come from 
a Chern-Simons-Witten theory with compact gauge group \cite{TelemanRemark}. It is important to bear in mind here the relevance of the notion of Witt equivalence
\cite{DavydovWitt1,DavydovWitt2}. 
One can ask whether the Haagerup MTC's are Witt equivalent to the theory with a compact gauge group. 
%
%
\footnote{Witt equivalence has recently been finding physical applications of fusion and modular categories. See, for example \cite{Freed:2020qfy} where it is used to discuss the existence of topological boundary conditions in three-dimensional topological theories or 
\cite{Geiko:2022qjy} where it is applied to the question of time-reversal symmetry in 3d Chern-Simons theory. }

Ironically, the first topological field theory to be discovered, 
Donaldson-Witten theory \cite{Witten:1988ze},
does not obey the official definition of a TQFT. There are difficulties making sense of the instanton Floer theory on arbitrary $3$-manifolds, and it is certainly not known how to define a fully extended TQFT corresponding to the Donaldson invariants.
(It certainly cannot be unitary \cite{Freedman:2005eqs} or semisimple.)
In a similar vein the  GL-twisted 
${\CalN}=4$ SYM \cite{Kapustin:2006pk}   has been extensively developed and intensively
studied, but it is not known how to define the topological theory on arbitrary four-manifolds.
Similar remarks apply to three-dimensional Chern-Simons theories with noncompact gauge groups. 
These are examples of theories that are only \emph{partially defined}, they have not been 
defined as fully extended (a.k.a. fully local) TQFTs. It is not known if a fully extended 
version exists, or if there are fundamental obstructions to such an extension. Clearly, there remain fundamental aspects to be understood even in some of the most basic examples of what physicists regard as topological quantum field theory. 

There are other examples of such partially defined TQFT's where, 
in fact, we do know of deformations or regularizations that still lead to tractable theories. One good example is a two dimensional relative of Donaldson theory, the topological Yang-Mills theory in two dimensions. It has an infinite set of states, the torus partition function is divergent and only a restricted set of correlation functions makes sense. However, it has 
several well-known regularizations: the gauged $G/G$ WZW theory \cite{Blau:1993tv,Gerasimov:1993ws, Witten:1993xi}, and a finite coupling two-dimensional Yang-Mills theory \cite{Moore-ICM,Witten:1992xu}. The latter is not, strictly speaking, a topological theory, as it depends on a choice of a measure, and the Wilson loop observables, although quite acceptable in the Yang-Mills theory, are not, strictly speaking, topological - they only depend on the areas of the components of the surface on the complement of the 
Wilson lines \cite{Cordes:1994sd,Gross:1993hu}. Another interesting ``deformation'' is $q$-deformed 2d YM and further generalizations which have shown up in a variety of contexts 
\cite{NikThesis,Aganagic:2004js,Gadde:2011ik,Rastelli:2014jja,Tachikawa:2015iba}.

\subsection{Algebraic Structures In Quantum Field Theory}

\subsubsection{Algebraic Structures Related To Operator Product Algebras}

The notion of an operator product expansion indicates that
algebraic structures of some sort play a fundamental role in
quantum field theory. This notion has been formalized and
generalized in many ingenious ways and the investigation of
algebraic structures in QFT has been a dominant and enduring theme of
research   and promises to continue
being an important topic.

The study of two-dimensional conformal field theory
led to the   mathematical study of vertex operator algebras (see e.g. \cite{Goddard:1986bp,Borcherds,FLM}),
a notion with a clear mathematical definition
which nicely captures and makes precise constructions which had previously been used in the physics literature.  Today, vertex operator algebra (VOA) theory is a well-established topic of mathematics
but many interesting open problems remain in this field. There is still much to learn about the representation categories of VOA's. For example, one natural question is whether there is an analog of Tannaka-Krein reconstruction.  In another direction, one of the beautiful recent results in traditional VOA theory is a complete classification
of the holomorphic VOAs of central charge 24
\cite{Moller:2021clp,Schellekens:1992db,Schellekens:1993va,vanEkeren:2017scl}. Another major gap in
our knowledge is any semblance of a classification of holomorphic
vertex operator algebras of central charge greater than 24. This
appears to be a very challenging problem, even more challenging than finding
a useful classification of even unimodular lattices. Virtually nothing is known
about what the general holomorphic CFT looks like. It is natural to speculate that there is an analog of the 
Smith-Minkowski-Siegel mass formula for holomorphic CFTs, but, so far as we know,  no work on this idea has been published. 

Remarkably, VOA's and structures similar to VOA's of various kinds have been discovered in a
wide variety of new contexts in higher dimensional QFT.
One generalization of the subject of vertex operator algebras is
the entire  framework of factorization algebras \cite{Costello:2016vjw}, as discussed in \S2.1.
(These generalize the $E_d$ algebras relevant to TQFT.) 
An important open problem is to connect these ideas to more traditional
ideas of quantum field theory based on Wightman and Haag-Kastler
axioms. The work \cite{Carpi:2015fga} discusses the relation between VOA and the Haag-Kastler axioms, and should provide a first step in this direction. 
It would also be important to understand better the relation
to the recent work \cite{Kontsevich:2021dmb}. Another rich source of nontrivial
algebraic structures are the supersymmetric theories discussed in the next section.

\subsubsection{Algebraic Structures Associated With Supersymmetric Sectors And BPS States}

Using supersymmetry and defects one can define important subsectors
of the full operator product algebra of certain quantum field theories.
This has been done with great success in the past ten years.
Examples include the chiral algebras appearing in $d=4, {\cal{N}}=2$
supersymmetric field theories \cite{Beem:2013sza,Beem:2014rza,Beem:2017ooy}.
Another beautiful example are the algebraic structures associated with holomorphic sectors of supersymmetric theories \cite{Johansen1995,Clash1997}, 
and with  hybrid holomorphic-topological twists of supersymmetric theories
with extended supersymmetry \cite{NikThesis, NikBaulieuLosev, Kapustin:2006hi}.  
This is currently a very active direction of research with many potential applications 
\cite{Aganagic:2017tvx,Costello:2017dso,Costello:2020ndc,Gwilliam:2021zkv,Oh:2019mcg,Garner:2022its}.

Another class of algebraic structures are the generalizations of
Gerstenhaber algebras that appear in TQFT's of cohomological type
such as those studied in \cite{Beem:2018fng,Getzler:1994yd,Lian:1995wa}, leading to connections to BV algebras and homotopical algebras.

Closely related to the above are algebraic structures associated to BPS
states.  One early hint that spaces of
BPS states should have an interesting algebra-like structure was inspired by the 
work of Nakajima \cite{Nak94,Nakajima-Heisenberg} and is described in \cite{Vafa:1994tf}. 
This led to various proposals for algebraic structures associated to BPS states in
\cite{Avatars,Harvey:1996gc,NikRome:2009,Kontsevich:2010px}. 
One curious application 
of these ideas is a   very suggestive approach to the
important problem of finding a conceptual basis for understanding
the role of genus zero groups in monstrous moonshine
\cite{Harrison:2021gnp,Paquette:2016xoo,Paquette:2017xui}.
More recently, the subject of quiver algebras associated to
BPS states on D-branes has just begun to be developed
\cite{Li:2020rij,Galakhov:2020vyb,Galakhov:2021xum,Galakhov:2021vbo,Noshita:2021ldl,Noshita:2021dgj},
giving a new physical interpretation of some old constructions of
Nakajima and Kontsevich-Soibelman with many promising directions
for generalizations and further development.

In a parallel development, an in-depth study of the mathematical
structures of BPS states in two-dimensional QFT with $(2,2)$ supersymmetry
has revealed a very rich mathematical structure similar to, but different
from, topological string theory. This involves the 2-category of theories,
interfaces, and boundary-condition-changing operators,  and involves various constructions in
homotopical algebra \cite{Gaiotto:2015aoa,Gaiotto:2015zna,Kapranov:2014uwa,Kapranov:2020zoa,Khan:2020hir}. For yet another set of parallel developments see the remarks at the end of the section \ref{sec:Defects}. 
This structure was developed in an effort to understand 
the ``categorification of wall-crossing formulae.''
Indeed these papers have explained a categorical version of the 2d Cecotti-Vafa wall-crossing formula 
\cite{Cecotti:1992qh,Cecotti:1992rm}, and of Stokes' phenomenon in general.
But the extension to four dimensions has remained elusive and is an important open problem.
In \cite{Kapranov:2014uwa} the ``web-formalism'' of \cite{Gaiotto:2015aoa,Gaiotto:2015zna} was
reinterpreted in terms of polygons. This led to a higher-dimensional generalization in terms of
polytopes. It might be very interesting to find a physical interpretation of the higher-dimensional 
version of the web-formalism developed in \cite{Kapranov:2014uwa}. Many of the above developments 
were motivated - in part - by the desire to find a categorification 
of the full $2d4d$ wall-crossing formula. This goal remains elusive, and is surely a 
good direction for future research. 

For a recent overview of $L_\infty$ structures in field theory see \cite{Hohm:2017pnh}.

\subsection{Classification And Deeper Understanding Of New Classes Of Supersymmetric Field Theories}

A well-known and very important problem in physical mathematics is
finding a satisfactory formulation of the six-dimensional superconformal
field theories with $(2,0)$ supersymmetry. Perhaps it should be formulated
axiomatically (an attempt was made in section 6 of  \cite{MooreKleinLectures}).
Or  perhaps there is some formulation in terms of some local
degrees of freedom, perhaps by making sense of a ``nonabelian 2-form gauge potential.'' There are many attempts in the literature taking the latter approach. For two examples see  \cite{Lambert:2010wm,Lambert:2010iw,Lambert:2012qy} and 
\cite{Baez:2004in,Fiorenza:2012tb,Fiorenza:2013kqa,Samann:2016ksp}.
String theory constructions, which led to the discovery of these theories
 \cite{Seiberg:1997ax,Strominger:1995ac,Witten:1995zh},
  strongly indicate we know what the examples are:   
  The basic components are made
of ``Abelian'' theories, and nonabelian theories classifed by an ADE classification.
Just assuming the theories exist has led to a host of very fruitful mathematical
constructions shedding light on hyperk\"{a}hler geometry, Hitchin systems, exact WKB theory, and low
dimensional topology, among other things.

One important aspect of the 6d theories are the various supersymmetric
defects they possess. Among these the half-supersymmetric codimension two defects are
among the most important. Their basic classification 
has been investigated in, for examples,
\cite{Balasubramanian:2018pbp,Balasubramanian:2020fwc,Chacaltana:2012zy}.
These defects have an interesting operator product structure whose ``structure constants'' are themselves four-dimensional quantum field theories  \cite{Gaiotto:2011xs}. Clearly they deserve continued study.

It would also be of great interest to understand the full range of possibilities
and map out the geography of four-dimensional theories with ${\CalN}=2$ supersymmetry. An overview of the current status is provided in the Snowmass paper \cite{Argyres:2022mnu}. 
There are three broad classes of known theories: 

\begin{enumerate} 

\item One class of such theories have 
Lagrangians ${\CalN}=2$ based on semi-simple Lie groups 
 and hypermultiplets. All such UV-complete theories 
are classified in \cite{Bhardwaj:2013qia}. 

\item Another class (which partly overlaps with the previous one) are the theories obtained as limits of ADE $(0,2)$ superconformal theories 
on spacetimes of the form $C \times {\BR}^{4}$ where $C$ is a (possibly punctured) curve. 
These are the class $\CalS$ theories  \cite{Gaiotto:2009we,Gaiotto:2009hg,Klemm:1996bj,Witten:1997sc}. (For some reviews see \cite{Tachikawa:2013kta,Akhond:2021xio,MooreKleinLectures}.) 

\item Geometrically engineered theories and models engineered using brane constructions. 
For the current status of these see section
\ref{sec:GeoQFT}.

\end{enumerate} 

Within each of these classes quite a lot is known. 
Some of the Lagrangian theories can be realized as the low energy limits of the theory on a stack of $D3$-branes probing an orbifold singularity, the quiver gauge theories \cite{Douglas:1996sw}. The requirements of asymptotic freedom constrains the choices of gauge groups
and matter multiplets of quiver theories. For example, if the gauge group is a product of (special) unitary groups, then the only choices
of quivers are the Dynkin diagrams of simple simply laced Lie algebras or affine Lie algebras.\footnote{ One can have also a more exotic possibility of finding a Lagrangian description for class ${\cal S}$ theories which manifests only ${\cal N}=1$ supersymmetry (see \cite{Maruyoshi:2016aim,Razamat:2019vfd,Zafrir:2019hps} for some examples). We will say more about this in Section \ref{sec:GeoQFT}. } In the latter case the gauge group and matter content theory is uniquely determined up to a single integer $N$. For such theories, possibly mass deformed in an ${\CalN}=2$ supersymmetric way, the Seiberg-Witten geometry has been computed in \cite{Nekrasov:2012xe} using the exact quantum field theory calculations
performed using localization on a suitably (partly) compactified moduli space of quiver instantons. The result of the limit shape calculation
identifies the Coulomb branch with a base of an algebraic integrable system whose phase space is a moduli space of either the ADE instantons of charge $N$
on ${\BR}^{2} \times T^{2}$, or the ADE monopoles with Dirac singularities on ${\BR}^{2} \times S^{1}$.
(Two pedagogical reviews of limit shapes are \cite{KenyonLimitShape,OkounkovLimitShapes}.)
In these results the charges are determined by the ranks of gauge groups and the numbers of fundamental hypermultiplets. The results are in broad  agreement with   earlier
conjectures based on string theory considerations \cite{Katz:1997eq,Cherkis:2000ft,Friedman:1997yq}, although some of the earlier conjectures require modification. The theories with eight supercharges, appropriately lifted to $k+4$ dimensions 
on a spacetime of geometry ${\BR}^{4} \times T^{k}$
lead to the moduli spaces of instantons (monopoles) on ${\BR}^{2-k} \times T^{2+k}$ (${\BR}^{2-k} \times T^{1+k}$), respectively \cite{Nekrasov:2012xe}. (In the lifted theories one can add Chern-Simons couplings and/or encounter new anomalies.) 

One might wonder if the problem simplifies if one tries to classify theories with a 
large $N$ limit using the AdS/CFT correspondence. 
 The affine type quiver theories were previously studied, in the large $N$ limit, in the context of the AdS/CFT correspondence, 
in \cite{Kachru:1998ys,Lawrence:1998ja}, 
while the large $N$ type $\CalS$-class theories were studied in \cite{Gaiotto:2009gz}. (The holographic picture gives an interesting
view on the renormalization group flow, which is sometimes possible, in the $\CalN=2$ case, to connect to the field theory computations \cite{NikitaKlebanov:2000}. This is a nice example of the geometrization of QFT discussed in more detail in section \ref{sec:GeoQFT} below.)

\subsection{Generalizations Of The Notion Of ``Symmetry'' }

Many of the most important developments in 20th century
physics, including relativity, quantum mechanics, and
gauge theory make fundamental use of the notion of symmetry.
In recent decades the notion of symmetry has been further enhanced
and extended and applied in surprising new ways. Further generalizations
of symmetry will surely be a theme for several years to come.

One example of an important new notion is that of ``generalized $p$-form symmetries.''
In various forms these have been used for a long time. For example the use of
``center symmetry'' played an important role in the early studies of phases in
nonabelian gauge theories \cite{PolyakovBook,tHooft:1979rtg}.
  In   \cite{Freed:2006yc} they played an important role in defining flux sectors
in arbitrary Abelian gauge theories (including self-dual theories)
based on generalized cohomology theories.
 In an important advance,
``generalized $p$-form global symmetry'' was formulated in terms of topological defect operators  \cite{Gaiotto:2014kfa} where 
these defect operators were put to
very good use. In the case of two-dimensional rational conformal field theories, defects as symmetries---and duality defects---appeared a decade earlier in~\cite{MR2115752}.
A hint on generalized symmetry and topology also appeared in \cite{Nussinov:2006iva,Nussinov:2009zz}. New insights into instanton effects, even in
textbook examples of quantum mechanics  were found and striking
impliciations for QCD at $\theta=\pi$ were pointed out \cite{Gaiotto:2017yup}.
Since then there has been widespread use of such generalized
symmetries. Combined with the theory of anomalies striking
applications have been made to determine dynamics of nontrivial
theories \cite{Gaiotto:2017tne,Cordova:2019uob,Cordova:2019jnf,Cordova:2019jqi,Cordova:2019bsd, Bhardwaj:2022dyt, Delmastro:2022pfo, Brennan:2022tyl}.
Motivated, in part, by these developments there has been some
exploration into the notion of ``two-group symmetries''  \cite{Cordova:2018cvg,Cordova:2020tij}
and these works clearly point the way to potentially interesting future developments.

Further developments along these lines have   tried to generalize even
further the notion of symmetry in terms of ``categorical symmetries'' and ``noninvertible symmetries.''  Algebras of symmetries are common in both mathematics and quantum theory, and algebras include noninvertible elements.  In mathematics various categorified generalizations of algebras are being developed.  For example, tensor categories, for which there is a robust theory~\cite{Etingofbook}, are a once-categorified version of an algebra, and this higher algebra underlies much research on Reshetikhin-Turaev-Witten and Turaev-Viro models.  There are also higher categorical versions, including ``$E_n$-algebras,'' which enter topological field theory and are destined to find applications beyond the topological case.  The whole subject and its application to physics is under active development.  Certainly there is no settled definition of what ``categorical symmetry" should mean in field theory. Roughly speaking, a categorical symmetry involves topological defects with sums of simple objects in
their fusion rules (in a semisimple setting) and couplings of $n$-dimensional theories to $n+1$-dimensional topological field
theories.
A Simons Collaboration is devoted to exploring these ideas further \cite{SimonsCollab:CategoricalSymmetry}. 
Some of the recent developments are summarized in the Snowmass contribution \cite{Cordova:2022ruw}. 
From a mathematical point of view the emergence of categorical structures in QFT is particularly noteworthy, with most recent developments in this direction in  
\cite{Ji:2019jhk,Komargodski:2020mxz,Kong:2020cie,Thorngren:2019iar,Thorngren:2021yso, Bhardwaj:2022yxj, Bhardwaj:2022lsg, Bartsch:2022mpm, Lin:2022dhv, Freed:2022qnc}. 
An in depth discussion of generalized categorical symmetries in the context of QFTs, string theory and holography can be found in sections \ref{sec:Defects} and \ref{sec:GeoQFT}.
Although not phrased as a ``symmetry'' very closely related ideas had been previously explored in
\cite{Brunner:2014lua,Chang:2018iay,Lin:2019hks}.

In another direction, the papers \cite{Banks:2010zn,Harlow:2018tng,Harlow:2018jwu}
have explored deeply the role of symmetry in black hole physics, quantum gravity
and the  AdS/CFT correspondence. It would be interesting to combine the
ideas of these papers with some of the emerging ideas on global categorical
symmetry. Some early developments in this direction can be found in
\cite{Heidenreich:2021tna} and in holography/string theory in \cite{Apruzzi:2022rei, GarciaEtxebarria:2022vzq}. See also section \ref{sec:Defects}.

Another very important emerging set of ideas concerns asymptotic symmetries in gauge theories.
These were, indeed the basis for an early version of the AdS/CFT conjecture \cite{Brown:1986nw}
and recently they have been much studied in the context of BMS symmetries and the
infrared structure of massless gauge theories \cite{Strominger:2017zoo}. Related
ideas concerning a hypothetical ``celestial conformal field theory'' holographically dual
to gravity in flat space have been discussed in a number of papers  \cite{Raclariu:2021zjz}.
The existence of such a celestial conformal field theory
would be quite remarkable since it is far from obvious why the Mellin transforms of scattering amplitudes should behave like the correlators of a local quantum field theory.

Finally, the Batalin-Vilkovisky formalism gives yet another, perhaps not unrelated, generalization of symmetry. The conventional Lie algebraic symmetry corresponds to a homomorphism ${\CalV}$ from some Lie algebra $\mathfrak{g}$
to the algebra $Vect(M)$ of vector fields on some (super)manifold, e.g. a space of fields (more generally, the space of fields and anti-fields in the BV sense). The classical action $S_{0}$, which is a function on $M$ is invariant under the ${\CalV}({\mathfrak{g}})$-action. It means that it can be extended to a function $S$ on $M \times \mathfrak{g}^{*} \oplus {\Pi}{\mathfrak{g}}$ in such a way, that the $\mathfrak{g}^{*}$ dependence of $S$ is linear ($\Pi\mathfrak{g}$-dependence is at most quadratic), and $S$ solves the classical master equation $\{ S, S \} = 0$. The extension of $S_0$ to $S$ corresponds to the familiar BRST formalism (except that BV allows for both gauged and global symmetries to be treated in a similar way). The $L_{\infty}$-symmetry is a non-linear formal generalization of this construction, where the requirement of linearity in anti-fields for ghosts  is dropped. The need for such generalization is provided by the renormalization group, as integrating out the charged degrees of freedom, produces the above-mentioned non-linear terms, \footnote{We thank A.~Losev for illuminating discussions on this point}. For related ideas see  \cite{Verlinde:1992qa}.

\subsection{The Space Of Quantum Field Theories And The Renormalization Group}

It has been noted on many occasions that one of the main barriers
to a more productive dialogue between physicists and mathematicians
is a rigorous understanding of scale transformations and the renormalization
group. Some important progress - in the perturbative regime - was made in
\cite{Costello:Renormalization}, which makes essential use of the 
Batalin-Vilkovisky formalism. Another, quite different approach to 
renormalization theory makes use of ideas of noncommutative geometry, 
and phrases renormalization in terms of a Birkhoff factorization problem 
in a suitable Hopf algebra \cite{Connes:1998qv,Connes:1999yr,Connes:2000fe}. 

In spite of these and other approaches,   a full nonperturbative
treatment remains open. Indeed, it is the subject of one of the Clay Millenium
Problems.

Although we do not have a completely satisfactory and universal
definition of what should be meant by a ``quantum field theory,''
it might not be too early to ask what can reasonably be said about the full space
of quantum field theories - although it must be admitted that there is no
really rigorous definition of the ``space of quantum field theories''.
Nevertheless, physicists have some intuitive pictures based on Wilsonian
ideas and   the crucial concept of renormalization group flow.

While understanding the space of 1d QFT's presents some interesting
issues the first really nontrivial case is the case of two-dimensions.
This has long been recognized as a very important question for
fundamental physics \cite{Douglas:2010ic,Friedan:2002aa}. Here there are
well-known and profound results on renormalization group flow relating it
to Einstein's equations \cite{Callan:1985ia,Friedan:1980jm}. In addition
to this there is the remarkable Zamolodchikov c-theorem \cite{Zamolodchikov:1986gt}.
An important analog appears in the case of conformal field theories
with a non-conformal boundary theory \cite{Affleck:1991tk,Kutasov:2000qp,Friedan:2003yc}.
This progress grew out of an interesting approach to background-independent
string field theory \cite{Witten:1992qy,Gerasimov:2000zp,Kutasov:2000qp,Kutasov:2000aq}.
In this approach one considers 
a two dimensional conformal field theory on a surface with boundary. The two-dimensional ``bulk'' is 
conformal but one allows boundary operators and boundary conditions that break the conformal invariance.  
The RG does not change the conformal 
field theory in the ``bulk'' but does alter the boundary conditions. The ``space'' of 
such theories  (with a fixed conformal theory in the two-dimensional bulk)
is an interesting example of a space of theories might be more tractable than the general case of all two-dimensional theories. It is possible
one can say something about the space of components of such a ``space of
quantum field theories'' using ideas from K-theory, an idea which has
not been very deeply explored, although it was mentioned in \cite{Moore:2003vf}.

 In the mathematical realm
a very interesting proposal for a space of $2d$ quantum field theories
with $(0,1)$ supersymmetry was made by Stolz and Teichner \cite{StolzTeichner1,Stolz:2011zj}.
The Stolz-Teichner program grew out of attempts to extend the ideas of Graeme Segal
on elliptic cohomology to a theory conjecturally equivalent to the theory of
topological modular forms. The subject has recently gotten a big boost
from several recent works applying it to interesting questions in
quantum field theory \cite{Gaiotto:2018ypj,Gaiotto:2019asa,Gaiotto:2019gef}
and anomaly cancellation in heterotic string theory \cite{Tachikawa:2021mvw,Tachikawa:2021mby,Yonekura:2022reu}. There is clearly a lot more to be learned in this subject. For example, the
generalized cohomology theory of topological modular forms (tmf) has a $576$-fold periodicity,
analogous to the   $8$-fold periodicity of Clifford algebras and real K-theory.
A clear physical explanation of this remarkable aspect of tmf might be highly interesting.

Moving on to higher dimensions, there are conjectural analogs of
the Zamolodchikov theorem, and there has been remarkable progress
on this topic \cite{Komargodski:2011vj,Elvang:2012st}. Moreover
it is related to questions of quantum information theory
\cite{Casini:2006es,Casini:2012ei,Casini:2015woa,Casini:2016fgb,Casini:2017vbe}.
Some bold conjectures about the global nature of the
space of conformal field theories were made by Seiberg in \cite{SeibergConjecture}. It remains
a challenge to see of any of the above ideas can shed light on Seiberg's
conjectures.

Some of the issues here are closely related to those touched on in the
condensed matter theory section \ref{sec:CondMat}.

\subsection{Fundamental Formulation Of Non-Lagrangian Superconformal Field Theories}

As was mentioned above, an important long standing open problem is the formulation of
the  six-dimensional superconformal field theories with $(2,0)$ supersymmetry. A key obstacle
here is the proper understanding of the formulation of self-dual theories. Note that these
include the important sub-case of chiral scalar fields in $1+1$ dimensions. The subject
has a long and extensive history.

Even the free self-dual Abelian theories - often dismissed as trivial - have not received a complete
and rigorous formulation as fully local (aka fully extended) field theories.
There is a long history of attempts to write action principles for 
these theories. 
See 
\cite{Andriolo:2020ykk,Andriolo:2021gen,Belov:2006jd,Belov:2006xj,
Sen:2019qit} and the many references therein. There is still no 
universally agreed-upon approach to defining these theories via an 
action principle. Moreover, even in the   Abelian case there are many subtle topological issues some of which 
 are discussed in \cite{Alvarez-Gaume:1987wwg,Alvarez-Gaume:1986nqf,
 Belov:2006jd,Belov:2006xj,Freed:2006yc,Freed:2006ya,
 Hopkins:2002rd,Hsieh:2020jpj,Witten:1996hc}.
 Finding a complete and definitive treatment of self-dual Abelian gauge fields based on an arbitrary self-dual generalized cohomology
theory remains an important open problem. These are free theories, so, although the topic is subtle and difficult,
there is no insuperable obstacle to giving such a treatment. And there is really no excuse for not doing so. 
To illustrate the incompleteness of our understanding, we remark that the anomaly theory of the basic self-dual field is constructed in~\cite{Hopkins:2002rd}, where an integral lift of the middle Wu class is used.  We inquire: What are the physical consequences of this choice?  For example, dimensional reduction of the 6-dimensional self-dual field to 4~dimensions yields Maxwell theory, and we can ask how this choice manifests there.
%
%
%

 For generic choices of compactification parameters the class ${\cal S}$ theories \cite{Gaiotto:2009we,Gaiotto:2009hg,Klemm:1996bj,Witten:1997sc} do not possess a known Lagrangian description. On the other hand, the semi-simple gauge group Lagrangians with manifest ${\CalN}=2$ supersymmetry and hypermultiplet matter, describing an interacting SCFT are classified in \cite{Bhardwaj:2013qia}.
 An interesting question is to understand whether Lagrangians manifesting less symmetry can be constructed which will flow to a generic ${\CalN}=2$ theory, i.e. a class $\CalS$ theory. This question can be generalized to whether any conceivable SCFT has a description in terms of weakly coupled fields. The reader will find more on this question in section \ref{sec:GeoQFT}.

\subsection{Nonperturbative Effects, Resurgence, Stokes Phenomena, And Exact WKB Methods}

\label{subsec:nonpert}

If QFT is understood in terms of the Feynman path integral, with Planck constant $\hbar$ measuring the strength of the action in the exponential, the behavior of physical quantities as analytic functions of $\hbar$ should be at least as complicated as that of an exponential integral. In a nice analytic situation the latter is a vector-valued function of parameters, the basis
being given by the periods of the corresponding form over the submanifolds known as Lefschetz thimbles. A critical point of the action may be in the
complex domain. The conventional perturbative expansion takes into account one, typically real, classical solution, i.e. the one that
asymptotes to some number of in-coming and out-going particles. The sum of loop corrections almost never converges, e.g. by Dyson's argument \cite{Dyson:1952tj}.  

However, this divergence is already present in finite dimensional exponential integrals, so one can try to isolate the divergence of the perturbative expansion of an exponential integral within that of a path integral. So far this wish was partly granted. One widely used technique is 
Borel resummation, which is a way to trade the factorially growing sum over diagrams for an exponential integral of a rational form. For some nice expository accounts of these ideas see  \cite{ArnoldGZVarchenkoV2,Marino:2012zq,Witten:2010cx}.

 The development of this idea leads to the notion of a trans-series and resurgence, see \cite{Pham1,Pham2,MR670417,MR670418,MR852210}, and \cite{Aniceto:2018bis} for a modern review. One class of problems on which these methods were tested since the mid-1970's \footnote{In works of e.g.
 Balian–Bloch, Dingle, Leray, Sibuya, Zinn-Justin.}, are the spectra of the one-dimensional Schr{\"o}dinger operators $- {\partial}^{2}_{q} + U(q)$. The convenient choices of 
 the potentials $U(q)$ are polynomials in $q$ (see \cite{Voros}
 for a beautiful discussion).

 The singularities of the Borel transform are, in finite dimensional examples, related to the critical points of the action, which, in turn, are telling us about the possible Lefschetz thimbles. To some extent this is confirmed in Lipatov's method
of evaluating large order perturbative contributions by a saddle point method \cite{Lipatov:1976ny}. However, it has been argued \cite{Hooft1977CanWM} there are singularities on the Borel plane not attributed to the instantons of \cite{Lipatov:1976ny}. Very little is known about these singularities - which are sometimes referred to as \emph{renormalons}. If a semiclassical approach is valid then it is possible that the missing critical points can be found in the properly understood complexification of the space of fields \cite{Nekrasov:2018pqq, NikitaKrichever1}. The practitioners are widely spread across the spectrum of optimism on this issue. For example it is not clear if renormalons can be approached semiclassically, as has been stressed by M.~Mari{\~n}o.

The plethora of exact computations made available by localization in supersymmetric gauge theories and topological strings, supplemented with dualities, mapping
$\hbar$ to some geometric parameters, gives
a huge playground for investigations of these ideas, including the ill-understood ``instanton-fractionalization'' \cite{Dunne:2012ae, Jeong:2017pai}, sometimes used to explain the renormalons.
(It would be desirable if the computations of 
\cite{Dunne:2012ae} could be phrased in a controlled 
truncation to a quantum mechanical problem.) 
It is also of interest to consider non-supersymmetric, 
asymptotically free, integrable theories in two dimensions such as $O(n)$ sigma models, the Gross-Neveu, model and so forth in this context. Recent results in this area 
\cite{Marino:2021dzn} indicate that some standard lore about renormalons  needs to be revised. 

The standard WKB method can be enhanced to give
exact results in nontrivial quantum systems using ideas explored
some years ago in \cite{Delabaere,Voros}. Again, embedding quantum systems in the supersymmetric context e.g. via Bethe/gauge correspondence gains some mileage. For example, the exact quantization conditions \cite{Jentschura:2004jg} were related \cite{Krefl:2014,Basar:2015xna}  
to equations \cite{Nekrasov:2009rc}  describing the vacua of the  $\Omega$-deformed four dimensional $\CalS$-class ${\CalN}=2$ theories, in the NRS coordinates\footnote{Sometimes called the complexified Fenchel-Nielsen coordinates in the literature, the NRS atlas \cite{NRS} of Darboux coordinates
interpolates between Fenchel-Nielsen, Goldman,  Kapovich-Milson, and Klyachko coordinate systems on various real slices and tropical limits of ${\CalM}_{\BC}$. They can also be viewed as a generalized system of Fock-Goncharov coordinates, associated to laminations with closed orbits.} 
on the moduli space ${\CalM}_{\BC}$ of flat $SL(2,{\BC})$-connections, cf. \cite{Teschner:2017djr}.

In an independent line of development, the study of wall-crossing and the counting of BPS states led to a new mathematical object, known as \emph{spectral networks} \cite{Gaiotto:2012rg} which generalizes exact WKB theory and leads to explicit constructions of BPS degeneracies of class S theories, as well as Darboux coordinates on moduli spaces of flat connections. These Darboux coordinates sometimes coincide with Fock-Goncharov and other cluster coordinates. The canonical line bundle on the spectral curve admits a flat connection whose holonomy provides these coordinates, which are thereby closely related to (and generalize) Voros symbols. Spectral   networks provide the data to give a converse   construction, namely a construction of the corresponding nonabelian flat connection from a flat connection on the spectral cover. These aspects of exact WKB theory were studied in  \cite{Gaiotto:2009hg,Gaiotto:2012rg,Gaiotto:2012db,Hollands:2013qza,Codesido:2017jwp,Hollands:2019wbr,Coman:2020qgf}, and more generally in \cite{kontsevich2021analyticity}.

The results of this development (in particular through the ``conformal limit and the relation to opers'' \cite{Dumitrescu:2016ick,Gaiotto:2014bza})
have implications beyond that of supersymmetric field theory, leading
to new insights in ordinary differential equations, such as the Schr{\"o}dinger equation. See \cite{Hollands:2021itj,Grassi:2021wpw,Yan:2020kkb} for a recent discussion. In particular \cite{Hollands:2021itj} has a useful review of several aspects of exact WKB analysis and spectral networks and the relation to opers. 
Very recently
the Borel summability of the WKB series in many quantum
problems has been rigorously established  \cite{Nikolaev:2021xzt}. 

The relation between Abelian and Nonabelian connections given by 
spectral networks has a $q$-deformed analog 
\cite{Galakhov:2014xba,Gabella:2016zxu}  
 which has been beautifully developed and extended in 
the recent works \cite{Neitzke:2020jik,Neitzke:2021gxr}. This subject should be further 
developed and should lead to useful insights about quantization of 
character varieties, connecting to the works \cite{fock2006moduli,quantumFG:2008,fock2009cluster} on quantum cluster varieties on the one hand, and the two \cite{Gukov:2008ve, Gaiotto:2021kma} and four dimensional  \cite{Nekrasov:2010ka} approaches to quantization of these spaces, on the other. There are also intriguing recent math
results on the representation theory of Skein modules
\cite{Bonahon-Wong} which deserve a physical explanation.
\footnote{Note added for v3: A physical explanation, and extension, of some of the results of Bonahan and Wong is given in \cite{Gaiotto:2023ezy}.}

The work of \cite{Gaiotto:2015aoa,Kapranov:2020zoa} can be interpreted as a categorification of Stokes phenomenon: The so-called $S$-walls are Stokes' lines. A 
construction of categorified parallel transport 
of a flat connection was given and 
the categorified $S$-wall factors in section 7.6 of \cite{Gaiotto:2015aoa} are categorified Stokes' factors. This categorification of exact WKB theory should be developed further. 

Considerations involving Stokes phenomena have also
lead to striking results about 3-manifold topology
in the context of complex Chern-Simons theory \cite{Gu:2021ize,Garoufalidis:2020xec,Witten:2010cx}.
See section \ref{subsec:Knot-3-Fold} below for more discussion about the relation to the topology
of three-manifolds and links.

\subsection{Defects}
\label{sec:Defects}

Traditional quantum field theory has been concerned with correlation
functions of local operators. However, going back to the work of Wilson on
lattice gauge theory it has been known that extended ``operators'' such as
Wilson lines (the trace of the holonomy of a gauge field in a representation)
play an important role in formulations and explorations of field
theory. These ``operators'' are more properly understood as ``defects,''
couplings of a quantum field theory to a lower-dimensional theory of fields
localized on a submanifold
of spacetime. Having phrased things that way, a host of generalizations emerge. 
Moreover, the concept of defects can be further broadened to include other 
local objects, and extended operators, defined by allowing singular behavior of fields in the neighborhood of specified subvarieites. In this way one can include disorder operators, 't Hooft lines and other generalizations such as 
surface defects, and higher dimensional defects.
 See \cite{Kapustin:2005py} for one careful attempt to give a definition of supersymmetric defect lines ('t Hooft-Wilson lines)  in ${\CalN}=4$ SYM. Likewise, a surface defect in a four-dimensional gauge theory with a gauge group $G$ could be described as a coupling of a two dimensional 
sigma model having $G$ as a global symmetry, to the gauge fields propagating in the ambient space, but pulled back to the surface. (But not all surface defects can be described in this way.)   See  
\cite{Avatars, Braverman:2004vv, NN2004:BPSCFT1, Gukov:2008sn, Gaiotto:2013sma, Gukov:2014gja, nekrasovbpscft4, nekrasovbpscft5} for a small sampling of studies and reviews of surface defects.

Anton Kapustin, in his talk~\cite[\S2]{Ka1} at the 2010 International
Congress of Mathematicians, clarified the relationship between defects and
extended field theory.  We remark that extended field theory is most
developed for topological field theories, as in section~\ref{sec:TQFT}, but
one can also contemplate extended general quantum field theories as well.  
Thus for
an $n$-dimensional Wick-rotated field theory~$\mathrsfs{F}$, the
value~$\mathrsfs{F}(Y)$ on a closed Riemannian $(n-1)$-manifold $Y$ is a
(topological) vector space, the value~$\mathrsfs{F}(Z)$ on a closed
Riemannian $(n-2)$-manifold $Z$ is a linear category (presumably with
topology), etc.  Suppose $X$~is an $n$-dimensional Riemannian manifold---a
Wick-rotated spacetime---and $W\subset X$ is a 1-dimensional submanifold,
thought of as the worldline of a particle.  The \emph{link} of~$W\subset X$
is a sphere~$S^{n-2}$, and defects supported on~$W$ are labeled by objects
in the category~$\mathrsfs{F}(S^{n-2})$.  More precisely, we should
take a limit as the radius of the linking sphere shrinks to zero.    These ideas extend to submanifolds
in~$X$ of lower codimension, to which the theory~$\mathrsfs{F}$ associates
higher categories.  Furthermore, one can extend to submanifolds with
boundaries and corners and other singularities.
One important aspect to understand better is the data on which a defect depends.
As a very simple example, the Wilson lines in 3d Chern-Simons theory depend on a
framing of the normal bundle. This can be seen quite clearly in the quantum correlators
of the Wilson line defects. The analogous data for other defects in most
contexts has not been properly investigated.

 In some situations, e.g.~with topological defects, or with supersymmetric
defects one can start to discuss their operator products.
From the extended field theory point of view, these products are based on the $E_k$-algebra structure of the sphere $S^{k-1}$ in the bordism category.
Here again, the framing data is not usually mentioned, 
but the OPE of Wilson line defects does depend on
such data \cite{MooreTASI}. The products of
line defects in Chern-Simons theory are closely related to the fusion rules of
anyons (and the framing data detects their spin). OPE's of line defects
 have also played an important role in the geometric Langlands program \cite{Kapustin:2006pk}.
Recently there have been some investigations into the
 ``operator product expansions'' of surface defects
 \cite{Gaiotto:2015aoa} and interfaces \cite{Dedushenko:2021mds}, finding interesting relations to homotopy
 algebra and infinite-dimensional quantum algebras. It seems clear that there is a
 rich mathematical structure to be discovered here.

In the recent literature there has been much discussion of ``non-invertible defects'' and their relations to ``non-invertible symmetries.'' Important examples of such noninvertible defects are defects which implement  Kramers-Wannier-like dualities in two and higher dimensions.   For some recent progress and the first construction of non-invertible symmetries in $d>3$ dimensional QFTs see 
\cite{Kaidi:2021xfk, Choi:2021kmx, Bhardwaj:2022yxj}. Many of these are related to so-called condensation defects, which can be understood as gauging of a (higher-form) symmetry on a defect, as opposed to the full theory, 
\cite{Gaiotto:2019xmp, Roumpedakis:2022aik, Choi:2022zal, Bhardwaj:2022lsg, Bartsch:2022mpm, Lin:2022xod}. The symmetry structure that emerges from these non-invertible defects should naturally fit into the structure of higher fusion categories (for fusion 2-categories see \cite{douglas2018fusion}) and has become a very interactive area of exchange between mathematics and physics, see \cite{Bhardwaj:2022yxj, Bhardwaj:2022lsg, Bartsch:2022mpm, Freed:2022qnc} for some recent progress on formulating symmetry  structures in terms of higher-categories. 

In lower dimensions the categorical structure was exploited successfully and numerous physically interesting results can be derived by utilizing non-invertible symmetries 
\cite{Fuchs:2002cm, Bachas:2004sy, Fuchs:2007tx,Kapustin:2010if, Fuchs:2012dt,Gaiotto:2015aoa, Bhardwaj:2017xup,Chang:2018iay, Thorngren:2019iar, Komargodski:2020mxz,Lin:2022dhv}. In $d=4$ some examples have emerged studying e.g. constraints on pion decays from non-invertible symmetries \cite{Choi:2022jqy, Cordova:2022ieu}, and the characterization of vacuum structure of 4d $\mathcal{N}=1$ SYM \cite{Apruzzi:2022rei}.

Some of the techniques here are related to old ideas of ``condensation of anyons''  \cite{Aasen:2017ubm,Bais:2002ye,Bais:2006bq,Bais:2008ni}, 
a topic which itself is closely related 
to simple current algebras 
\cite{Gato-Rivera:1990lxi} 
and extensions of chiral algebras \cite{Moore:1988ss}. 
This ``condensation''  is reminiscent of tachyon condensations of D-branes used construct D-branes of various dimensions
with nontrivial torsion K-theoretic RR charge \cite{Gopakumar:2000zd,Harvey:2000jt,Harvey:2000te,
Horava:1998jy,Kraus:2000nj,
Martinec:2002wg,Sen:1998sm,Witten:1998cd, Schafer-Nameki:2004ewn}. 
It is an interesting open issue whether ideas 
related to tachyon condensation, K theory, and stable non-supersymmetric branes can be related to the modern work on defect/anyon condensation. Some evidence for this was recently provided in the construction of non-invertible topological defects using branes in string theory \cite{Apruzzi:2022rei}, where the fusion is replicated by brane-dynamics such as the Myers effect.

Another potentially useful set of ideas in the theory of defects 
goes back to \cite{NikRome:2009, Nekrasov:2009uh}.  
The space of supersymmetric ground states of a 
supersymmetric field theory in finite volume admits an action of a pencil of commutative associative rings of local operators, commuting with some supercharge. In addition, in Euclidean space-time, one can also act on the space of ground states by the non-commutative algebra of supersymmetric interfaces. Interestingly enough, compositions of ``raising'' and ``lowering'' interfaces may produce local protected operators, in the cohomology of some supercharge. The whole structure resembles a Lie algebra, or its Yangian, or quantum, deformation, as seen in the context of gauge theories with $(4,4)$ supersymmetry in two dimensions. 
For the recent progress see \cite{Dedushenko:2021mds}. These developments
are partly based on the mathematical discoveries in \cite{nakajima1999quiver,maulik2018quantum}. 
The computations \cite{nekrasovbpscft4} of expectation values of supersymmetric defects, relating them to local operators in the cohomology of some supercharge, reveal fascinating connections to both geometric \cite{Frenkel:2004LP} and analytic Langlands program \cite{Etingof:2019pni,Etingof:2021eub, Etingof:2021eeu} and two dimensional conformal field theory, e.g. \cite{Nekrasov:2021tik}.

\section{String Theory And M-Theory}
\label{sec:STMT}

\subsection{What Is The Definition Of String Theory And  M-Theory?}

\bigskip
\bigskip
\begin{center} \framebox[1.2\width]{\framebox[1.1\width]{We don't know.}} \end{center}
\bigskip
\bigskip

This is a fundamental question on which relatively little work is currently 
being done, presumably because nobody has any good new ideas. 
Nevertheless, we should
never lose sight of the fact that it is one of the central unanswered problems in fundamental theoretical physics.
M(atrix) theory \cite{Banks:1996vh,Banks:1997mn,Banks:1999az,Bigatti:1997jy} and AdS/CFT \cite{Aharony:1999ti}
are profound insights and give partial answers, 
but they are tied to specific backgrounds,
and do not give satisfactory definitions of M-Theory/String Theory. 
 See the Snowmass whitepaper \cite{Gopakumar:2022kof} discussing applications of the bootstrap philosophy to derive facts about string theory and \cite{Gopakumar:2003ns,Das:2003vw,Gopakumar:2005fx,Razamat:2008zr,Eberhardt:2019ywk,Aharony:2020omh} for progress towards deriving string theory from field theory following the ideas of holography.

One notable attempt to give 
a fundamental formulation of string theory is
the subject of string field theory. A solid foundation was given in \cite{Zwiebach:1992ie,Taylor:2003gn}
building on important previous work such as \cite{Witten:1985cc}.  Unfortunately, the theory,
as currently understood, is intractable. 
Important progress was made, nevertheless, 
with exact results on tachyon condensation \cite{Schnabl:2005gv,Fuchs:2008cc}, an action for open superstring field theory using $A_\infty$ structure\cite{Erler:2016ybs}
and most recently distinct progress on $D$-instanton amplitude computations 
in \cite{Sen:2019qqg,Sen:2020cef,Sen:2020eck}, which would probably benefit from the more precise matching to the instanton counting developed 
in the field theory context \cite{Nekrasov:2002qd,Nekrasov:Japan}.

\subsection{Topological String Theory}
\label{subsec:topstrth}

We will be brief in this section because much more 
discussion can be found in \cite{Commando}. 
Topological string theory is both a metaphor of the physical string theory, and a substructure of it. 
One central reason for studying topological string theory 
is that if we are ever going to make progress on the question 
of ``what is string theory,'' it is reasonable to expect that 
the much simplified version of the question 
``what is topological string theory?'' will be a more tractable
and a useful first case study. Topological string theory was 
invented by Witten in \cite{Witten:1988xj,Witten:1991zz}.
Some basic material for this subject is covered in 
\cite{HoriMirrorSymmetryBook,AspinwallMirrorSymmetryBook}.

In the effort to understand topological string theory much attention has
focused on the topological string partition function. In the $A$-model this is defined, as a
formal (possibly asymptotic)  series by a collection of functions ${\CalF}_g$ on Calabi-Yau
moduli space, which are themselves generating functions for
Gromov-Witten invariants of genus $g$ curves. This topological string partition function has been the focus of intense research. The famous work of Bershadsky, Cecotti, Ooguri, and Vafa \cite{BCOV:1994} almost
provides a recursive definition. In the compact case there is  a crucial ``holomorphic ambiguity,''
which has only been fixed up to $g\leq 51$ in some of the most favorable cases \cite{Huang:2006hq}. The open topological string has the form of Chern-Simons theory, perhaps with the addition of some Wilson loop observables \cite{Witten:1992fb}.

The computation of ${\CalF}_g$ for all $g$
has essentially been solved by using techniques adapted from the
theory of matrix models
for the $B$-model \cite{Bouchard:2007ys,Eynard:2014zxa}
 and by using the topological vertex for the $A$-model in the
case of non-compact Calabi-Yau manifolds
\cite{Aganagic:2003db}. The status of non-perturbative definition of the topological string theory is discussed in \cite{Commando}.
Another review of the relevant issues can be found in \cite{Marino:2012zq}. 

A suggestion
to define nonperturbative topological string theory based on Fredholm
determinants associated to ``quantum Riemann surfaces''  was made in
\cite{Grassi:2014zfa,Codesido:2015dia,Marino:2015nla,Grassi:2019coc}.  Curiously, similar ideas have recently been applied
in the context of nonperturbative definitions of JT gravity \cite{Johnson:2021zuo}.
Some important progress in the theory of quantum curves and the relation
to topological string theory was made in 
\cite{Coman:2018uwk, Coman:2019eex, Coman:2020qgf,Grassi:2022zuk}. The relation between the quantization of a classical system based on a spectral curve, and the topological string, a $B$ model on a local Calabi-Yau which is a complexification of a handlebody whose boundary is that curve, or an $A$ model on a local Calabi-Yau mirror, involves a version of a blow-up formula and is discussed in \cite{Commando}. 
The topological string partition function is closely related 
to BPS state counting for compactification of strings on 
Calabi-Yau manifolds.  Various  generating functions of BPS state multiplicities, 
or twisted Witten indices of supersymmetric
backgrounds of string and M-theory can often can be mapped
to the free-energy of some version of a topological string,
i.e. genus zero Gromov-Witten prepotential accounts for the
one-loop effects of the bound states of instantons and $W$-bosons in $M$-theory on a local Calabi-Yau manifold, 
\cite{Lawrence:1997jr}, while subjecting it to the constant graviphoton background on ${\mathbb R}^4$ reveals all-genus
topological string partition function \cite{Gopakumar:1998ii, Gopakumar:1998jq}. 
These results predict  nontrivial integrality properties
of the Gromov-Witten invariants (the mathematical definition of those via stable maps only guarantees their rationality), which can be proven mathematically, using, curiously, $p$-adic analysis \cite{Kontsevich:2006an}. For physics 
justification of these results see \cite{Dedushenko:2014nya}. 

The case of compact Calabi-Yau threefolds seems much more difficult and
there are no examples where we can compute all the $F_g$, let alone give a
nonperturbative definition. One potential way forward is through an  
intriguing connection to BPS statecounting known as the OSV conjecture (named after Ooguri-Strominger-Vafa). 
If properly understood, it might give a road to a nonperturbative definition of 
topological string theory using the physics of supersymmetric black holes \cite{Ooguri:2004zv}.
The most thorough attempt to formulate and prove a sharp version of the OSV
conjecture was made in \cite{Denef:2007vg}, but a completion of the project
requires the solution of a number of unsolved problems addressed in the conclusion of \cite{Denef:2007vg}. 
Some mathematical progress on these issues has been achieved in 
\cite{Toda-BologmolovGieseker,Feyzbakhsh:2022ydn}.

One of the problems one must solve to use an OSV-like formula is an 
effective way of computing BPS degeneracies for compact Calabi-Yau manifolds. One important outcome of
\cite{Denef:2000nb,Denef:2001xn,Denef:2002ru, Denef:2007vg} 
is an interesting formula for BPS degeneracies based on ``attractor flow trees.'' The formula of \cite{Denef:2000nb,Denef:2001xn,Denef:2002ru, Denef:2007vg}   has been refined and 
clarified in \cite{Manschot:2010xp,Andriyash:2010yf,Alexandrov:2018iao}. 
Recently there has been some rigorous mathematical work confirming the 
physical conjectures \cite{Arguz:2021zpx,Mozgovoy:2021iwz}. Another notable approach to computing 
BPS degeneracies associated to non-compact Calabi-Yau manifolds uses the mapping 
to associated quivers and employing fixed point formulae to obtain explicit results
\cite{Manschot:2010qz,Manschot:2011xc,Manschot:2012rx,
Manschot:2013sya,Manschot:2014fua}. Other recent progress on the BPS spectrum for string compactification on non-compact Calabi-Yau manifolds can be found in \cite{Mozgovoy:2020has,Descombes:2021snc,Alexandrov:2022pgd}.
A technique known as ``exponential spectral networks'' has been devised which might lead to a systematic way of computing  BPS degeneracies on non-compact Calabi-Yau manifolds \cite{Eager:2016yxd,Banerjee:2018syt,Banerjee:2019apt,Banerjee:2020moh}. 
In spite of all this, finding the full BPS spectrum of 
the compactification on any compact Calabi-Yau other than (an orbifold of) a torus remains an important 
open problem.

\subsection{String Perturbation Theory}

Superstring perturbation theory remains a fundamentally
important subject, central to the claims that string theory
provides a truly UV finite theory of quantum gravity.
In \cite{Witten:2013cia,Witten:2019uux} Witten revisited, summarized, and clarified
a great deal of work that had been done on superstring perturbation
theory in the 1980's. In particular \cite{Donagi:2013dua}  demonstrated conclusively
the important fact that supermoduli space is not split.
Recent investigations  \cite{Felder:2019iqj,Felder:2020odc}  have raised the possibility of
unexpected divergences in the type II string theory. These
appear to be unphysical and it is important to clear up
this point. It is important to incorporate $B$-fields in the
consideration of summing over spin structure, perhaps using
the methods of \cite{Distler:2010an,Kaidi:2019pzj,Kaidi:2019tyf}.

Most of the work on high-genus superstring perturbation theory
has been done in the context of the ``RNS string formulation.''
There is widespread belief that this formulation is not optimal -
for example spacetime supersymmetry is not manifest (but it is there \cite{Friedan:1985ge}).
Related to this, the incorporation of nontrivial RR backgrounds, crucial to most
forms of the AdS/CFT correspondence,  is nontrivial.
Subsequently,   much work and progress has been made on alternative
worldsheet formulations. See \cite{Berkovits:2021xwh} 
for the state of the art on such
alternative formulations, and \cite{Commando}  for further remarks. 
This remains an important research
direction for the future.

In the context of the AdS/CFT correspondence, important advances
using worldsheet perturbation techniques have been made, opening up a host of interesting future
directions for research, and making contact with other
current trends in thinking about quantum gravity. Some remarkable new results on
the $SL(2,\mathbb{R})$ WZNW model have recently been achieved \cite{Dei:2021xgh,Eberhardt:2019ywk,Eberhardt:2021jvj}
pointing to a number of interesting directions in the application of string perturbation theory
to the AdS/CFT correspondence.

\subsection{LEET}

A very important aspect of string compactification is 
the construction of low energy effective actions. 
From the mathematical viewpoint some of the main interest 
here are relations to hyperk\"ahler and quaternionic k\"ahler 
geometry. Some aspects of this are discussed in 
section \ref{sec:InteractionsLowDimensional} below, and other 
aspects, related to   the theory of automorphic forms
are discussed in section \ref{sec:NumberTheory}
below.  

\subsection{Noncommutative Spacetime}

String theory clearly calls for some kind of emergent
spacetime based on an algebraic structure. One time-honored
idea is that spacetime is a noncommutative manifold, an idea which can
be made fairly concrete in the context of string theory
\cite{Douglas:2001ba,Konechny:2000dp,Konechny:2001wz,Seiberg:1999vs}, following the earlier ideas \cite{Connes:1997cr, Douglas:1997fm,Schomerus:1999ug}. One of the precursors of the interest to the realization of noncommutative geometry in string theory is the identification
of the moduli space of torsion free sheaves with that of instantons on a noncommutative space \cite{NekrasovS:1998}.  
The idea of an emergent spacetime from a more algebraic structure 
is part of the framework of both M(atrix) theory \cite{Banks:1996vh} and AdS/CFT \cite{Aharony:1999ti}. A simple concrete baby example can be 
given in the case of 2d open-closed TQFT, which may be viewed as a string theory with $0$-dimensional target space. From a semisimple open string Frobenius algebra one may recover the spacetime, along with the entire category of branes \cite{Moore:2006dw}.
In the closely related subject of noncommutative field theory
\cite{Douglas:2001ba}
many open problems remain. The extension of noncommutative
field theory to include nontrivial background NS flux (background gerbe connection with nonzero curvature) is a very interesting
problem which might involve operator algebra theory at a deeper
level than thus far has been employed. One early attempt is
the paper \cite{Harvey:2000te}. This subject has been somewhat neglected in
recent years but there are surely many treasures left to uncover.

\section{Anomalies}
\label{sec:Anomalies}

The subject of anomaly cancellation blossomed in the early-to-mid 1980's,
in part driven by important advances in
understanding the formal structure of perturbative anomalies \cite{Zumino:1983rz}
and by the renewed interest in string theory and higher dimensional
supergravity theories. The main idea was to use anomalies as a systematic way to distinguish consistent
from inconsistent field theories and supergravity theories
\cite{Alvarez-Gaume:1983ihn,Witten:1985xe}. This period was crowned with the
discovery of the   renowned Green-Schwarz anomaly cancellation mechanism
in 10-dimensional string theories \cite{Green:1984sg}.

\subsection{Anomaly Cancellation In String Compactification}

Much work has been done on anomaly cancellation in string theory
and M-theory but only recently has a full consistency check of global anomaly
cancellation in smooth M-theory backgrounds, not necessarily
orientable (but endowed with some other tangential structures),
been performed in \cite{Freed:2019sco}, building on much previous work
\cite{Witten:1996md,Diaconescu:2000wy,Diaconescu:2003bm,Freed:2004yc,Witten:2016cio}.
In \cite{Freed:2019sco} it is observed that there 
are two distinct trivializations of the anomaly, and it remains a
puzzle to decide if one is more physically relevant than the other.
The generalization of these results to include the presence of branes and singularities remains
to be done. Many issues remain from the  examination of ``frozen singularities''
in M-theory \cite{deBoer:2001wca} and it would seem the time is ripe for
a re-examination using the tools of equivariant differential cohomology,
now that differential cohomology is becoming a more familiar tool
for physicists.

In string theory, a full discussion of anomaly cancellation on the
worldsheet in the presence of orientifolds remains to be done,
and similarly a comprehensive discussion of
global anomaly cancellation in the presence of
branes and orientifolds has not been done.
(Some significant unpublished work can be found in
\cite{FreedOrientifolds}.)

In general, there is no proof that global anomalies
are canceled in compactifications of string theory.
For example, this is fairly nontrivial in compactifications
of F-theory to 6d 
\cite{Kumar:2010ru, Taylor:2011wt, Grassi:2011hq,  Monnier:2017oqd,Monnier:2018cfa,Monnier:2018nfs}. 
Some recent very intriguing work on anomaly cancellation in heterotic string theory can be found in  \cite{Tachikawa:2021mvw,Tachikawa:2021mby,Yonekura:2022reu}. These latter papers are particularly notable for their heavy use of the   Stolz-Teichner conjecture
relating the topology of the space of (0,1) supersymmetric 2d QFTs to topological modular forms.

Finally, an old and vexing issue, closely related to anomalies, is that of the proper quantization
conditions for the various $p$-form gauge fields in M-theory and
string theory. In the large distance weak-coupling limit it would
appear that one proper model of the $M$-theory $C$-field is based on
differential cohomology whose underlying generalized cohomology theory is   
singular cohomology \cite{Diaconescu:2003bm}. On the other hand, 
in type II string theory the RR fields should be based on differential $K$-theory
\cite{Minasian:1997mm,Witten:1998cd,Witten:2000cn,Freed:2000ta,Freed:2006ya,Freed:2006yc}.
The consistency of these two viewpoints is only partially understood
\cite{Diaconescu:2000wz,Diaconescu:2000wy}, and appears to rely on
remarkable and subtle topological facts. 
As emphasized in \cite{Diaconescu:2003bm} the $M$-theory Chern-Simons
term is really a \underline{cubic} refinement of the triple intersection
product on differential $H^4$ and a full study of the implications of
summing over torsion fluxes for this cubic refinement has not been
carried out. (Some recent work on the cubic form in this context has appeared in  
\cite{Freed:2019sco,Han:2020lxd}.) Moreover the compatibility of the use of twisted
differential cohomology in IIB string theory with the expected
S-duality has been a long-standing puzzle \cite{Diaconescu:2000wy}.
These old puzzles have
recently been revived in a provocative paper \cite{Debray:2021vob}
investigating duality symmetry anomalies in type IIB string theory; see also \cite{Gaberdiel:1998ui,Mukhi:1998gi,Minasian:2016hoh, Assel:2016wcr,Pantev:2016nze,Tachikawa:2018njr} for earlier work on duality anomalies in Type IIB. One possible way forward in addressing this old puzzle concerns reconciling the K-theoretic classification of RR charges in orientifolds \cite{Distler:2009ri} with Witten's discussion of S-duality for the baryon vertex in AdS/CFT \cite{Witten:1998xy}.

\subsubsection{Global Anomalies In 6d Supergravity Theories}

It has long been recognized that not all LEETs are UV complete,
or even internally consistent. This idea is at the heart of
the 't Hooft anomaly matching conditions and the use of anomalies
for separating ``good'' theories from ``bad'' theories. In modern
parlance these conditions are related to ``landscape'' and
``swampland'' conjectures, and whether well-defined low energy 
theories with gravity can be derived from consistent string theory and/or M-theory 
compactification. One well-defined program for using
anomalies to shed light on such issues is reviewed in  \cite{Taylor:2011wt}. Considerations of global
anomalies lead to further constraints which have been
explored in \cite{Monnier:2017oqd,
Monnier:2018cfa,Monnier:2018nfs,Hsieh:2020jpj}
but those papers leave some unresolved issues. For example, 
the discussion of \cite{Monnier:2018cfa,Monnier:2018nfs} reduced 
global anomaly cancellation to the triviality of a certain 
7-dimensional spin topological field theory. Roughly speaking, this 
theory is the difference between the invertible anomaly theory of the 6d 
supergravity fields and the invertible theory where the Green-Schwarz 
counterterm is valued. It would be good 
to have an effective way to compute this theory given the defining 
data of the supergravity. Moreover, \cite{Monnier:2018cfa,Monnier:2018nfs} 
identified new theta angles and torsion anomaly coefficients whose 
physical role has not been fully understood. Finally, it would be good 
to clarify which is the proper generalized cohomology theory underlying 
the differential cohomology used to formulate the self-dual fields in the 
context of 6d supergravity. For example, it is natural to wonder, 
given the experience from Appendix B of \cite{Freed:2000ta} whether the 
cleanest story would not come from using differential KO theory. This 
idea has not been seriously explored yet. 
It seems fair to say that a full and complete discussion of global anomaly cancellation in
6d supergravity remains to be given, although this is not a view shared by all 
experts in the field. Indeed, even in 8d supergravity important subtleties arise as discussed in 
\cite{Lee:2022spd}.

\subsection{Anomalies And Invertible Field Theories}

The geometrical interpretation of anomalies originated in the mid 1980's as
mathematicians such as Atiyah, Bismut, Freed, Quillen, and Singer formulated a theory of
anomaly cancellation in terms of determinant line bundles. This has evolved into a modern
interpretation \cite{Freed:2014iua,Freed:2016rqq} using the concept of invertible field theories \cite{Freed:2004yc}. This modern interpretation is a generalization of the anomaly-inflow mechanism that goes back to \cite{Callan:1984sa,Faddeev:1984ung}. 

Briefly, field theories have an associative composition law with unit.  Namely, two field theories can be ``stacked'' or tensored with no interaction.  If $\mathrsfs{F}_1,\mathrsfs{F}_2$ are field theories, then the state space of $\mathrsfs{F}_1\otimes \mathrsfs{F}_2$ on a spatial manifold~$Y$ is the tensor product $\mathrsfs{F}_1(Y)\otimes \mathrsfs{F}_2(Y)$.  Correlation functions and evolution operators are similar tensor products.  The unit theory~$\boldsymbol{1}$ has 1-dimensional state space~$\mathbb{C}$ on all spatial manifolds, and all correlation functions equal~1.  A field theory~$\alpha$ is invertible if there exists~$\alpha'$ such that $\alpha\otimes \alpha'$ is isomorphic to $\boldsymbol{1}$.  It follows that all state spaces of~$\alpha$ are 1-dimensional, but they do not come with a basis element.  Perhaps at first surprisingly, such theories may be nontrivial and furthermore have applications in physics beyond anomalies.  For example, they  capture short range entangled (SPT) phases in condensed matter theory, if one is willing to jump from discrete models to effective field theories.  Invertibility takes us full circle from the bordism categories~\cite{Milnor} of the Segal~\cite{SegalCFT} and Atiyah~\cite{Atiyah} axiom systems for field theory to the classical bordism of Thom~\cite{Thom} and a modern variant~\cite{GMTW}; see~\cite{FHT,Freed:2016rqq}.  Invertible theories are not necessarily topological. A good example of nontopological invertible field theories are those defined by an exponentiated eta invariant. These are related to parity anomalies 
\cite{Redlich:1983kn,Redlich:1983dv,Niemi:1983rq,Alvarez-Gaume:1984zst} and are known as Dai-Freed theories. For two recent reviews summarizing the status of these theories see \cite{Witten:2019bou,Freed:2021wqa}.
The general theory of nontopological invertible theories is under rapid development; a small sample of very recent papers is \cite{YaYo,GY}.

This viewpoint has been a very flexible way of understanding anomalies but it is not clear that it extends to all anomalies, such as duality symmetry anomalies or conformal anomalies, so there is more to understand here.
(For example, the conformal anomaly involves the Euler characteristic. It is not clear how to obtain this from the usual descent formalism two dimensions higher.) Moreover, as  mentioned above there are new ideas involving extensions of the notion of symmetry to
so-called ``categorical symmetry'', and these ``symmetries'' are also expected to have anomalies.
This subject is in its infancy.

\subsection{Anomalies And Dynamics}
\label{sec:tHooftAnomalies}

As first stressed by 't Hooft \cite{tHooft:1979rat}, anomalies
provide powerful constraints on RG flow. These ideas were
used to great advantage in advances in understanding the
dynamics of supersymmetric gauge theories in the mid-1990's
\cite{Intriligator:2007cp}. In the past few years they have
been combined with the new ideas based on invertible
topological field theories and generalized symmetries
to discover new constraints on nontrivial RG flows in
quantum field theories, see e.g. \cite{Gaiotto:2017yup,Gaiotto:2017tne,Cordova:2019uob,Cordova:2019jnf,Cordova:2019jqi,Cordova:2019bsd, Bhardwaj:2022dyt, Delmastro:2022pfo, Brennan:2022tyl}.


\section{Mathematics Resulting From  Holography And Quantum Gravity}
\label{sec:HoloQG}

Sometimes general conjectures about quantum gravity, and the role 
of string theory in quantum gravity, can motivate the discovery 
of new mathematical results. In this section we focus on some of these 
aspects of physical mathematics.

\subsection{General Quantum Gravity Conjectures With Precise Mathematical Consequences}

The past few years have seen some heightened activity in deriving general properties of quantum gravity by studying generic properties of  string compactifications and their global consistency conditions. Much of this originates in the weak gravity conjecture. 
Here we will focus on the connection to mathematics and refer to the reviews on this topic for a broader overview 
\cite{Taylor:2011wt, Brennan:2017rbf, Palti:2019pca, vanBeest:2021lhn, Harlow:2022gzl}.  
In particular,  we will focus on consequences in algebraic and differential geometry, 
especially in relation to moduli spaces of manifolds of special holonomy.

Sometimes the constrained structure of supergravity Lagrangians can
lead to remarkable mathematical predictions. 
The reasoning is that the existence of a suitably 
covariantly constant spinor on a compact manifold 
implies a certain amount of unbroken supersymmetry when 
superstring theory or supergravity is compactified on that 
manifold. But then the scalar fields in the supergravity 
are related to moduli of the compactifying manifold. The 
constraints on how these scalar fields can appear in a 
supergravity Lagrangian have direct mathematical implications 
for the moduli space. 
The phenomenon of mirror symmetry was in fact anticipated 
by exactly this kind of reasoning \cite{Dixon:1987bg,Dixon:1989fj,Lerche:1989uy}.
 Another nice example of this reasoning is   \cite{Seiberg:1988pf} 
where the constraints of the 
LEET coming from ${\cal{N}}=4$ supergravity predicted the form of the moduli 
space of K3 surfaces with complexified K\"ahler class to be a double coset of $O(4,20)$, a fact later confirmed 
by precise mathematical analysis 
\cite{Aspinwall:1994rg,Aspinwall:1995td}. 

A more speculative aspect of quantum gravity concerns 
aspects of the ``landscape'' of possible string compactifications. 
One might ask what principles should prefer particular vacua, 
especially when they come in continuous families. 
One approach to this question is to posit that there should be a well-defined probability distribution on a suitable set of 
string vacua. For perturbative string compactifications 
there is a natural metric on the moduli spaces of string 
vacua - the Zamolodchikov metric \cite{Zamolodchikov:1986gt}. 
It is tempting 
to use this metric to define a natural measure on string 
compactifications, but that will only make sense if the 
total volume is finite, a point first emphasized in 
\cite{Horne:1994mi}. In this way one is led to conjecture 
that, for example, the total volume of moduli spaces of  Calabi-Yau three-folds is finite
\cite{Horne:1994mi}. For moduli spaces with a fixed topological 
type the Zamolodchikov metric on the complex structure moduli space is the Weil-Peterson metric. Thus, the general quantum gravity conjecture leads to a very precise mathematical prediction. 
For moduli spaces with a fixed topological type this   conjecture was shown to be correct in \cite{Douglas:2005hq}. Analogous statements for, say,
$G_2$ compactifications of $M$-theory remains interesting and open. 
 
There are a number of general conjectures in quantum gravity 
which have generated a huge amount of interest in the past 
several years. Again, we refer to \cite{Brennan:2017rbf, Palti:2019pca, vanBeest:2021lhn, Harlow:2022gzl} for reviews. Some of these conjectures, when 
made suitably precise, can lead to impressive mathematical predictions. 
One example is the so-called ``swampland distance conjecture'' (SDC)
of \cite{Ooguri:2006in}. In rather general terms the conjecture states the following: A string compactification can have a moduli space $M$, which is parametrized by massless scalar fields. This moduli space can have a natural metric on it, and two points in $M$ can be separated by an infinite distance. The conjecture states, that an infinite tower of states become exponentially light, when moving an infinite distance in the moduli space. 
More intuitively the conjecture can be rephrased as stating that an infinite distance limit either corresponds to a decompactification, or to a tensionless limit of a weakly-coupled string (which in turn furnishes the infinite tower of massless states). The latter refinement is sometimes 
called the emergent string conjecture (ESC) \cite{Lee:2019wij}. 

When the SDC is made mathematically precise in the context of Calabi-Yau compactification some very nontrivial mathematics emerges.  
For example tests of the conjecture using the spectrum of D-branes 
as a function of the complex structure moduli space makes detailed 
and precise use of the Schmid Nilpotent Orbit Theorem and the theory 
of degenerations of Hodge structures
\cite{Eguchi:2005eh,Grimm:2018ohb,Grimm:2018cpv,Joshi:2019nzi,Bastian:2021eom}. 
A key role in this analysis is that infinite 
distance in the Zamolodchikov-Weil-Peterson metric is characterized 
by infinite order monodromy of the periods. The ESC  has motivated 
refined studies of the ways K3 surfaces can degenerate \cite{Lee:2021qkx}. 
There is room to run in this subject: 
Mathematically the challenge is to find a characterization of the behavior at infinite distance limits for general Calabi-Yau three-fold  geometries, as well as elliptically fibered Calabi-Yau manifolds of other complex  
dimensions. 

A much harder version of these questions concern the mathematical predictions following from precise versions of the SDC in the context of other non-Calabi-Yau compactifications, e.g. for compactifications on manifolds of exceptional holonomy.   Some preliminary results for a special class of $G_2$ manifolds has appeared in \cite{Xu:2020nlh}. 
Constructions of such exceptional holonomy manifolds are central to the Simons Collaboration on Special Holonomy \cite{SimonsCollab:SpecialHolonomy}, and will be discussed in section \ref{sec:ExHol}. 

Some extensions of tests of the SDC to AdS compactifications 
have been studied in \cite{Lust:2019zwm,Baume:2020dqd,Perlmutter:2020buo}. The AdS-distance conjecture implies in particular that in an 
AdS $\times X$ compactification of string theory, the 
AdS scale cannot be separated from the scale of the internal manifold $X$. 
In turn this implies bounds on the conformal dimension of the first non-trivial operator in the dual CFT. 
Mathematically this problem can be rephrased as implying a bound on the eigenvalue of the (scalar) Laplacian on $X$. Recently this was proven in
\cite{Collins:2022nux} for a large class of holographic models where $X$ is a  Sasaki-Einstein five-manifold, and a conjecture was put forward, that such a bound should exist and be universal (only dimension dependent). 

In the past year there has been exciting progress in the application of the theory of von Neumann algebras to questions about the entropy of black holes and de Sitter space
\cite{Leutheusser:2021qhd,Leutheusser:2021frk,Witten:2021jzq,Witten:2021unn,Longo:2022lod,Chandrasekaran:2022cip,Chandrasekaran:2022eqq}. Notably some nontrivial constructions in operator algebra theory relating type II and type III von Neumann algebras turn out to have natural physical interpretations in terms of black holes. It seems clear that this direction will lead to some interesting future progress

A completely different approach to landscape and swampland issues
has been initiated in \cite{Freedman:2020isd,Freedman:2021skw} and
this is a potentially promising direction.

\subsection{Holography}

For well over 25 years the ideas of holography have been 
of central importance to the string theory and quantum gravity 
communities. Yet, in relation to its immense importance and conceptual 
depth, the direct connections between holography and pure mathematics have been 
relatively meager\footnote{Connections between solutions to supergravity theories, and the geometry of Sasaki-Einstein manifolds has of course led to very important developments in mathematics, see section \ref{sec:GeoQFT}}. That situation has started to change.

A very significant advance in the theory of holography was recently made in \cite{Saad:2019lba,Stanford:2019vob,Mertens:2022irh},
where the path integral of a form of two-dimensional quantum gravity was related
to both matrix models and the mathematical work of M. Mirzakhani. (For some background
on the relation to topological gravity see  \cite{Dijkgraaf:2018vnm}.)    This work, 
combined with some older results known as the ``factorization problem in AdS/CFT''  \cite{Witten:1999xp,Maldacena:2004rf}, 
has in turn raised some key conceptual questions about the AdS/CFT correspondence,
such as whether topological 
effects in quantum gravity (e.g. wormholes)
require one to ``average'' over boundary quantum field theories.  
As mentioned above, moduli spaces of string theories come with a 
natural Zamolodchikov measure (provided the volume is finite), 
and the concept of Zamolodchikov measure can be extended to moduli 
spaces of superconformal field theories (provided the volume is finite). 
 When this is done for conformal field theories
of free bosons on a torus (a.k.a.~ Narain theories) the resulting ensembles bear some resemblance to three-dimensional Chern-Simons gravity \cite{Afkhami-Jeddi:2020ezh,Maloney:2020nni}.
 These recent papers have intriguing results, 
 but puzzles remain \cite{Benjamin:2021wzr} and it would be good
to clarify the status of these theories and to understand them better.
(As an example of one puzzle, it would be nice to clarify whether it is 
consistent, in a 3d theory of gravity, to sum over topologies which are 
just handlebodies.) 
A recent proposal for solving the ensemble-average puzzle
has in fact led to new results in mathematics proper concerning (regularized) volumes of hyperbolic manifolds \cite{Schlenker:2022dyo}. 
\footnote{Another, very different, approach to these puzzles makes use of the notion of ``non-invertible symmetries'' mentioned above \cite{Benini:2022hzx}. } 

A related set of observations concern certain natural ensembles of two-dimensional 
$(4,4)$ supersymmetric sigma models \cite{Benjamin:2015hsa,Moore:2015bba}.
The main result here is that these can only have  a weak-coupling gravity dual 
on a set of Zamolodchikov measure zero.  Interestingly 
the papers \cite{Afkhami-Jeddi:2020ezh,Maloney:2020nni} and 
\cite{Benjamin:2015hsa,Moore:2015bba}  made nontrivial use of the work of C.L. Siegel in analytic number theory. It would be very interesting to understand if there are deeper relations between averaged QFT and analytic number theory. For example, it would be natural to expect that there are connections to Kudla-Millson theory \cite{KudlaMillson}. Currently other ensemble averages of higher dimensional superconformal theories are under investigation with a view to possible holographic interpretations \cite{Collier:2022emf}. If the ensembles have string-theoretic 
interpretations as spaces of vacua one might expect the total 
Zamolodchikov volume again to be finite, but this remains to be 
seen in all but the simplest examples. This line of development 
might be a promising avenue for interesting new interactions between analytic number theory and QFT/ST.

A curious related  recent development, 
inspired by questions in quantum gravity, including the 
ensemble averaging question, 
are the ``topological models of baby universes,'' investigated in
\cite{Balasubramanian:2020jhl,Banerjee:2022pmw,deMelloKoch:2021lqp,Gardiner:2020vjp,Marolf:2020xie}. These involve the well-defined problem of summing over bordisms with fixed
boundaries in a topological field theory.
The model of \cite{Marolf:2020xie} was interpreted as supporting the view that holographic duality involves ensemble averages, but other interpretations are possible \cite{Banerjee:2022pmw}.
As far as we know there is no previous mathematical work in this direction and it presents a potentially interesting opportunity for some interactions between physicists and mathematicians. One important question is whether there are any analogs of these models in $d>2$ dimensions. Naive generalizations will not work. It has been suggested by M. Kontsevich that it might be important to consider nonstandard ways of weighting the sum over topologies using the results of 
\cite{nariman2021finiteness}.

\section{Interactions With Number Theory}\label{sec:NumberTheory}

There are several
intriguing connections between string theory and supersymmetric QFT 
and number theory. Many potentially fertile seeds been planted, but have not
yet blossomed, perhaps solely due to lack of cultivation. 
We survey a few of these developments here. There are a number of other intriguing relations to number theory that have arisen in the study of mathematical aspects of scattering amplitudes. These have been covered in a separate Snowmass document \cite{Arkani-Hamed:2022rwr}.

\subsection{Automorphic Forms And Partition Functions}

One of the primary connections to number theory is through
automorphic functions and analytic number theory.
An important historical example is the
explanation of modularity in Monstrous Moonshine Conjectures via the
relation of the Monster group
to an orbifold conformal field theory \cite{Borcherds1,Borcherds2,Dixon:1988qd,FLM}
\footnote{The role of various kinds of Moonshine in physical mathematics
is an important and rich subject in and of itself. We will not pursue
it here because it is being addressed in a separate
Snowmass document in \cite{Harrison:2022zee}.}.

Partition functions - understood in very broad terms - can be automorphic for
three (interrelated) reasons: They can be covariant under diffeomorphism symmetries,
they can be covariant under duality symmetries such as S-duality, or they can be
terms in low energy effective actions, whose duality symmetry implies automorphy
of the functions entering the effective action. Often, but not always, these partition
functions are associated with generating functions of invariants of BPS states.
We will just list here a small set of examples of this phenomenon. It is not possible
to give a complete list because the literature is quite vast.

\begin{enumerate}

\item Two dimensional CFT, especially RCFT, is a rich source of automorphic
functions thanks to modular (or mapping class group) covariance of partition functions, conformal blocks, and correlation functions. These considerations continue to play an important role in the modern era.
For example it is an important tool in investigations
of spectral aspects of conformal field theory, where many of the 
questions are motivated by aspects of the AdS/CFT correspondence
\cite{Ganguly:2019ksp,Hartman:2014oaa,Kaidi:2020ecu,Keller:2014xba}.
(A related set of ideas applies the conformal bootstrap to four-point 
functions and this has led to interesting results on the spectrum of the Laplacian on hyperbolic surfaces \cite{Kravchuk:2021akc}. Note that the bootstrap equations follow from the mapping class group symmetry of the four-punctured sphere and are thus a form of modular invariance.)
 In a related
set of developments  modularity plays an important role in discussions of
``extremal CFTs'', i.e.  CFTs with small numbers of low lying states 
\cite{Hohn-Extremal,Witten:2007kt,Gaberdiel:2008xb}. Several important questions concerning
existence of ${\cal{N}}=0$ extremal CFTs remain open and it would be 
extremely interesting to clarify them \cite{Witten:2007kt,Ferrari:2017kbp,Gaberdiel:2008xb,Gaiotto:2007xh,Gaiotto:2008jt,Harrison:2016hbq}.
Higher dimensional theories involving
general $p$-forms will illustrate similar phenomena. The higher dimensional generalizations
are not well-studied. To choose but one example: the conformal blocks of self-dual six-dimensional
theories are expected to be interesting automorphic objects for groups of disconnected diffeomorphisms
of six-manifolds.

\item Terms in the action for the LEET of string compactifications with duality symmetries provide a rich source of examples.
For example toroidal compactification of type II string theory naturally leads to Eisenstein series. For a review see \cite{Obers:1998fb}. If we combine this observation with the S-duality of type IIB string theory,
then, in the hands of \cite{Green:2010kv,Green:2011vz,Green:2014yxa,Gourevitch:2019knu} 
nontrivial observations about some of the automorphic forms appearing in the Langlands program, and conjectures about constant terms, can be addressed.
The book \cite{Fleig:2015vky} has an extensive review of the relation of 
string theory scattering amplitudes and automorphic functions.  
These developments have also led to the development of the whole
subject of modular graph functions (which are related to the integrands of expressions giving terms in the LEET).
This is an active area of 
development
\cite{Pioline:2015qha,Broedel:2018izr,DHoker:2015wxz,DHoker:2017pvk,DHoker:2017zhq,DHoker:2020uid}.
The paper  \cite{Pioline:2015qha} points out an interesting connection to the work of Faltings and the Kawazumi-Zhang invariant.

\item Another notable use of the automorphy of the LEET is the T-duality symmetry 
(via heterotic/typeII duality) of the prepotential of type II theories compactified on K3-fibered Calabi-Yau manifolds. Study of these prepotentials led to a simple proof of the Borcherds lifts of automorphic forms \cite{Harvey:1995fq} in what is a nice example of the math-physics dialogue. The work of \cite{Harvey:1995fq} applied some standard techniques from string perturbation 
theory \cite{Dixon:1990pc} in a fertile setting that explained the
remarkable phenomenon of Borcherds lifting in a conceptual way. The method was generalized to the construction of automorphic forms for  $SO(p,q)$ in \cite{Borcherds:1996uda}. 
The resulting theory of singular theta lifts, and the related results of \cite{Dijkgraaf:1996xw}
have led to several ingenious extensions and generalizations in both the math 
and physics literature 
\cite{Bryan:2018nlv,
BruinierFunke,
Carnahan:2009mjm,
Dabholkar:2006xa,Eguchi:2011aj,
GritsenkoNikulinLorentzKM,
GottscheKoolRank2DMVV,
Klemm:2005pd,
Persson:2013xpa,
Pioline:2015qha,Angelantonj:2015rxa}, and have also been of use in physics-inspired 
investigation of four-manifold invariants, as discussed elsewhere 
in this paper.

\item The terms in the LEET mentioned above 
are often generating functions for invariants
of BPS states. The automorphy of such BPS statecounting functions played extremely important roles in establishing fundamental string dualities 
\cite{Bershadsky:1995qy,Ferrara:1995yx,GottscheHilbScheme,
Sen:1999bqa,Sen:2001di,Vafa:1994tf,Vafa:1995bm}  and
in accounting for microstates of black holes \cite{Strominger:1996sh}. 
The breakthrough paper \cite{Strominger:1996sh}
has led to a vast literature on BPS statecounting and automorphy, which 
continues to inspire and direct research to this day.  
Among the many modern issues resulting from BPS state counting 
the question of the automorphic nature of the
generating functions of  certain DT invariants  of
torsion sheaves on compact Calabi-Yau manifolds  (``D4-D2-D0 counting functions'')  remains an important 
open one. Physical considerations suggest that the 2d $(0,2)$ elliptic genus associated with 5-branes should exhibit interesting automorphic properties 
\cite{deBoer:2006vg,Denef:2007vg,Gaiotto:2006wm,Gaiotto:2007cd,Ooguri:2004zv,Alexandrov:2019rth,Alexandrov:2022pgd}.  Mathematically rigorous
confirmation of some of these physical predictions have been made in the last several
years and will probably be the source of interesting mathematical research into the   
future \cite{Bouchard:2016lfg,klemm2008noetherlefschetz,
Gholampour:2013hfa,maulik2012gromovwitten,TodaFlop,toda2013sduality,
toda2014generalized}.
(For related, but slightly different appearances of modular forms in 
mathematically rigorous investigations of 
enumerative geometry of Calabi-Yau manifolds see 
\cite{Bryan:2018nlv,Bryan:2019qex,Maulik-Pandhari-Thomas}.)

\item Another example of an important BPS counting function is the elliptic genus of ${\cal N}=2$ theories. In the context of AdS/CFT this led to the application of Rademacher expansions \cite{Dijkgraaf:2000fq,Manschot:2007ha,Cheng:2011ay,Cheng:2012qc,Cheng:2012rca,Cheng:2012tq}
\footnote{The mathematics here is closely related to a mathematical subject known as 
Eichler cohomology \cite{KnoppEichlerCoho,Knopp-Rademacher,NieburAutomorphic}. }.
That in turn has led to important results in 3d quantum gravity \cite{Maloney:2007ud} raising questions regarding unitarity, and the existence of a Hilbert space interpretation in 3d gravity,  that have yet to be resolved. It also motivated some important developments in Moonshine. For example \cite{Duncan:2009sq}  observed an important connection between
Rademacher summability and the renowned genus zero property of the Moonshine groups.
(The relation between Rademacher summability 
and genus zero goes back at least to 
\cite{Knopp-Rademacher} and references therein.)
In the past several years there have been extremely beautiful proofs of such
``Fareytail expansions'' of BPS counting functions via localization in the supergravity path integral, and the development
of some of these localization tools have illuminated interesting issues in strongly
coupled four-dimensional QFTs \cite{Dabholkar:2011ec,Dabholkar:2014ema,Gupta:2012cy,Jeon:2018kec,Murthy:2015yfa}.

\item Partition functions of topologically twisted field theories 
are a natural source of automorphic functions thanks to nontrivial duality symmetries such as $S$-duality. A famous and paradigmatic example are the partition functions of Vafa-Witten twisted ${\CalN}=4$ SYM \cite{Vafa:1994tf}.
Extensions to higher rank go back to \cite{Minahan:1998vr} and have been discussed in \cite{Manschot:2011dj,Manschot:2014cca,Manschot:2017xcr,Alexandrov:2019rth}. 
The verification and extension of the physical predictions of  
have proven to be a rich source of inspirations in enumerative 
algebraic geometry and have led to important advances such as 
\cite{Sheshmani:2019tvp,Tanaka:2017jom,Tanaka:2017bcw,Thomas:2018lvm}.
The partition functions of other $S$-dual symmetric quantum field theories
will lead to similar counting functions for other enumerative invariants. 
See, for example,   \cite{Manschot:2021qqe}.

\end{enumerate}

\subsection{Generalizations Of Automorphy}

Mathematicians and physicists have been exploring various 
extensions of the notion of automorphic forms in the past 
few years. A number of promising directions have emerged
that might well lead to important results in the future. 

One of the most striking generalizations of modularity 
are the mock modular forms. See \cite{Dabholkar:2021lzt}  for a 
recent review of some of the ways they have appeared in physics. 
 A precise mathematical definition 
was codified in \cite{zwegers2008mock} but in fact many 
examples have been known since the work of Ramanujan, and, somewhat later, in the work of D. Zagier 
\cite{ZagierNombre,HirzebruchZagier}.
Mock modularity has been appearing in many distinct ways 
in the literature on physical mathematics, and we will 
indicate some of the occurrences below. 
It would be desirable to understand if there is a common concept
underpinning these various instances of mock modularity in physics.
Perhaps related to this, modularity has a beautiful representation-theoretic interpretation in terms of $SL(2,\mathbb{R})$. It would be very desirable to have an analogous representation-theoretic basis for mock modularity, but such an interpretation is at present lacking. 

Mock modular forms showed up in the study 
of automorphic properties and holomorphic anomalies of 
twisted supersymmetric gauge theory partition functions 
on four-manifolds \cite{Vafa:1994tf,Moore:1997pc}. More recently 
the appearance of mock modularity in this context has 
been understood more systematically in 
\cite{Dabholkar:2020fde,Korpas:2017qdo,Korpas:2019ava,
Korpas:2019cwg,Manschot:2021qqe}  and indeed the study of some examples 
lead to open questions about mock modular forms 
\cite{Manschot:2021qqe}.

Mock modularity also shows up naturally in the study of 
the Fareytail expansions of elliptic genera and holomorphic 
anomalies in the context of AdS/CFT \cite{Manschot:2007ha}.  
Another, related, source is the appearance of these forms in 
BPS state counting functions for ${\CalN}=4$, $d=4$ supersymmetric 
black holes. This has been extensively investigated in 
the major work \cite{Dabholkar:2012nd}.   Similar considerations 
are expected to occur in ${\CalN}=2$, $d=4$ examples, but this 
has been much less investigated and is a natural direction 
for future research. For some relevant results see 
\cite{Manschot:2010sxc,Cheng:2017dlj,Alexandrov:2018lgp,Alexandrov:2019rth,Alexandrov:2022pgd}.
  Another appearance of mock-modularity 
can be found in the effects of D3 instantons on low energy 
effective actions  \cite{Alexandrov:2016tnf,Alexandrov:2017qhn}.

Yet another source of mock modularity is non-compactness 
of the target space in a sigma model
\cite{Eguchi:2008gc,Eguchi:2009cq,Eguchi:2010cb,Harvey:2013mda,
Harvey:2014cva,Murthy:2013mya,Sugawara:2011vg,Troost:2010ud,Troost:2017fpk}.
Recently a beautiful physical 
derivation of the APS index theorem based on path integrals 
made use of this phenomenon \cite{Dabholkar:2019nnc}. 
It should be possible to understand, conceptually, 
the appearance of mock modular objects in  
noncompact sigma models by applying similar considerations 
to the Dirac operator on loop space, to get an ``APS theorem'' 
for such Dirac operators, but this has not been 
done. The role of mock modularity in noncompact sigma models 
has recently played an unexpected and important role in the connection of topological modular
forms to the classification of supersymmetric field theories 
\cite{Gaiotto:2019gef}. 

Mock modular forms can be generalized to ``mock modular forms of finite depth.'' 
A mock modular form - very roughly speaking - is a nonholomorphic function 
$f(\tau, \bar \tau)$  on the upper half plane which has the property that 
$\bar\partial f$ is simply related to a holomorphic modular form. One can 
then use this idea to define iteratively a series of generalizations where 
a mock modular form of depth $d$ is related, in an analogous fashion, to  
a mock modular form of depth $(d-1)$, where holomorphic forms are of depth $0$. 
Such functions have appeared in 
partition functions of twisted higher rank supersymmetric theories as well as in 
D-instanton expansions. They have been studied mathematically in, for examples, 
\cite{Alexandrov:2016enp,Funke-Kudla,Nazaroglu:2016lmr}, which also make use of 
various generalizations of the error function. These ideas have been 
applied to the counting functions associated with M-theory black holes in 
\cite{Alexandrov:2018lgp} and to $D$-brane BPS counting functions in 
\cite{Chattopadhyaya:2021rdi}. The paper \cite{Chattopadhyaya:2021rdi}
raises some unresolved puzzles related to the scaling BPS black hole solutions 
of \cite{Denef:2007vg}. (In general, the scaling solutions tend to present 
puzzles \cite{Denef:2007vg,Bena:2012hf,Manschot:2013sya,Manschot:2014fua} and deserve to be understood 
better. A general condition for their existence can be found in \cite{Descombes:2021egc}.)

Another direction in which the notion of automorphy is 
being extended is to the domain of ``quantum modular forms'' and 
``false theta functions.''  At present, the precise definition  ``quantum modular 
form'' by D.~Zagier leaves open the possibility of several variations 
\cite{Lawrence-Zagier,Zagier-QMF}.
It is possible that physical applications will guide the way to the ``best'' definition. One important recent source of quantum modular forms is via $3$-manifold invariants associated to the partition functions of $d=3$ ${\CalN}=2$ field theories
on plumbed 3-manifolds. These ``$\hat Z$-invariants'' have illustrated 
curious relations to mock modular forms and false theta functions 
\cite{Bringmann:2018ddv,Cheng:2018vpl,Cheng:2019uzc}. A great deal needs to be clarified concerning these and related intriguing observations. Another source of ``quantum modularity'' in physics appears through 
the presence of MacMahon, and MacMahon-like functions in black hole 
microstate counting \cite{Dabholkar:2005dt,Jafferis:2008uf,Mozgovoy:2020has,Okounkov:2003sp}.

Yet another context in which modularity and its generalization play a role is the realm of supersymmetric partition functions/indices in various dimensions \cite{Pestun:2016zxk}. Computing such indices one can obtain explicit expressions which often exhibit interesting (mock/quasi/etc) modular behavior and its generalizations. For example, in
the supersymmetric index in four dimensions \cite{Kinney:2005ej,Romelsberger:2005eg,Dolan:2008qi} one can see hints of $SL(3,{\mathbb{Z}})$ modularity \cite{MR1800253,Spiridonov:2012ww,Gadde:2020bov,Jejjala:2021hlt,Jejjala:2022lrm}. Specializing to ${\CalN}=2$ theories and to Schur indices \cite{Gadde:2011ik,Gadde:2011uv}  these hints of 
$SL(3,{\mathbb{ Z}})$ modularity become hints of $SL(2,{\mathbb{Z}})$ modularity \cite{Razamat:2012uv}. Some of these relations have been intensely studied: {\it e.g.} in the context of the relation of the index to chiral algebras \cite{Beem:2013sza} the Schur indices can be thought of as components of a vector valued modular form and are solutions to certain logarithmic modular differential equations \cite{Beem:2017ooy}. There are also intriguing connections between these indices and the work of Kontsevich and Soibelman on wall crossing phenomena \cite{Cordova:2016uwk,Cordova:2017mhb,Kontsevich:2008fj,Dimofte:2009tm}. Moreover as the Schur indices are often given as integrals over Jacobi modular forms they can exhibit quasi and mock modular properties \cite{Beem:2021zvt,Pan:2021mrw} which would be very interesting to explore in further.

\subsection{Geometric Langlands Program}

\label{subsec:GLP}

The famous Langlands program in mathematics relates - very roughly speaking - arithmetic questions related to Galois groups and motives on the one hand and automorphic function theory (and harmonic analysis) associated with number fields, on the other.  The Langlands program can be viewed as a generalization of class field theory (and is sometimes described as ``non-abelian class field theory''). 
 Some of the most important advances in number theory can be interpreted within the framework of this program, which, moreover,  promises to unify many areas within number theory and perhaps even resolve some of the longstanding and outstanding questions within that field. When one replaces number fields by function fields of curves the questions can be stated more geometrically, and this has led mathematicians to formulate an analogous program, the renowned Geometric Langlands Program (GLP). When the curve $C$ in question is over $\mathbb{C}$ the GLP, in particular, connects\cite{BeilinsonDrinfeld-HitchinQuantization,BD:2005opers} the holomorphic quantization of Hitchin's moduli space 
${\CalM}_{H} (C, G)$ for gauge group $G$ and the classical
holomorphic symplectic geometry of the moduli space of 
$^{L}G$-local systems (flat connections, for physicists), see
\cite{Fren95}. Moreover, Hitchin's space plays a role in Ngo's proof of the ``fundamental lemma of the Langlands program'' thus leading to one route from the modern work on the GLP back to the original Langlands program. Other ideas have been inspired by the 
connection to physics and have led to progress purely within the mathematical activity in this area. See  \cite{Frenkel:2004LP,Frenkel:2015LP,BenZvi-Frenkel,Frenkel-LecturesGLP} for introductory reviews 
on the Langlands program and its geometric cousin. 

%
%

It was understood early on that the GLP is intimately connected to two dimensional conformal field theory, more specifically to 
the theory of chiral algebras. This follows from the fundamental result of \cite{FF:1992cl} realizing the center 
$Z ({\widehat{\mathfrak{g}}})$ of
the universal enveloping algebra of a current algebra
$\widehat{\mathfrak{g}}$ associated
to a simple Lie algebra $\mathfrak{g}$ at the critical level $k = -h^{\vee}$, 
as a limit of the associated $W$-algebra. In turn, $W$-algebras appear naturally in 2d CFT \cite{ZW}. The elements of the center become the differential operators on $Bun_{G}(C)$ twisted by the square root of its canonical bundle (${\CalM}_{H} (C, G)$ is a hyperk{\"a}hler manifold, which in one of its complex structures is birational to $T^{*}Bun_{G}(C)$). There was a little setback, because unitarity of conformal field theories with current algebra symmetry requires $k > 0$. 
Despite this, other hints connecting GLP and physics were showing up, as described in \cite{Frenkel:2004LP}. In \cite{Feigin:1994in} a relation was found between
the Bethe ansatz, Sklyanin separation of variables, and the Beilinson-Drinfeld view of GLP adapted to the case of
genus zero curves with punctures, at least in the case ${\mathfrak{g}} = {\mathfrak{sl}}_{2}$. 
{}In \cite{Gorsky:1999rb} the separation of variables was interpeted in the language of $D$-branes 
wrapping various cycles in a hyperk{\"a}hler manifold, as an example of mirror symmetry or $T$-duality. 
An example of a quantum integrable system associated with a genus one curve $E$, an elliptic Calogero-Moser (eCM) system of $A_{N-1}$-type, was studied in \cite{Etingof:1993wp}, while
the classical construction was given in  \cite{Gorsky:1994dj}. The full set of quantum integrals of motion of $A_{N-1}$ eCM can be identified with the $\CalD$-module part of GLP adapted to the
case of a genus one curve with one so-called minimal puncture. The dual side is represented by the variety $L$ of opers, 
which in the present case are the order-$N$ differential operators on the elliptic curve with a regular singularity at the puncture. The symbol of an oper is a meromorphic function on $T^{*}E$, which is a degree $N$ polynomial along the cotangent directions, whose coefficients are meromorphic functions on $E$ of specified poles at one point $z=0$ at $E$. The vanishing
locus is the spectral curve $\Sigma$.

A year later this algebraic integrable system, in the very guise of  \cite{Gorsky:1994dj}, 
was proposed \cite{Donagi:1995cf} as a candidate
description of the Seiberg-Witten geometry of the
four dimensional ${\CalN}=2^{*}$ with the gauge group $SU(N)$. The elliptic curve determines the gauge coupling of the
ultraviolet theory (which has ${\CalN}=4$ supersymmetry), on the one hand, 
the coupling constant is related in \cite{Gorsky:1994dj}
to the residue of the Higgs field at the puncture, on the one hand, 
and to the mass of the adjoint hypermultiplet, on the other. Finally, the abovementioned spectral curve $\Sigma$ is the Seiberg-Witten curve of the effective ${\CalN}=2^{*}$ theory. 

Going back with the GLP story, the opers are, sometimes, called the quantum Seiberg-Witten curves.  We will get to their meaning for the ${\CalN}=2^{*}$ theory (or its higher genus analogues) later. 
In general it is hard
to understand the physics of the variety of opers within the realm of the conventional
two dimensional conformal field theory with $\widehat{\mathfrak{g}}$-symmetry with a notable exception of Liouville (or Toda) field theory, where the
so-called BPZ differential equations, which are obeyed
by the conformal blocks of correlators involving the so-called
degenerate fields, do approach the opers on genus zero curves with punctures, in the $b \to 0$ or $b \to \infty$ limits
(the large central charge, or quasiclassical, limit of Liouville), see \cite{Teschner:2010je} for review. 
The subsequent progress came from the physical realization of ${\CalM}_{H}(C, G)$ as an effective target space of a two dimensional 
sigma model. This is to be contrasted with the way the moduli space of (compact group) flat connections are related to the two dimensional conformal field theory on that same Riemann surface $C$. 
Given Riemann surfaces $D$ and $C$, a twisted version of ${\CalN}=4$ super-Yang-Mills theory 
on $D \times C$, in the limit where $C$ is much smaller than $D$ 
becomes a sigma model with $D$ as a worldsheet and ${\CalM}_{H}(C, G)$ as a target. Moreover, the Montonen-Olive $S$-duality of the ${\CalN}=4$ super-Yang-Mills is then related to $T$-duality in the sigma model
\cite{Bershadsky:1995vm,Harvey:1995tg}. Then, some ten years later, a few more crucial ingredients became available: the
generalized complex structures, the extension of the Fukaya category to hyperk{\"a}hler manifolds \cite{Kapustin:2001ij}, the identification of Hecke operators and Hecke eignesheaves 
of the GLP in terms of the actions of 't Hooft-Wilson lines on branes, 
and special branes on Hitchin moduli space, respectively. Thus, the connection of ${\CalN}=4$ super-Yang-Mills theory to GLP started to emerge \cite{Kapustin:2006pk}
in a powerful way. In addition to the strong-weak coupling duality symmetries of supersymmetric gauge theories 
and sigma models, the GLP was thus connected to various geometric symmetries 
\cite{Balasubramanian:2017gxc,Gukov:2006jk,
Kapustin:2006pk,Teschner:2010je,Witten:2007td,
Witten:2008ep,Witten:2009at,
Witten:2009mh,Witten:2015dta}. This led to important questions of the classification of supersymmetric boundary conditions of ${\CalN}=4$ SYM 
\cite{Gaiotto:2008sa,Gaiotto:2008sd,Gaiotto:2008ak}, which, 
in turn, 
plays an important role in the gauge-theoretic interpretation 
of knot homology, as described above. 
These insights also gave a large impetus to the development of the theory of supersymmetric defects \cite{NN2004:BPSCFT1}, related to the
ramified version of the GLP \cite{Gukov:2006jk, Witten:2007td, Gukov:2008sn, Gukov:2014gja}. In particular, the geometric Satake theorem has a natural 
interpretation in terms of extended observables, i.e. line operators 
\cite{Ben-Zvi:2013zda,Ben-Zvi:2016mrh, Ben-Zvi:2017jfi,Frenkel:2007tx,HiRa} and surface defects. More generally, the ideas of the GLP, and its gauge-theoretic interpretation have played an important role in the development of geometric representation theory  
\cite{ChrissGinzburg-GRT} 
of both finite-dimensional and infinite-dimensional groups (such as loop groups). Also, thinking about the kind of quantization 
of ${\CalM}_{H}(C, G)$  involved in the GLP 
raised interesting questions in the general 
theory of quantization \cite{Gukov:2008ve, Nekrasov:2009rc, Etingof:2019pni,Nekrasov:2010ka, NRS, Frenkel:2018dic,Gaiotto:2021kma,Gaiotto:2021tsq,Teschner:2017djr}.

The insight into the GLP which follows from assuming the $S$-duality of the ${\CalN}=4$ super-Yang-Mills theory can be 
further strengthened (sometimes leading to the actual proofs) using the input from string theory.  In particular, the realization
of the ${\CalN}=4$ super-Yang-Mills as a six dimensional $(2,0)$ superconformal theory compactified on an elliptic curve $E$
of vanishing area, brings yet another tool in the game. Namely, take the $(2,0)$ theory
on $E \times C \times D$, but start with the compactification on $C$. This gives, at low energy, an ${\CalN}=2$
superconformal field theory in four dimensions, compactified on $E \times D$. Moreover, there is a deformation of the latter,
the so-called $\Omega$-deformation \cite{Nekrasov:2002qd}, 
which allows $E$ to vary over $D$, possibly contracting at the boundaries and the corners of $D$, permitting for a smooth (partial) compactification to a manifold with torus isometry, while preserving some supersymmetry \cite{Nekrasov:2010ka}. 
In this way some of the peculiar branes, discovered in \cite{Kapustin:2001ij} and used in \cite{Kapustin:2006pk}, acquire (conjecturally) a natural geometric origin, as ``no-boundary'' condition in the geometry of the form
${\rm cigar} \times {\BR} \times S^{1}$, where the cigar is viewed as an $S^1$ fibration over ${\BR}_{+}$, degenerating over $0 \in {\BR}_{+}$. 

If, by some lucky guess, we can realize this ${\CalN}=2$ theory as a quiver gauge theory with unitary gauge group factors, 
localization \cite{Nekrasov:2002qd,LocalizationReview2017}
lets one perform some efficient exact
calculations 
\cite{nekrasovbpscft1, nekrasovbpscft2, nekrasovbpscft3,nekrasovbpscft4,nekrasovbpscft5,Nekrasov:2021tik}, once the supersymmetry of $\Omega$-background
is identified with the rotation-equivariant de Rham differential
acting on the space of gauge fields. In this way the ingredients of geometric Langlands correspondence are mapped to the expectation values of certain supersymmetric observables in ${\CalN}=2$ theory, such as surface defects of various kinds (the monodromy, or orbifold, or regular, defects, vortex string defects, $Q$-observables, etc.). Then, 
in case of genus $0$ and $1$ curves $C$ with punctures, both the expectation values and the Dyson-Schwinger equations they obey, can be 
computed and hence mapped to the equations of \cite{Knizhnik:1984nr, Bernard:1987df}, producing the opers and $\CalD$-modules generalizing those in\cite{Feigin:1994in}. 
In addition one finds an extension of the GLP away from the critical level. This is the so-called quantum Langlands correspondence, whose early incarnation is the relation between the $\widehat{\mathfrak{sl}}_{2}$-current algebra conformal blocks and Virasoro conformal blocks with degenerate
fields \cite{FZ,sto}, and whose $k \to \infty$ limit gives 
the well-known Painlev{\'e} form of the Schlesinger equations
\cite{Teschner:2010je, Litvinov:2013sxa}. 

Now, having the four dimensional ${\CalN}=2$ theory perspective
on GLP, it is a natural next step to a $q$-deformed
version of the latter, by lifting the theory to five dimensions
\cite{NikFive, NikBaulieuLosev, NPS}
and compactifying on a circle with a twist, which in string theory
terms corresponds to replacing the $(2,0)$-superconformal field
theory by a little string theory
\cite{Aganagic:2017smx}. 

The three-fold view on the six-dimensional theory
on $D \times C \times E$ and other manifolds is very powerful, especially if supplemented by
additional string dualities. These become operational
once the six dimensional theory is related to the theory of NS fivebranes of IIA string theory. These ideas imply the BPS/CFT correspondence \cite{NN2004:BPSCFT0}
between the conformal field theories (in the extended sense) in two dimensions (and their $q$-deformations) and supersymmetric sector of four dimensional theories (and their uplifts to higher dimensions). At the origin of this correspondence are the chiral tensor fields propagating in the six dimensional $(2,0)$ theory
\cite{Nekrasov:2002qd}, or their cousins living at the intersections of branes (which could be obtained by T-dualizing the cigar-shaped geometries subject to $\Omega$-deformation \cite{Gaiotto:2017euk}, or by taking their $S$-duals   \cite{nekrasovbpscft3}).

{}The typical localization computations in ${\CalN}=2$ gauge theories in four dimensions produce functions matching the conformal, or chiral, blocks of some two dimensional chiral algebras. The most celebrated example is the AGT correspondence
\cite{Alday:2009aq}, relating Liouville theory and class $\CalS$
gauge theories of $A_1$-type. 

However, conformal blocks are not yet the physical observables
of the two dimensional conformal field theory, one needs to couple chiral blocks with the anti-chiral blocks. For example, 
the $4$-point function on a sphere can be expressed as a one-dimensional integral of a square of an absolute value of a conformal block, where the integration variable parametrizes the
conformal dimension of an intermediate field \cite{Zamolodchikov:1995aa}. This convolution can be mapped
to a partition function of a class ${\CalS}$ theory corresponding to a $4$-punctured sphere, which happens to be an $SU(2)$ theory with $N_f = 4$ fundamental hypermuliplets, compactified on an ellipsoid \cite{Hama:2012bg}. 
The shape of the ellipsoid is determined by the $b$ parameter
of the Liouville theory, so that for $b =1$ one gets the supersymmetric partition on a round four-sphere \cite{Pestun:2007rz}.

Thinking about the correlation functions of WZNW theory, 
analytically continued in the level $k$ of the current algebra, presumably along the lines of \cite{Witten:2010cx}, one naturally
encounters questions of real analysis on $Bun_G$. At the critical level $k \to - h^{\vee}$ the correlation functions become the eigenfunctions of both sets of quantized Hitchin hamiltonians, twisted holomorphic and anti-holomorphic differential operators on $Bun_G$, acting on half-densities. The choice of an eigenvector, and the corresponding eigenvalue are determined by the choice of a classical limit of $r = rk (G)$ fields, presumably the $T$-duals of the subset of Cartan-valued ``free fields,'' the $\varphi$-fields of the Wakimoto realization \cite{Wakimoto:1986gf, FF1,Gerasimov:1989ng}.
Presumably \cite{Frenkel:2022}, this relates the ``analytic Langlands correspondence'' of \cite{Etingof:2019pni,Etingof:2021eeu, Gaiotto:2021tsq} to 
the bosonic instantonic theory, also known as the curved $\beta\gamma{\bar \beta}{\bar\gamma}$-system  \cite{Nekrasov:2005wg,Losev:2005pu,Frenkel:2006fy,
Frenkel:2008vz} on the complete flag variety $G/B$, and to the higher rank generalizations of \cite{Ribault:2005wp,Hikida:2007tq,Teschner:2017djr}. Finally, 
the main ingredient of both analytic and classical GLP, 
Hecke operators \cite{Etingof:2021eub} (which commute with quantized Hitchin hamiltonians), show up in the ${\CalN}=2$ description of the
theory, as the so-called $Q$-observables, the specific surface defects, which can be engineered using the folded instanton construction \cite{Jeong:2021rll, Jeong:2022}. In this way 
explicit formulae for the eigenvalues of the Hecke operators
of \cite{Etingof:2019pni, Etingof:2021eub} can be obtained by localization. This example could become a specific application in mathematics
of the six dimensional view on the GLP. For other applications see \cite{Frenkel:2007tx}. 
It is tempting to conjecture that the paper \cite{GinzburgKapranovVasserot} also fits the six dimensional perspective. 

{}The subject is clearly
very deep and  will surely be 
an inspiring  source of much future progress.

\subsection{Attractors And Arithmetic}

The beautiful attractor mechanism for construction of
supersymmetric black holes \cite{Ferrara:1995ih}
leads to some interesting questions about arithmetic
aspects of Calabi-Yau manifolds and special Hodge
structures on these manifolds \cite{Moore:1998pn,Moore:1998zu,Moore:2004fg}.
\footnote{A very beautiful generalization of the 
attractor mechanism based on defect lines was introduced 
in \cite{Brunner:2010xm}. The ideas of this paper have not 
been developed very far, but would seem to be promising.}
The attractor conjectures  have recently received some
attention by physicists and mathematicians. See \cite{Benjamin:2018mdo} 
for an overview.  An important distinction 
must be drawn between ``rank one'' and ``rank two'' attractors. 
Recently important evidence has been presented for the (not unexpected) result that 
rank one attractors are not arithmetic varieties \cite{Lam:2020qge}. 
Although the arguments in \cite{Lam:2020qge} are based on some (well-founded) conjectures, 
strictly speaking they do not apply to the three-dimensional case. 
Clearly there are loose ends to be tied up here. 
\footnote{M. Elmi has informed us that there is numerical evidence that the generic rank 1 attractor point is not arithmetic.} 

The rank two attractors have especially nice
properties and, thanks to standard conjectures in arithmetic 
geometry are almost certainly arithmetic. 
Recently some beautiful new examples of rank two attractors
with curious properties have been discovered \cite{Candelas:2019llw}. Interestingly, rank two attractors can be found by searching for suitable factorization of zeta functions. In examples these zeta functions are related to modular forms, but there is no known physical meaning of these modular forms. 

Some interesting related work on aspects of 
complex multiplication in the context of RCFT can be found 
in \cite{Kondo:2018mha}, which takes note of   
connections between the weight two modular 
forms associated to L-functions of rational elliptic 
curves and traces in the R-sector of RCFTs

There are potential connections to the modular 
Calabi-Yau manifolds \cite{yui2012modularity,Yui:2017ixa},  but a clear physical interpretation 
has not been given. A potentially important contribution in 
this direction is the   recent observation relating special flux compactifications to 
modular Calabi-Yau manifolds \cite{Kachru:2020sio}.

\subsection{Other Directions Relating Number Theory And String Theory}\label{subsec:OtherNumberTheory}

We mention a number of other intriguing directions in which research 
is making new and potentially fruitful connections to number theory. 

Some years ago  \cite{Ashok:2006br} 
some interesting observations were made about the 
relation of ${\cal{N}}=1$ points in the Coulomb branch of 
$d=4$ ${\cal N}=2$ theories with Grothendiecks' \emph{Dessin's d' Enfants}. This 
direction seems fascinating, but remains unexplored. 

There are interesting ideas concerning a possible 
notion of ``Hecke operators'' on rational conformal field theories \cite{Bae:2020pvv,Harvey:2018rdc,
Harvey:2019qzs}. This direction is relatively unexplored. One of the main 
open problems, in our view, is to extend the observed Hecke action to modular tensor categories. 
Progress in this direction has been initiated in \cite{Harvey:2019qzs}.

In a beautiful series of papers written over the past 20 years 
\cite{Candelas:2000fq,Candelas:2004sk,Candelas:2007mb,
Candelas:2019llw,Candelas:2021mwz,Candelas:2021tqt}
P. Candelas and X. de la Ossa and collaborators have explored 
various arithmetic aspects of Calabi-Yau manifolds, including 
relations between point-counting and periods for Calabi-Yau manifolds
over finite fields. Recent achievements of this program have included 
discoveries of remarkable properties of zeta functions of Calabi-Yau manifolds 
over finite fields \cite{Candelas:2021tqt} and 
intriguing new results relating attractor points to special values of 
$L$-functions \cite{Candelas:2021mwz}. 

There is a famous analogy between number fields and compact oriented 
three-manifolds, in which knots are analogous to primes. A nice 
exposition can be found in \cite{morishita2009analogies}. Starting 
with these ideas the research program 
\cite{Chung:2016gxx,Chung:2019ekx,Kim:2015dzy,Kim:2017eyv}
seeks to define an arithmetic version of Chern-Simons theory.  

Special values of $L$-functions are of interest to mathematicians 
for a number of reasons. In a very interesting recent work \cite{Bonisch:2022mgw} it is noted that in some  $d=4$ ${\cal N}=2$ CY compactifications 
the ${\cal N}=2$ central charges of BPS states at special points can be written 
naturally in terms of  special values of $L$-functions. It would be 
of interest to generalize these relations to a broader class of theories.

\section{Interactions With Condensed Matter Physics}\label{sec:CondMat}

Although many of the techniques and motivations in physical 
mathematics have their origins in theoretical particle physics, 
the connections to condensed matter physics have been increasingly 
important. The use of topology in condensed matter physics goes 
back a long way (it can be traced back to a paper of Einstein's 
on semiclassical quantization \cite{EinsteinChaos,DougStone}) 
and is now a major aspect of modern condensed matter physics. A proper discussion should include a least: 

\begin{enumerate} 

\item General theory of the Berry connection. 

\item Relations to the quantum Hall effect and Chern-Simons theory. 

\item Aspects of anyon physics. 

\item Applications to quantum information theory and 
topological quantum computing, and their relation to 
fusion categories and modular tensor categories. 

\item Topological insulators, topological superconductors, and topological aspects of band structure, 
including relations to  (twisted, equivariant) K-theory and several distinct 
  ``10-fold ways.'' 

\item Weyl semimetals.

\item Topological phases more broadly, including what 
is known about their classification, and how the classification 
is modified by conditions such as symmetries and/or 
short-range entanglement. 

\item Relations of this classification program to 
the theory of anomalies. 

\item Nonsupersymmetric dualities and higher dimensinal bosonization. 

\item Emergent gauge fields. 

\end{enumerate} 

Covering all this material properly would more than double the length of 
this panorama. Just a few of the very many relevant reviews include 
\cite{Mermin:1979zz,Shapere:1989kp,Fradkin:2013sab,Stone:1992ft,Hasan:2010xy,Qi:2010qag,bernevig2013topological,Frohlich:2018cdi,Rowell:2018wnv,Tong:2016kpv,Brauner:2022rvf,McGreevy:2022oyu}.
We will content ourselves with just a 
few of the most recent connections most relevant 
to other parts of this essay.

As we mentioned above, one of the most important open problems in physical mathematics,
and a root cause of the occasional chasm separating the intuition
of mathematicians and physicists, is the proper rigorous definition
of quantum field theory with nontrivial RG flow.
There is a significant body of mathematical work on quantum mechanical theories---from both condensed matter and quantum field theory---which incorporate energy and length scales.  It is highly desirable to find closer connections of this work with the advances in the algebraic and topological sides of quantum theory. 

 It is often said that the best route to a
rigorous definition of quantum field theory 
is to use a lattice regularization and find a
careful definition of the continuum limit.
An interesting new wrinkle on this old problem is the
 discovery of lattice models involving, for example, fracton phases. 
 For one recent review, see \cite{Pretko:2020cko}.
The investigation of fractor phases has become an important  research direction
 in condensed matter physics, as well as  in high energy physics, because
 these lattice examples  highlight how subtle taking the continuum limit can be.
Recent work  \cite{Gorantla:2021svj,Seiberg:2020bhn,Seiberg:2020cxy,Seiberg:2020wsg}
 has clarified the continuum interpretation
of some of these models and has demonstrated that, from traditional
viewpoints of quantum field theories, the models are somewhat exotic.
Certainly, the considerations in these papers need to be further explored
in several directions. For just one example, nontrivial interactions should be
included - the above works are based entirely on quadratic actions.
It is possible that a theory of ``quantum field theory with foliations''
is relevant to these fracton phases.  An exploration along these lines can be
found in \cite{Ma:2020svo,Shirley:2017suz,Shirley:2019uou} as well as in the related work \cite{Aasen:2020zru} on topological defect networks. This is
new and relatively unexplored territory. It would be desirable to extend these
models to include more nontrivial foliations. Even in the framework of topological
field theory it might be interesting to enhance the usual bordism categories to
include foliations, or to explore whether, for example, the Godbillon-Vey class
could serve as an interesting action principle. Note, that the abelian five dimensional 
Chern-Simons theory has the space of codimension two foliations with a flat connection on its leafs
as a phase space \cite{FNRS}\footnote{The theory with the action $\int A \wedge dA \wedge dA$ associates to a four manifold $M^4$ the phase space ${\CalP}_{M^{4}}$ which is a quotient of the space of solutions to the Gauss law $F \wedge F = 0$ by the action of the group generated by the kernel of the form $\int_{M^{4}} F \wedge {\delta}A \wedge {\delta}A$, which is the semi-direct product of the gauge group and the group of diffeomorphisms
of $M^4$. The foliation is spanned by the kernels of $F$}. Also, a ``Wick-rotated'' Kodaira-Spencer theory describes the three dimensional foliations
of a six dimensional manifold equipped with a volume form
(this is an interpretation of Sect 2.1 of \cite{Gopakumar:1998vy}, referring to \cite{BCOV:1994}),
alternatively, one can also describe it by the ``Wick-rotated'' version of the Polyakov-like formulation of the six dimensional Hitchin theory, obtained by 
replacing $+6$ by $-6$ in the Eq. (11) of \cite{Zth}. There are also interesting relations between fracton phases and supersymmetric quiver theories \cite{Razamat:2021jkx, Franco:2022ziy} (as well as brane constructions \cite{Geng:2021cmq}).

While fracton phases are currently popular we should not forget that even understanding rigorous formulations of continuum
limits of lattice formulations of more standard theories remains
an important open problem. Therefore it is important to be open to
other novel ways to formulate theories with scale. For example, in \cite{Freed:2018cec} the two-dimensional Ising model is placed in the framework of extended topological field theory creating a potential new pathway to incorporate the renormalization group. Other new approaches to the problem of scale appear in   \cite{Elliott:2017ahb,Grady:2021kii}.

One important question in this general set of ideas 
is when a topological phase must necessarily
have gapless modes when formulated on a manifold with boundary.
A  fairly general result in this direction, in $2+1$ dimensions,
can be found in \cite{Freed:2020qfy,Kaidi:2021gbs}. 
See also \cite{Cordova:2019bsd} for related obstructions to the 
existence of gapped boundaries in invertible theories. One potentially 
interesting direction for   future research is to generalize these ideas and 
results other dimensions 
and to classify the various boundary (and defect) 
theories that can exist in the presence of topological phases.  
An alternative set of ideas addressing these kinds of problems can be found in 
\cite{Prodan-Schulz-Baldes}.

A long-term project of Alexei Kitaev's has been the classification
of short-range-entangled topological phases of matter in terms of a
spectrum (in the sense of algebraic topology)
in a generalized cohomology theory. This approach attempts
to classify phases of matter not by classifying spaces of (local!)
Hamiltonians but rather by examining the spaces of ground states of
such local Hamiltonians. Kitaev has given precise definitions of
short-range entangled states and a roadmap to their classification.
A successful outcome of this program is likely to have a wide impact. 
\footnote{Michael Freedman has, for some time now, suggested that a useful mathematical framework for
this program is coarse geometry and associated invariants, as developed by
John Roe~\cite{Roe1,Roe2,RoeHigson}.}

For some recent progress in the mathematical theory of SPT phases in quantum spin chains, see \cite{Ogata}.
If one allows passage to continuum field theory, then short-range entangled states go over to \emph{invertible} field theories~\cite{Freed:2014eja}, for which there is now a well-developed theory---at least in the topological case---in the framework of stable homotopy theory.  This leads to a general and computable formula for the abelian group of invertible phases in all dimensions and for all symmetry types~\cite{Freed:2016rqq,Kapustin:2014tfa,Yonekura:2018ufj}.  One can also use homotopical methods to compute groups of invertible phases on a fixed space or stack (space with symmetry)~\cite{Shiozaki:2018yyj,Freed:2019jzd,Debray}.  This includes \emph{interacting} ``crystalline phases", a subject of current interest in the condensed matter community; a very small sample of literature includes \cite{Thorngren:2016hdm,Else:2018eas}.

In the noninteracting case the classification of topological band 
insulators is fairly well-understood. It is based on K-theory 
\cite{Kitaev:2009mg,Schnyder:2008tya,Schnyder:2009klk,Ryu:2010zza}, 
and, when crystalline symmetry is taken into account, twisted 
equivariant K-theory \cite{Freed:2012uu,Kruthoff:2016ver,Stehouwer:2018xfs,Okuma:2018sfw}.
\emph{Some} of the invariants implicit in twisted equivariant K-theory (obtained by localizing the K-theory) have been used 
to give extensive lists of materials which might possibly be crystalline topological insulators 
\cite{VishwanathClassification1,VishwanathClassification2,BradlynCanoEtAl,BernevigClassification}. 
An interesting open question is whether there is topological information in the 
twisted equivariant K-theory which has not yet been taken into account in the classification 
schemes of 
\cite{VishwanathClassification1,VishwanathClassification2,BradlynCanoEtAl,BernevigClassification}.

The connection between topological phases and TQFT's has inspired new efforts in 
the classification of TQFTs. Recently, 4d TQFTs were classified in \cite{Lan:2018bui,Zhu:2018kzd,Lan:2018vjb, Johnson-Freyd:2020usu}, and most recently in \cite{TheoTBA}, where it was shown that not all 4d TQFTs are necessarily gauge theories of Dijkgraaf-Witten  (for a finite group, or higher-group) but can have a non-trivial so-called Majorana-layer, that obstructs mapping this to a gauge theory. 
5d topological order was discussed furthermore in \cite{Johnson-Freyd:2021tbq}.

Another fruitful point of contact between condensed matter theory
and high energy physics is the exploration of the notion of
higher Berry curvatures and their relations to transport coefficients
\cite{KitaevSPT,Hsin:2020cgg,Kapustin:2020eby,Kapustin:2020mkl,
Kapustin:2022apy}.
 Among the many open questions these developments raise is the
 construction of a full differential cohomology class corresponding 
 to the higher Berry curvatures. Another important problem is the 
 generalization to include thermal transport coefficients.

Underlying many of the ideas about topological phases of matter
is quantum information theory, a subject which has also been
playing a central role in modern developments in quantum gravity.
An interesting generalization of finite depth quantum circuits are
the quantum cellular automata  \cite{Arrighi:2019uor,Farrelly:2019zds,Freedman:2019ucy}
which have interesting relations to $C^*$-algebras and index theory.
It is natural to conjecture that they too will play an interesting
role in fundamental physics.

\section{Connections To Geometry And Low-Dimensional Topology}\label{sec:InteractionsLowDimensional}

\subsection{Two-Dimensions: Moduli Spaces Of Curves And Hurwitz Theory}
 
The development of matrix models of two-dimensional gravity 
\cite{Brezin:1990rb,Douglas:1989ve,Gross:1989vs} 
(see \cite{Ginsparg:1993is} for a review) led to a physical proposal 
for computing intersection theory of Mumford-Morita-Miller classes on the moduli spaces of 
curves \cite{Witten:1990hr}, a proposal which was spectacularly 
confirmed in \cite{Kontsevich:1992ti}. See 
\cite{Looijenga-ModuliIntersectionReview}  for a review. 
A much less well-known 
example are the generalizations studied in \cite{WittenAgebraicGeometryMatrixModel}, 
which remain largely unexplored (see, however, \cite{Polishchuk-Spinr-Curves} for some developments). One interesting aspect of this work is that it uses an idea closely related to the definition of Bauer-Furuta invariants of four-manifolds \cite{Bauer-Furuta}.  
\footnote{The general idea behind the Bauer-Furuta 
invariant can in principle be applied to many other topological field theories of cohomological type. This is largely unexplored territory. Some aspects have been explored in the context of Floer homotopy theory
\cite{CohenJonesSegal1,CohenJonesSegal2,Manolescu}.
For a nice review, see the talk of C. Manolescu at the WHCGP, October 3, 2022. }

A different approach to the Witten conjecture was 
pursued in \cite{okounkov2001generating,Okounkov:2000gx} using Hurwitz moduli spaces. 
These moduli spaces also appear quite naturally in large $N$ expansions of two-dimensional Yang-Mills theory \cite{Gross:1993hu,Cordes:1994sd,Cordes:1994fc,
Moore-ICM94}. More discussion
can be found in \cite{Commando}.  Yet another approach to
the Witten conjecture makes use of the ideas of geometric recursion 
\cite{andersen2019geometric,Andersen:2020ylr}. See also \cite{Bouchard:2007hi} for appearance of Hurwitz number in topological string theory.

{}More recently some remarkable new observations about Hurwitz theory motivated by the AdS3/CFT2 correspondence have been observed in  
\cite{Dei:2021yom} (See also \cite{Lunin:2000yv,Pakman:2009zz}.). 
It would be quite interesting to know whether (and if so, how) 
these appearances of Hurwitz theory are related to each other.

\subsection{Knots, Links,  And Three-Manifold Invariants}\label{subsec:Knot-3-Fold}

In the past few decades there have been remarkable advances in
the topology of 3-manifolds and knots. Some of these advances 
have been deeply related to explorations of quantum field theory. 
One of the most spectacular applications of quantum field theory 
to low dimensional topology is Witten's interpretation of the Jones 
polynomial in terms of Chern-Simons field theory 
\cite{Witten:1988hf} and the associated 
discovery of the WRT invariants of three-manifolds \cite{Reshetikhin:1990pr,Reshetikhin:1991tc}.
The ramifications of these discoveries continue to the present day. 

Many modern investigations of the relation between 
three-manifold topology and supersymmetric quantum field 
theory are ultimately related to the $6d$ $(2,0)$ theory. 
Typically, one considers this theory on a product of 
two 3-manifolds, compactifies on the factors in different 
orders and learns something interesting. 
An important example of this general idea is provided by the study of theories $T[M]$ associated to considering the 6d theory on a 3-manifold $M$ \cite{Dimofte:2011ju}.

One can study, for example, partition 
functions of theories $T[M]$ on $S^3$ or more generally on lens spaces. 
Somewhat surprisingly, these 
are closely related to Chern-Simons theories 
with noncompact gauge groups such as complex gauge groups. See \cite{Cordova:2013cea}
for one example of such a computation. 
A great deal of work has been done on localization computations of supersymmetric quantum field theories on spaces admitting a torus action. For example, \cite{NikThesis} rederived the Verlinde formula using the embedding of Chern-Simons theory into a twisted version of supersymmetric Yang-Mills theory. In \cite{NikBaulieuLosev, beasley2005nonabelian} the formalism applicable to lens spaces,  
$S^1 \times \Sigma$, and more general Seifert manifolds was developed. See \cite{Willett:2016adv}
for a review. The techniques have been extended to larger classes of three-manifolds such as Seifert-fibered spaces  
\cite{Closset:2016arn,Closset:2017zgf,
Closset:2018ghr,Closset:2019ucb,
Closset:2019hyt, Eckhard:2019jgg}, as well as to five dimensional theories \cite{Kallen:2012cs}.

Through the above route (and others) the 
study of the 6d (2,0) theory leads to the study of Chern-Simons theory for 
noncompact gauge groups, a subject which  has been vigorously studied for decades now. 
A very incomplete set of references includes
\cite{Andersen:2014aoa,Dimofte:2009yn, Dimofte:2016pua,Witten:1989ip,Witten:2010cx}. 
The subject continues to present important open problems. 
It cannot be a topological field theory
in the strictest sense, because state spaces 
are infinite dimensional, and gluing is not 
obvious. Nevertheless, it is ``almost'' a TQFT. 
For some approaches to solving this problem 
see \cite{Andersen:2018pnw,Gukov:2015sna,Gukov:2017kmk,Closset:2019hyt}.  

A very important example of a 
``non-compact group Chern-Simons theory'' 
is the  ``Teichmuller TQFT.'' 
\cite{Andersen:2018pnw,
EllegaardAndersen:2011vps,
EllegaardAndersen:2013pze,
Mikhaylov:2017ngi,
Teschner:2005bz}. 
In particular, the extremely beautiful 
paper of V. Mikhaylov \cite{Mikhaylov:2017ngi} relates the Teichmuller 
TQFT to the 6d (2,0) theory, and formulates 
the hyperbolic volume conjecture of Andersen-Kashaev in terms of a conjecture on the nature of solutions of the Kapustin-Witten equations on $M \times \mathbb{R}_+$ with Nahm pole 
boundary conditions at the finite boundary. 
It was recently pointed out
\footnote{In unpublished work of A. Khan, D. Gaiotto, G. Moore and F. Yan} 
that this conjecture gives a natural explanation of some remarkable discoveries relating three-dimensional indices to Stokes matrices of three-dimensional complex Chern-Simons theory 
\cite{Garoufalidis:2020xec}.

One concrete form of the partition function of a complex Chern-Simons theory on a 3-fold are the state-integral invariants, which
can be defined for a 
subclass of 3-manifolds that admit simplicial decompositions with a so-called ``positive angle structure.'' They take the form of a contour integral of products of factors, involving the quantum dilogarithm, and the precise contours are determined by the angle structure
\cite{Andersen:2014aoa,Andersen:2015tma,
Dimofte:2009yn,
Dijkgraaf:2010ur,
Dimofte:2011gm, 
Dimofte:2014zga,
Dimofte:2015kkp,
Garoufalidis:2013axa,
Hikami:2006cv}.

If, instead one considers the partition function on a product of a 
disk and a circle, with a suitable 2d QFT on the boundary, and when $M$  
has a presentation given by a ``plumbing presentation'' one is led to contour integral formulae for the ``homological blocks''  discussed in  
\cite{Gukov:2016gkn,Gukov:2017kmk}. (Precursors to these blocks, sometimes called ``holomorphic blocks'' in this context were described in 
\cite{Beem:2012mb,Witten:2011zz}.)  
The derivation of these formulae could use 
further clarification - this is a good problem for the near-future. 
The resulting formulae are  clearly 
of great interest. They give interesting $q$-series related to mock modular forms, resurgence and quantum
modular forms \cite{Cheng:2018vpl}. Moreover, as shown in  \cite{Beem:2012mb} 
they lead to many interesting identities and results on $q$-series.

Another source of three-manifold invariants based on twisted supersymmetric QFT are the Rozansky-Witten invariants associated to a choice of holomorphic symplectic manifold
\cite{Kapranov-RW,Rozansky:1996bq}. (These are also known as B-twisted 3d ${\CalN}=4$ sigma models.)  Recently interesting progress has been made on such theories associated with noncompact manifolds such as Coulomb branches of 3d ${\CalN}=4$ theories \cite{Brunner:2022rpd,Creutzig:2021ext,Gukov:2020lqm}. These investigations are part of a program to give a physical framework for the 
non-semisimple 3-manifold invariants discussed in \cite{Costantino:2012it,Hennings,kashaev2002invariants,Lyubashenko:1994tm}.
\footnote{These works are in turn closely related to ``logarithmic CFT,'' 
perhaps better called ``non-semisimple conformal field theory.'' 
If one generalizes the representations of the Virasoro algebra provided by 
primary fields to reducible, but indecomposable, representations then the usual conformal Ward identities lead to correlators that involve logarithms. 
It is now recognized that there should be an important generalization of RCFT based on allowing such indecomposable representations and this is an active area of current research. For reviews see \cite{Creutzig:2013hma,Flohr:2001zs,Fuchs:2019xkv,
Gaberdiel:2001tr,Kawai:2002fu}. The subject is difficult. It is analogous to generalizing 
the representation theory of compact groups to the representation theory of noncompact groups. 
Nevertheless, it is now showing up in the physics of percolation, noncompact Chern-Simons theory, instantonic field theories \cite{Frenkel:2006fy, Frenkel:2008vz},
and the geometric Langlands program, so it is clearly important. 
}

Yet another application of the $6d$ $(2,0)$ theory is to link homology   
\cite{Bar-Natan-KnotHom,KhovanovHomology}. One approach to a physical understanding of 
link homology is based on statecounting and the M5 brane 
\cite{Gorsky:2013jxa,Gukov:2004hz,Gukov:2010agg, Gukov:2012jx,Gukov:2016mqo,Gukov:2016gkn}. 
A second (but closely related) approach to a physical basis for knot homology was initiated in \cite{Witten:2011zz,Witten:2014xwa}. The latter approach is intimately connected to the physical interpretation of the geometric Langlands program and the theory of supersymmetric boundary conditions in ${\cal{N}}=4$ SYM. This approach relies on Morse complexes based on the difficult Kapustin-Witten and Haydys-Witten equations
\footnote{Whose moduli spaces are beginning to be understood from a solid mathematical framework
\cite{Mazzeo:2013zga,Mazzeo:2017qwz,
Taubes:2017pzr,Taubes:2019zhd,Taubes:2021clf}.}
but under some conditions important simplifications can be achieved and the formulation of knot homology can be expressed in terms of two-dimensional Landau-Ginzburg theories \cite{Gaiotto:2011nm,Gaiotto:2015aoa,Galakhov:2016cji}. 
A related, but distinct, program has been vigorously pursued by M. Aganagic  \cite{Aganagic:2020olg,Aganagic:2021ubp}
in work that makes interesting use 
\cite{Aganagic:2016jmx,Aganagic:2017smx,Aganagic:2017gsx} of the 
$q$-deformed Knizhnik-Zamolodchikov equations of 
\cite{Frenkel:1991gx}.

In a rather different direction, an important source of progress in understanding the topology 
of 3-manifolds has been the study of Ricci flow 
\cite{Hamilton-RicciFlow,Morgan-Tian-RicciFlow,Morgan-Tian-Geometrization,
Perelman:2006un,Perelman:2006up,Perelman:2003uq}  which itself is closely related to, the RG equations of nonlinear sigma models \cite{Callan:1985ia,Friedan:1980jm}. 
It is natural to wonder whether there can be useful technology transfer from the recently developed mathematics back to physics. See 
\cite{Frenkel:2020dic} for a very interesting recent development in this direction.

The theories $T[M]$ for $M$ a non-hyperbolic three-manifold flow to topological theories in the IR. It was recently conjectured that this provides an interesting, new way to classify both bosonic and fermionic phases of matter in 3d \cite{Cho:2020ljj}. Fermionic phases arise by including refined global information in the dimensional reduction, which account for 1-form symmetries in the 3d theory \cite{Eckhard:2019jgg}. This potentially provides a complementary, geometric, way to classify gapped phases, distinct from the approach using modular tensor categories (MTCs); see \cite{WenTopologicalOrder, Etingofbook,Rowell:2018wnv,Kong:2022cpy} for a discussion of bosonic topological order. Exploring this connection between non-hyperbolic three-manifolds and MTC seems an intriguing direction for the future.

\subsection{QFT And Four-Manifold Invariants}

One of the paradigmatic examples of physical
mathematics is the interaction between supersymmetric
field theory and four-manifold invariants. After the
discovery of instanton solutions of Yang-Mills equations \cite{Belavin:1975fg,Atiyah:1978ri}
the work of the Oxford school, led by M. Atiyah and N. Hitchin, culminated in the discovery
of Donaldson invariants of smooth four-manifolds \cite{DonaldsonKronheimer,FriedmanMorganDonaldsonInvt}. Motivated by
questions posed by M. Atiyah, Witten produced a quantum field theory interpretation
of the Donaldson invariants and in the process invented topological quantum field theory
\cite{Witten:1988ze}. The desire to use this formulation to give an effective
evaluation of Donaldson invariants was one of the motivations for the development
of Seiberg-Witten theory  \cite{Seiberg:1994aj,Seiberg:1994rs}  culminating in the
 introduction of Seiberg-Witten invariants \cite{Witten:1994cg},
a development which revolutionized the study of four-manifolds \cite{DonaldsonReview}. Given this spectacular success
one naturally wonders if further study of (topologically twisted) supersymmetric field theory
can lead to further insights into the differential topology of four- (and three-) manifolds.
In recent years there have been several developments along these lines.

From the physical viewpoint the Seiberg-Witten theory 
can be used to give 
an evaluation of the Donaldson invariants via the study of a certain 
integral along the Coulomb branch, sometimes referred to as the ``$u$-plane 
integral''. This family of integrals was first studied in \cite{Moore:1997pc,Losev:1998}
and it leads to many remaining open problems. In the case of pure 
$SU(2)$ theory the wall-crossing behavior of the Coulomb branch integral 
leads to a rather direct derivation of the famous ``Witten conjecture'' 
expressing the Donaldson invariants in terms of the Seiberg-Witten invariants 
for compact oriented four-manifolds with $b_{2}^{+} > 1$ \cite{Moore:1997pc}. 
(One good aspect of this approach is that the four-manifold need not be toric, or 
even complex. But it does require the existence of 
an almost complex structure, as is standard in Donaldson theory.) 
The integral can be generalized to include other Lagrangian ${\CalN}=2$, $d=4$ theories
\cite{Moore:1997pc,Losev:1998,Marino:1998bm}. The formulation of 
$u$-plane integrand for general $d=4$ ${\CalN}=2$ Lagrangian theories 
involves a crucial - but difficult - issue of determining ``contact terms'' in the presence of observables. The determination of these terms was 
described in \cite{Losev:1998}  as a theory of 
deformations of special K{\"a}hler manifolds, yet to be developed. 
This remains an important open problem. As explained in 
 \cite{Losev:1998} in the general Lagrangian case the path integral 
 localizes to the moduli space of the nonabelian version of the Seiberg-Witten 
 equations which, when localized further to the moduli space of instantons, involves 
densities different from those used to compute the Donaldson invariants. 
The developments described here are nicely reviewed in detail in the 
book \cite{LabastidaMarinoBook}.  

The direct evaluation of $u$-plane integrals is an efficient way to use physics to derive the invariants for manifolds 
with  $b_2^+ =1$ such as $\mathbb{C}\mathbb{P}^2$ and $S^2 \times S^2 $. It turns out that an effective way to evaluate these integrals
involves the use of modular and mock-modular forms. This was   recognized 
early on for the special cases of $\mathbb{C}\mathbb{P}^2$ and $S^2 \times S^2 $
\cite{Moore:1997pc,Malmendier:2008db,Malmendier:2012zz,Griffin:2012kw}, but
the relation to mock modular forms is quite deep and quite general, as   has been clarified recently in \cite{Korpas:2017qdo,Korpas:2019ava,Korpas:2019cwg,Manschot:2021qqe}. The latter papers highlight as well  the use of Jacobi-Maass forms. The 
technique has proven useful for the pure $SU(2)$ theory and 
for the $SU(2)$ ${\cal{N}}=2^*$ theory. 
 In the case of $SU(2)$  with 
fundamental flavors new important technical issues arise 
since the Coulomb branch is no longer a modular curve 
\cite{Aspman:2021vhs,Aspman:2021evt}. 

The case of the $SU(2)$ ${\CalN}=2^{*}$ theory presents a number of interesting variations 
on the above ideas \cite{Manschot:2021qqe}. First, it interpolates 
nicely between the Donaldson and Vafa-Witten invariants. Being superconformal 
it presents the added 
novelty that not only the integrand but the value of the Coulomb branch 
integral (as a function of the ultra-violet coupling) is a mock modular form. 
In fact,   the evaluation 
  on general $b_{2}^{+}=1$ manfifolds for general spin-c structure 
presents some new open problems concerning mock Jacobi forms which have
yet to be resolved \cite{Manschot:2021qqe}. Finally, and perhaps most 
interestingly, there are very close connections to a beautiful   
series of works on enumerative algebraic
geometry by  G{\"o}ttsche, Kool, Nakajima, and Williams \cite{Gottsche:2010ig,Gottsche:2020ale}.

The case of pure $SU(N)$ theory was studied in \cite{Losev:1998,Marino:1998bm}, 
and in \cite{Marino:1998bm} the wall-crossing technique was employed to give 
explicit formulae for the $SU(N)$ Donaldson invariants in terms of Seiberg-Witten 
invariants. In principle the formulae can be generalized to all higher rank compact 
semisimple gauge groups. This has not been done, and it would be interesting to do so
since it would lead to nontrivial predictions about Floer homology following the line 
of reasoning in \cite{Marino:1998eg,AliakbarDaemi}. 
The mathematical challenges to verifying these physical predictions are formidable but nontrivial progress has been made \cite{KronheimerHigherRank,AliakbarDaemi}. Generalizing the work of 
\cite{Korpas:2017qdo,Korpas:2019ava,Korpas:2019cwg,Manschot:2021qqe}  to the case of 
higher rank theories on $b_{2}^{+}=1$ manifolds is completely open, and presents an 
interesting challenge that will probably lead to new generalizations of the notion of 
automorphic functions. It would be very interesting to extend the relation of the results of \cite{Manschot:2021qqe} with \cite{Gottsche:2010ig,Gottsche:2020ale} to the higher rank case.

There are a number of other interesting and valuable generalizations of Coulomb branch 
integrals which should be pursued in the future, and we will describe some of these briefly.

One  important generalization is to consider suitable couplings to supergravity
to give field theoretic interpretations of Donaldson and Seiberg-Witten invariants associated
with \emph{families} of four-manifolds. This should connect to some old ideas of Donaldson concerning
generalizations of the Donaldson invariants to give interesting equivariant cohomology classes on
${\rm BDiff}(X)$ of a four-manifold $X$ \cite{DonaldsonYangMillsInvariants,DonaldsonReview}. 
There has been  some limited mathematical 
investigation of this topic. See \cite{LiuLi,NakamuraFamilies1,NakamuraFamilies2,RubermanPolynomial,BaragliaFamilies1,BaragliaFamilies2,BaragliaFamilies3,KonnoCharacteristic} for some examples. 
There has been some 
recent definite progress in formulating the relevant physical theories 
for addressing these problems \cite{CushingEtAl} (which explores in great detail a suggestion made in \cite{Moore:1997pc}). 
Nevertheless,  a great deal remains to be done. One key issue is 
finding a suitable generalizations of the $0$- and $2$-observables 
of Donaldson theory to the families case. It is natural to expect that there will be a family version of the ``Witten conjecture'' relating Donaldson and Seiberg-Witten invariants, but no concrete results on this have been published. Recently there have been some exciting developments on the mathematical side concerning somewhat related problems (i.e. new results on the diffeomorphism groups of four-manifolds) 
\cite{Watanabe-1,Watanabe-2,Watanabe-3,Watanabe-4}. It would be very interesting to see if these are connected to the invariants coming from SYM.

There should be new Coulomb branch integrals associated to 5d and $6d$  susy gauge
theory, a topic which has been explored in
\cite{Nekrasov:1996cz,Losev:1998,Closset:2021lhd,GottscheNakajimaYoshioka,Gottsche:2021ihz,Closset:2022vjj,KimManschotEtAl}. Mathematically,
the $5d$ invariants are related to ``$K$-theoretic generalizations of Donaldson invariants.'' 
Roughly speaking, one includes an insertion of the $\hat A$ genus in the intersection 
theory on the moduli space of instantons. 
Note, that on the physics side, for $4d/5d/6d$ theories with unitary gauge groups the
relevant contact terms were computed in \cite{Nekrasov:2002qd, losev2003small, nekrasov2003seibergwitten,Marshakov:2006ii,Nekrasov:2012xe,Kimura:2015rgi}, using several new ideas: $\Omega$-deformation \cite{Nekrasov:2002qd}, noncommutative field theory
\cite{NekrasovS:1998}, and localization. Very recently the Coulomb branch integral 
for such $5d$ invariants has been under very careful and intense scrutiny from 
several points of view \cite{Closset:2022vjj,KimManschotEtAl} with a view to reproducing 
and generalizing the results of \cite{GottscheNakajimaYoshioka}.

Moving on to the six-dimensional approach to the   Donaldson invariants 
and their generalizations, many new issues arise.   From one point of view, one would like 
to generalize the 4d theories from Lagrangian theories to all theories of class $S$. 
In \cite{Manschot:2021qqe}, section $8$, a number of open problems related to the formulation of general ``$u$-plane integrals'' for such theories  were spelled out, 
and they will not be repeated here - but they constitute an interesting set of directions for future progress.  Moreover, ever since the work of Vafa and Witten on $S$-duality of twisted ${\CalN}=4$ theory \cite{Vafa:1994tf}  a six dimensional perspective on four-manifold invariants has been pursued.  In particular, a six dimensional lift of Donaldson theory has been considered in section 5.6 of \cite{NikThesis}. A dual viewpoint emerges:  two dimensional $(0,2)$-supersymmetric sigma models with instanton moduli spaces as targets on the one hand, and the double periodic version of Seiberg-Witten theory, on the other \cite{NikFive, HIV:2003, BH:2003}. The first viewpoint upgrades Donaldson polynomials to the
elliptic genus of instanton moduli space with some bundle of matter zero-modes, mathematically viewed as moduli space of Higgs sheaves on a complex surface \cite{Gottsche:2019vbi}. It is related to the attempts to
find the four dimensional analogues of RCFT and the related ``four-dimensional Verlinde formula'' of \cite{Avatars}.

{}The 6d viewpoint is also related to attempts to understand a 
field-theoretic origin of Nakajima algebras. 
The cohomology of the six dimensional counterpart of the Donaldson supercharge contains an infinite dimensional chiral algebra. 
(In this context it is worth recalling that elliptic genera of instanton moduli spaces are used
in the attempts to build the second quantized string theory from Matrix theory, 
and in the calculations of black hole entropy \cite{DMVV:1997}, suggesting 
further intriguing potential connections.)

There are several more recent, but closely related, developments. References 
\cite{Dedushenko:2017tdw,Feigin:2018bkf,Gukov:2018iiq} suggested that there might be new four-manifold invariants associated
with the compactification of a $6d$ theory on a four-manifold. Roughly, one considers
the \emph{tmf} class of the resulting $2d$ susy QFT.
However, it remains to be seen if the \emph{tmf} class really captures new topological information beyond the homotopy type of the four-manifold. 
It is not obvious that the tmf class will depend on more than the homotopy type since the 
IR theory on the Coulomb branch depends only on the homotopy type of the four-manifold. 
Relations to vertex operator algebra theory, generalizing the profound and remarkable
results of Nakajima relating the cohomology of moduli spaces of instantons to
representations of affine Lie algebras, are also found in 
\cite{Dedushenko:2017tdw,Feigin:2018bkf,Gukov:2018iiq}.

Finally, as has been stressed by Witten, one of the most promising directions
for the discovery of new four-manifold invariants is to make good mathematical
sense of a categorification of Vafa-Witten invariants.

\subsection{Hyperk{\"a}hler and Quaternionic K{\"a}hler Geometry}

Hyperk{\"a}hler metrics play an important role in investigations of
moduli spaces of vacua of supersymmetric field theories. They
have long been a source of fascination within mathematics.
Indeed the hyperk{\"a}hler quotient construction grew out of an
extremely fruitful collaboration between physicists and mathematicians in \cite{Hitchin:1986ea}.

More recently, investigations into the BPS spectrum of field theory and their
wall-crossing properties led to a new construction of hyperk{\"a}hler metrics
based on integral equations formally equivalent to those appearing in the
thermodynamic Bethe ansatz (TBA) \cite{Gaiotto:2010okc}.  The new technique
might lead to useful exact results in hyperk{\"a}hler geometry.
There have been interesting rigorous mathematical \cite{Fredrickson:2018vok} and numerical \cite{Dumas:2018qpt,Dumas:2020zoz}
checks of  \cite{Gaiotto:2010okc}, but some basic questions (like existence and
uniqueness of solutions of the TBA equations) remain open.
The techniques can, in principle, be extended to give exact solutions of
Hitchin's equations on surfaces  \cite{Gaiotto:2011tf} but this
has been much less explored.

One of the initial hopes for the construction \cite{Gaiotto:2010okc}
was the idea that one could use
the construction to produce exact formulae for $K3$ metrics. For example,
one could combine these ideas with the discussion of the
$D3$ probe in $F$-theory as a deformation of a system consisting of
four copies of the $SU(2)$, $N_{f}=4$, theory \cite{Banks:1996nj}. The
main difficulty in this approach is that of determining the relevant
spectral networks and BPS spectrum, since the $D3$ probe theory is not a standard quantum
field theory. Recently this idea has been pushed further in the context of
little string theory in  \cite{Kachru:2020tat,Tripathy:2020mjj}. It would
be a major achievement of the mathematics-physics dialogue
 if these developments led to a tractable and exact formula
for the metric on any smooth nondegenerate $K3$ surfaces.

There has been some effort to extend the new constructions of
hyperk{\"a}hler metrics to quaternionic K{\"a}hler metrics. The two
are closely related by the Swann construction. These metrics
are of importance in deriving the hypermultiplet moduli spaces of
the supergravity LEET of four-dimensional ${\CalN}=2$ string compactifications.
Progress has been made in   
\cite{Alexandrov:2008gh,Alexandrov:2010qdt,Alexandrov:2010ca,Alexandrov:2011ac,Alexandrov:2012pr,Alexandrov:2013yva}
although obstacles remain because of the large growth of BPS black hole
entropy as a function of charge, and the related issue of how to account for NS5-brane instanton contributions.

Recently there have been some very interesting developments
concerning complex hyperk{\"a}hler geometries associated with
BPS states. These results shed light on some old work of
D.~Joyce on wall-crossing invariants and perhaps will
lead to progress in the construction of quaternionic
K{\"a}hler metrics \cite{Alexandrov:2021prq,Alexandrov:2021wxu,Alim:2021gtw,Alim:2021mhp,Bridgeland:2019fbi,Bridgeland:2020zjh}.

Finally, the physics of BPS states makes nontrivial predictions 
about the cohomology and differential geometry of certain hyperk{\"a}hler 
manifolds. Examples of the relevant manifolds include moduli spaces of magnetic 
monopoles on $\mathbb{R}^3$ in Yang-Mills-Higgs theory. A remarkable 
paper of A. Sen used the predictions of $S$-duality of ${\CalN}=4$ SYM to derive 
nontrivial conjectures about the existence of self-dual normalizable harmonic forms on monopole moduli space \cite{Sen:1994yi}. Historically, this paper provided a large impetus for the fundamental use of strong-weak duality symmetries in investigations of supersymmetric field theory and string theory. The essential technique used in \cite{Sen:1994yi} is the collective coordinate description of BPS states in weakly coupled field theories. 
Some of Sen's predictions were rigorously proven in \cite{Segal:1996eb}.  
When the technique of collective coordinate quantization is applied 
to describe BPS states in general ${\CalN}=2$, $d=4$ field theories physical 
statements about BPS states translate into nontrivial generalizations of 
Sen's predictions and can be phrased as general predictions about the $L^2$ kernels of Dirac operators coupled to hyperholomorphic bundles over monopole moduli space \cite{Gauntlett:1995fu,Moore:2015qyu}. In some analogous situations the physical predictions based on BPS states about the delicate questions regarding the kernels of Dirac operators on noncompact spaces have been rigorously verified \cite{Sethi:1997pa}. 

The topology - specifically the homology and cohomology - 
of Hitchin moduli spaces and character varieties has been the subject of much work beginning with a study of 
SYZ-type \cite{Strominger:1996it} mirror symmetry and its relation to Langlands duality \cite{Hausel:2002ap}. 
In another direction, a very general conjecture known as the ``$P=W$'' conjecture 
has been extensively studied in 
\cite{hausel2008mixed,HauselVillegas2011,
hausel2016arithmetic,
decataldo2011topology,decataldo2010exchange,
shen2018perverse,mauri2021geometric,szabo2019simpsons,
decataldo2021hitchin,decataldo2020pw,felisetti2022pw,
szabo2021pw,davison2021nonabelian}. The results can be 
interpreted very nicely  using  
BPS states associated with  string theory compactification on 
local Calabi-Yau manifolds 
\cite{Chuang:2012dv,Chuang:2013wpa,Diaconescu:2017tga,
Chuang:2018fks}.

\section{Geometrization Of Quantum Field Theory}
\label{sec:GeoQFT}

Geometric engineering of quantum field theory (QFT) broadly refers to the various methods used to define QFTs from a 
certain geometric scheme. This provides a definition of QFTs beyond the standard paradigm of Lagrangian theories and gives access to strongly-coupled regimes. Broadly speaking, the following approaches fall under the header of geometric engineering: 
\begin{enumerate}
    \item {\bf Geometric engineering:} The original framework that establishes the paradigm of geometric engineering is the construction of QFTs by decoupling the low-dimensional non-gravitational degrees of freedom of string or M-theory on a non-compact (and not necessarily smooth) space $X$. Some constructions of this type can be thought of as the limits of large volume  of the standard string compactifications on $X$, where gravity has been decoupled. 
    The requirement of supersymmetry in the resulting QFT usually implies that $X$ has reduced holonomy. 
    \item {\bf Branes:} QFTs can be geometrically defined by also studying the low energy dynamics of branes and various non-perturbative states of string and M-theory.  These can be obtained in diverse backgrounds which can be flat, have reduced holonomy, be singular, and/or be supported by flux.  
    \item {\bf QFT$_d \rightarrow {\text{QFT}_{d-n}}$:} Another powerful scheme of geometric engineering is to obtain QFTs in $d$-dimensions by reducing higher-dimensional field theories on manifolds $\Sigma$, possibly with boundaries.  In this paradigm, the QFT in $d$ dimensions is defined by the higher-$d$ theory, the geometry and topology of $\Sigma$, and the specific data for the reduction such as choice of topological twist, fluxes for flavor symmetry, and boundary conditions on $\Sigma$ for the higher-$d$ theory.  This approach is complemented by the previous methods when the higher-$d$ theory admits a geometric definition.  The reduction of the theory has a dual geometric picture where specific brane setups are wrapped on $\Sigma$.  
     \item {\bf Holography:} QFTs can also be defined geometrically (though maybe the terminology of geometic engineering is less standard in this context) by constructing $AdS_{d+1} \times M$ solutions in string and M-theory supergravity.  The reduction of string and M-theory on $M$ provides a supergravity theory on $AdS_{d+1}$ which then defines a conformal field theory in $d$ dimensions by the AdS/CFT dictionary.  This method is intimately related to the previous one since the $AdS$ solutions often describe the near-horizon limit of brane configurations.  
\end{enumerate}

Many of the above geometric realizations are inter-connected. Numerous theories have descriptions in several of the above constructions, which are related by string dualities. An example is the 4d $\mathcal{N}=4$ Super-Yang Mills theory with gauge algebra $\mathfrak{su}(N)$: 
it has a constuction in terms of Type IIB on  $\mathbb{C}^2/\mathbb{Z}_N \times T^2$, which can be thought of as a non-compact K3-surface times a torus; as theory on a stack of $N$ D3-branes (after decoupling the center of mass mode), as the 6d $(2,0)$ compactified on a $T^2$, and finally it has as holographic description in terms of the Type IIB supergravity solution AdS$_5\times S^5$. Different realizations can provide at times better control of certain aspects. E.g. some symmetries may be  manifest in some descriptions and not in others.

The holographic setting provides a unique laboratory to explore strongly coupled QFTs via weakly coupled gravitational theories.  Indeed, this has been most fruitful in cases where the QFTs have large number of degrees of freedom where the dual gravitational theory is in its classical regimes.  There is a very close connection between brane-engineering and holography in terms of the underlying geometies. The prime example here are the Sasaki-Einstein manifolds $X_5$ that appear in the constuction of AdS$_5$ duals to 4d $\mathcal{N}=1$ SCFTs. In turn these SCFTs can be constucted by probing the singular Calabi-Yau three-folds, that are the cones over the Sasaki-Einstein manifolds $X_5$ \cite{Gauntlett:2005ww}.

Geometric engineering provides new perspectives in the study of QFTs.  First it often provides a way to study both  weakly-coupled and strongly-coupled QFTs using one geometric framework. 
It provides a natural setting to extract and explore topological data (such as the generalized symmetries) and invariant quantities of RG-flows associated to QFTs  such as 't Hooft anomalies. Geometric structures of QFTs such as moduli space of vacua, conformal manifolds, duality relations, and RG flows are also naturally realizable in geometric engineering.  This paradigm also provides a way to make precise a classification program for QFTs and sharpens the notion of spaces of QFTs. 

This is often aided by the fact that the QFT classification program can be mapped to a geometric one, where it finds a mathematically well-defined formulation. 
 These can be for example geometries $X$ with reduced holonomy and certain singularities, geometries $M$ from solutions of supergravity equations or choices of boundary conditions and topological twists on compact manifolds $\Sigma$.  The interconnection of QFTs and geometry is also a point of contact between the mathematics and physics communities in the seminar series \cite{QFTandGeo}. 

\subsection{Geometric Classification of Superconformal Field Theories}
\label{sec:GeoSCFT}

The canonical setup of geometric engineering will preserve some supersymmetry, which in the simplest case implies that $X$ has reduced holonomy. Examples of this are 
Calabi-Yau $n$-folds as well as exceptional holonomy $G_2$ and $\text{Spin}(7)$ manifolds. 
The use of geometrically engineering QFTs is three-fold: 
1. it allows the construction of strongly-coupled QFTs, 2. important properties such as symmetries and their 't Hooft anomalies are encoded in the string theory construction, 3. the geometrization enables a classification program. 

In the recent past, the utility of this approach was realized in the context of a geometric classification of 6d superconformal field theories (SCFTs) \cite{Heckman:2013pva,Heckman:2015bfa, Bhardwaj:2015xxa}, and geometric exploration of 5d SCFTs starting with \cite{Jefferson:2017ahm, Jefferson:2018irk, Bhardwaj:2018yhy,Bhardwaj:2018vuu, Apruzzi:2019enx, Apruzzi:2019opn, Apruzzi:2019vpe, Bhardwaj:2019fzv}. SCFTs in dimensions 5 and 6 are automatically strongly-coupled, UV fixed points \cite{Seiberg:1996bd, Morrison:1996xf}, and not accessible with perturbative QFT methods. It is the existence of such strongly-coupled SCFTs, which is predicted by string theory. 

\subsubsection{6d SCFTs}
6d SCFTs have a geometric engineering realization in Type IIB string theory. The theories with $\mathcal{N}=(2,0)$ supersymmetry are classified by an ADE gauge algebra, and have a construction from $\mathbb{C}^2/\Gamma_{ADE}$, where $\Gamma_{ADE} \subset SU(2)$ is a finite subgroup of ADE type. 
Note, that these SCFTs are not absolute theories, see section \ref{sec:tHooftAnomalies}. The absolute theories in 6d with maximal supersymmetry are tabulated in \cite{Gukov:2020btk}. 

In contrast, the classification of 6d $(1,0)$ theories is geometrically much more intricate: these theories have a realization in terms of F-theory compactifications on elliptically fibered Calabi-Yau three-folds. F-theory is Type IIB string theory, where the axio-dilaton $\tau = c_0 + 1/g_s$ is not constant, but non-trivially fibered over the type IIB compactification space. For elliptic Calabi-Yau threefolds the base is a K\"ahler surface $B_2$ (and only for trivial fibrations is the IIB compactification space Ricci-flat). 
In the simplest instance when the elliptic fibration has a zero-section, it can be written in Weierstrass form 
$y^2 = x^3 + fx + g$, 
where $f$ and $g$ are sections of the canonical bundle of $B_2$. The elliptic fibration develops singularities, which follow (at least in codimension one in the base) the Kodaira classification of singular fibers, which is characterized in terms of the vanishing order  along a discriminant component $z=0$:
$(\text{ord}_{z} f,\text{ord}_{z} g, \text{ord}_{z} \Delta)  < (4,6,12)$, where $\Delta = 4 f^3+ 27 g^2$. 
Here $z$ is a local coordinate on the base $B_2$. SCFTs arise when the vanishing orders satisfy $\geq (4,6,12)$, and are so-called non-minimal fibers. 
The classification result in \cite{Heckman:2013pva,Heckman:2015bfa, Bhardwaj:2015xxa} states that the base geometries for 6d SCFTs are $B_2 = \mathbb{C}^2/\Gamma$ where now $\Gamma \subset U(2)$. For a review see \cite{Heckman:2018jxk}.
Resolving these singularities in the base, replaces $B_2$ with $\tilde{B}_2$, which is resolved by a collection of rational curves. This characterizes the tensor branch of the 6d SCFT, i.e. the moduli space that is characterized by giving vevs to scalars in the tensor multiplet. 
The Kodaira type of the singular fiber above compact (non-compact) curve in the base, maps in $F$-theory to the gauge (flavour symmetry) group of the $6d$ theory. Additional codimension two singularities occur when two components of the  discriminant  intersect. These correspond in the 6d theory to hypermultiplet matter. 

\paragraph{Frozen Phase of F-theory.}
In the realm of 6d SCFTs there are in addition so-called frozen singularities \cite{deBoer:2001wca, Tachikawa:2015wka, Bhardwaj:2018jgp}, which correspond to adding $O7^+$-planes into the F-theory compactification. One of the main challenges in the 6d classification program is a first principle understanding of these frozen phases, the anomaly cancellation and dual M-theory description. 
To provide a precise definition of the frozen phase of F-theory one would need to determine what corresponds to a ``frozen" F-theory background, when no dual IIA description exists. This is an important outstanding problem, which plays a central role in establishing a mathematically complete classification of 6d SCFTs.

\subsubsection{5d SCFTs}
A closely related question is the classification program of $5d$ ${\CalN}=1$ SCFTs. Again, these are strongly-coupled fixed points, whose existence is argued using string theory \cite{Seiberg:1996bd, Morrison:1996xf,Intriligator:1997pq,Douglas:1996xp}. The conjectured classification of $5d$ SCFTs is in terms of canonical singularities: i.e. non-compact Calabi-Yau three-folds $X$, with singularities that admit a resolution $\pi:\widetilde{X} \rightarrow X$, such that the canonical classes  
$
K_{\tilde{X}} =\pi^{-1} K_{X} + \sum_i  a_i D_i$, where $a_i \geq 0$,
where $D_i$ are the exceptional divisors of the blowup. For $a_i=0$ these admit so-called crepant resolutions. When $a_i >0$ the singularity is terminal. 
Canonical singularities also admit complex structure deformations. 
There are two types of moduli spaces for 5d SCFTs: the Coulomb branch and Higgs branch. The latter has the structure  of  a  hyper-K{\"a}hler cone. 

The precise dictionary for canonical singularities to $5d$ SCFTs is extremely well developed and understood for the case when crepant relations exist. 
The resolved geometry describes the Coulomb branch of the 5d SCFT: this is an abelian gauge group with matter and Chern-Simons couplings. At special subloci of the Coulomb branch there can be non-abelian gauge theory descriptions -- but not all theories have such a Lagrangian description. 
When the geometries have terminal singularities, the interpretation of the 5d SCFTs is not fully understood, although recent progress in understanding their deformations (and thereby the Higgs branch of the associated theories), has been made \cite{Closset:2020scj, Closset:2020afy, Closset:2021lwy}.

A classification program of 5d SCFTs was initiated using various approaches. The most systematic thus far is starting with the geometry underlying 6d SCFTs, and applying geometric deformations to these geometries. In particular $M$-theory on elliptically fibered Calabi-Yau three-folds gives rise to $5d$ KK-theories, which UV complete in $6d$ \cite{Jefferson:2017ahm}. Starting from these theories a systematic exploration can be carried out \cite{Bhardwaj:2018yhy,Bhardwaj:2018vuu, Apruzzi:2019enx, Apruzzi:2019opn, Apruzzi:2019vpe, Bhardwaj:2019fzv}, including salient features such as the symmetries and dualities of such theories. It remains a challenging geometric question to prove that these constructions from 6d result in a full classification of $5d$ SCFTs.

\paragraph{Future Directions.}

Mathematically canonical singularities for $3$-folds are in theory understood, based on the Mori minimal model program in algebraic geometry (for an introduction see \cite{Matsuki}). For $3$-folds this is developed in much detail to the extent that an algorithm exists, how to perform crepant and small  resolutions, and identify terminal singularities. The issue however is that the explicit classification of canonical three-fold singularities does not exist (as in a list, or a concrete realization of all such singularities). Clearly, the classification of $5d$ SCFTs provides some possible approach: starting with the gluing of compact surfaces and collapsing these to singularities as in \cite{Jefferson:2018irk}. However this approach has not resulted in a characterization of the Calabi-Yau singularities, nor does it incorporate terminal singularities. Combining the abstract insights from the minimal model program, combined with the  concrete construction of $5d$ SCFT from collections of compact surfaces, should be a very promising starting point to provide a concrete classification of canonical singularities. Note that for $n$-fold canonical singularities with $n>3$, the minimal model program is far less explicit than for $3$-folds, so that further insights from geometric engineering of QFTs (e.g. $3d$ theories or $4d$ theories for elliptic Calabi-Yau four-folds) could provide some guidance.

An important geometric question is the precise connection between canonical singularities and the quantum Higgs branch of the associated $5d$ SCFT.  What is the interpretation of terminal singularities? Some aspects of this are discussed in \cite{Closset:2020scj, Collinucci:2021ofd}.
Another important question is the inter-relation of SCFTs accross  dimensions: e.g. it is not known whether all $5d$ SCFTs be obtained from $6d$ SCFTs, using a circle-reduction and deformations (such as flavor Wilson lines, automorphism twists). Likewise an even more challenging question is the classification of $4d$ SCFTs with $8$ or $4$ supercharges. 

Developing a unified framework -- e.g.  mapping geometric realizations to brane-constructions, and vice versa -- will be essential in developing a complete classification of 5d SCFTs. Currently the descriptions have a large overlap, but there are theories, which evade constructions in one of the two frameworks. 
In particular, toric Calabi-Yau three-folds are dual to certain $5$-brane brane-webs. However, webs where multiple $5$-branes can end on a single $7$-brane -- so-called "white dots"  \cite{Benini:2009gi} --  are not understood in the geometry. Physically this is again interlinked by Higgsing. Mathematically this will lead to a generalized notion of toric geometry, as anticipated in \cite{Cabrera:2018jxt, VanBeest:2020kxw}.

There are numerous constructions for $4d$ SCFTs and many approaches to their characterization and classification have been put forward. Here we focus on the geometric engineering constructions. 
 We discuss in section \ref{sec:DimRedux} constructions starting from higher dimensional QFTs, such as class ${\CalS}$.
 Using geometric engineering they can be obtained in Type IIB string theory on canonical singularities -- i.e. the same spaces that underlie the construction of 5d SCFTs in $M$-theory \cite{Shapere:1999xr,Xie:2015rpa, Chen:2016bzh, Wang:2016yha, Chen:2017wkw}. See \cite{Akhond:2021xio} for a recent pedagogical overview.

\subsubsection{Moduli Spaces of Theories with $8$ Supercharges}

A beautiful connection between theories with $8$ supercharges has emerged and made precise in the last few years, which interconnects the theories by dimensional reduction (with added benefits), and the relation between their moduli spaces. Progress in the recent years has shown that the Higgs branches of higher-dimensional theories (5d, 4d, in particular) can be characterized in terms of an auxiliary 3d theory -- the magnetic quiver theory \cite{Cabrera:2019izd,Bourget:2019aer}. 

The Higgs branch of a 5d SCFT is related to the versal deformations of the Calabi-Yau singularity that engineers the theory in M-theory. However the Higgs branch receives quantum corrections. 
For isolated toric or hypersurface singularities \cite{Closset:2020scj} the deformation theory is well-defined, and in some instances the full quantum Higgs branch can be determined. In general this is however not understood, in particular for non-isolated singularities. Using an alternative description of 5d SCFTs, applicable for a large class of theories albeit not all, based on $5$-brane webs, an algorithm for determining the Higgs branches using auxiliary $3d$ quiver gauge theories   \cite{Bourget:2019rtl, Bourget:2019aer}. These so-called magnetic quivers have $3d$ ${\CalN}=4$ supersymmetry and conjecturally, their Coulomb branch gives rise to the Higgs branch of the 5d SCFT. The Coulomb branch of 3d ${\CalN}=4$ quivers can be given some  characterization by various $3d$ partition functions (see {\it e.g.} \cite{Benini:2015noa,Willett:2016adv,Closset:2017zgf,Benini:2016hjo}) with the Hilbert series being the simplest and most useful example. The expression for the Coulomb branch Hilbert series  first obtained in \cite{Cremonesi:2013lqa} is often called {\it the monopole formula} and for Lagrangian theories it can be obtained as a limit of the supersymmetric index \cite{Razamat:2014pta}.
 The Coulomb branches were also studied  using the algebra of protected local operators in  by Bullimore, Dimofte, Gaiotto \cite{Bullimore:2015lsa}, and 
 from more   mathematical perspective in the  works of Braverman, Finkelstein and Nakajima \cite{Nakajima:2015txa,Braverman:2016wma}. This links the geometric engineering program of SCFT to developments in 3d mirror symmetry and symplectic singularities.

Deriving the Higgs branch from a geometric description, and proving the conjecture connecting it to the Coulomb branch of the $3d$ magnetic quiver is one of the outstanding question in this area. This program of developing the quantum moduli space (including its global stratification structure into Higgs and Coulomb branches) will develop in parallel to the classification of theories with 8 supercharges.

\subsubsection{Generalized Symmetries and Anomaly Theories}

Not only does geometric engineering construct QFTs, it also allows the computation of physical properties of the theories. 
This is of particular utility when these theories are strongly-coupled and non-Lagrangian. Generalized global symmetries, higher-group symmetries and their anomalies, as discussed in section \ref{sec:Anomalies}, can be derived from the geometry. 

Global symmetries are encoded in particular relative homology classes of the space $X$, relative to its boundary $\partial X$: non-compact (relative) $q$-cycles in $H_{q} (X, \partial X, \mathbb{Z})$, can be wrapped by branes, and give rise to defect operators. 
The defect operators modulo screening by local operators  is defined as the set of such relative cycles, modulo screening branes wrapping compact $q$-cycles, which can be obtained by wrapping branes and their magnetic duals \cite{DelZotto:2015isa,Morrison:2020ool, GarciaEtxebarria:2019caf, Bhardwaj:2020phs, Albertini:2020mdx, Apruzzi:2021mlh,  Cvetic:2021sxm, Hubner:2022kxr, DelZotto:2022fnw, Cvetic:2022imb, DelZotto:2022joo}. 

Anomalies for generalized higher-form symmetries are formulated in terms of the background fields: for $q$-form symmetries these are $B\in H_{q+1} (M_d, \Gamma^{(q)})$, where $M$ is the spacetime. In M/string theory these are realized from supergravity form-fields expanded on the relative cohomology cycles of the compactification space $X$. The symmetry topological field theory is the $d+1$ dimensional TQFT, which upon imposing boundary conditions for the background fields, gives rise to the anomaly theory of the $d$-dimensional QFT. 
More generally anomalies for discrete symmetries require a refinement to generalized cohomology theories: the background fields for discrete symmetries descend from torsion cycles, which can be incorporated by treating the string/M-theory fields as cocycles in a differential cohomology theory
 \cite{CheegerSimons, Hopkins:2002rd}. See 
 \cite{Freed:2000ta,Freed:2006yc} for introductions to differential cohomology aimed at physicists. 
In M-theory it appears that the differential cohomology theory has an 
underlying quantization based on ($w_4$-shifted) singular cohomology theory, 
but in type II string theory the relevant cohomology theory appears to be 
differential $K$-theory 
\cite{Belov:2006xj,Freed:2000ta,Freed:2006ya,Freed:2006yc,Moore:1999gb}.
 This differential cohomology approach to anomaly theories and symmetry TFTs was developed for higher-form symmetries in 
M-theory compactifications in \cite{Apruzzi:2021nmk}:
The reduction on $\partial X$ results in the SymTFT, which after choosing boundary conditions, gives rise to the anomaly theory for 0- and 1-form symmetries.

The computation of anomaly theories can receive several contributions.  The canonical ones come from reducing the topological interactions of string or M-theory.  Another contribution comes from St{\"u}ckelberg mechanisms that exist from the reduction of string or $M$-theory.  In addition to providing new anomaly terms, these mechanisms also reduce the continuous symmetries from string or $M$-theory to discrete symmetries.  The existence of such mechanisms occur when the various q-form gauge symmetries in string or $M$-theory fail to admit equivariant extensions with respect to the isometry group of the manifold where the theory is being reduced on.  A systematic study of anomaly theories for continuous symmetries and the various obstructions that can lead to St{\"u}ckelberg mechanisms have been initiated in \cite{Bah:2018jrv,Bah:2019jts,Bah:2019rgq,Bah:2019vmq,Bah:2020jas,Bah:2020uev,Bergman:2020ifi}.

Clearly the field of generalized symmetries, higher-groups and categorical symmetries has great potential for more applications in the context of geometric engineering. Higher-group symmetries can be computed from the geometry, but anomalies for these should also fall into the framework of symmetry TFTs and should arise from string theory. Non-invertible symmetry in higher dimensions have recently been constructed within string theory \cite{Apruzzi:2022rei, Gaiotto:2020iye, GarciaEtxebarria:2022vzq, Heckman:2022muc}, and  play a  role in the context of the swampland program \cite{Heidenreich:2021xpr}.

\subsection{Manifolds with Exceptional Holonomy}
\label{sec:ExHol}

Special holonomy manifolds and calibrated geometry
have played a central role in string compactification
and in the study and application of D-branes and flux
compactifications ever since Calabi-Yau manifolds were
recognized as being of great importance in supersymmetric
string compactification  \cite{Candelas:1985en}.  In particular,
Calabi-Yau manifolds and holomorphic curves and bundles
have been objects of central interest, and have been at the
focus of developments in enumerative algebraic geometry.
Indeed, the entire subject of mirror symmetry - currently
an important part of modern mathematical research grew
out of these developments.

It is known from the Berger classification that there are
fascinating exceptional cases of special holonomy manifolds
in seven and eight dimensions with holonomy groups $G_2$ and
${\rm Spin}(7)$, respectively. Moreover, such manifolds admit
natural generalizations of the instanton equations.   Construction of
such manifolds and instantons, and of their moduli spaces present serious
challenges to differential geometers and analysts, but would be highly desirable in physics.

Enumerative problems in exceptional holonomy spaces could quite possibly see substatial progress. One such question is related to the counting of associative three-manifolds in manifolds of $G_2$ holonomy and the related problem of counting of $G_2$ instantons \cite{Donaldson:2009yq, Joyce:2016fij, Doan:2017qbq}, and a clearer understanding of their moduli spaces.
The main mathematical challenge is that associative three-cycles can degenerate, by either forming singularities or splitting. A consequence of this phenomenon is that the signed count of associative cycles is not an invariant on the moduli of special holonomy metrics.  Several proposals have been put forward to deal with this challenge \cite{Joyce:2016fij, Doan:2017qbq}.
Complementing this, there is an $M$-theory proposal of interpreting the superpotential of the 4d QFT, which is obtained by compatifying M-theory on a $G_2$-holobnomy manifold, as the counting function of associative cycles based on the observation that the effective superpotential obtains non-perturbative contributions from M2-branes wrapping associative three-cycles \cite{Harvey:1999as}. An open problem raised by \cite{Harvey:1999as} is the computation of the contributions of non-isolated associative cycles. The formula for isolated cycles involves the Ray-Singer torsion. Since torsion is most naturally interpreted as a measure it is reasonable to conjecture that the formula of \cite{Harvey:1999as} extends to non-isolated cycles, but this remains to be demonstrated from M-theory. There is evidence for the existence of infinitely many associative cycles in certain twisted connected sum $G_2$-manifolds \cite{Halverson:2014tya, Braun:2018fdp, Acharya:2018nbo}. 
A Simons Collaboration on Special Holonomy in Geometry, Analysis, and Physics \cite{SimonsCollab:SpecialHolonomy}
has focused on problems in this general
area and has been making excellent progress in the multitude
of nontrivial problems presented by these geometric structures. See \cite{DonaldsonICM2018} for a relatively recent review of some of the mathematical progress that has been made.

Geometric engineering with non-compact  $G_2$ holonomy spaces has a relatively long-standing tradition starting with \cite{Atiyah:2001qf}. M-theory compactification on $G_2$ holonomy manifolds gives rise to 4d $\mathcal{N}=1$ theories. In particular, such models have interesting particle physics applications: starting with the construction of pure Super-Yang-Mills theories \cite{Atiyah:2000zz}, and supersymmetric Standard-Model type models \cite{Acharya:2001gy}. 
The basic dictionary between gauge theory and geometry is as follows: codimension 4 singularities give rise to gauge groups (4d $\mathcal{N}=1$ vectors), 
matter in chiral multiplets arises from codimension 7 singularities, whereas non-chiral matter comes from codimension 6. 
The mathematical challenge, which to this moment is unsolved, is the construction of singular $G_2$ holonomy manifolds, which have 
codimension $4$ and $7$ singularities, thus giving rise to 4d gauge theories -- coupled to gravity -- which model (semi-)realistic models in 4d. 

In the past few years, much progress in the context of the Simons Collaboration on special holonomy \cite{SimonsCollab:SpecialHolonomy} has been made. Two areas in which there has been progress are addressing more global issues and the study of   compact geometries with $G_2$ holonomy. These new compact geometries go beyond the class of torus orbiolds known as Joyce orbifolds
\cite{Joyce1,Joyce2}. 
The results in \cite{Corti:2012kd} and unpublished work by Kovalev, suggested an infinite class of compact $G_2$ holonomy manifolds, which admit a fibration by K3-surfaces over three-manifolds (mostly $S^3$ and quotients thereof).  

The geometries found in \cite{Corti:2012kd} have already had numerous implications within physics: 
in \cite{Braun:2017uku} models with conjectural codimension 4 and 6 singularities were constructed as singular limits of these geometries. 
Such K3-fibered $G_2$ holonomy compactifications have dual descriptions in string theory, F-theory and heterotic strings, where the generation of chiral matter is well-understood, and could guide the mathematical efforts in the construction of compact $G_2$-geometries with chiral matter in 4d \cite{Pantev:2009de,Braun:2017uku,Braun:2018vhk,Barbosa:2019bgh,Cvetic:2020piw}. 

Another interface between geometry and physics arises in the context of M-theory/Type IIA limits: M-theory compactified on a circle gives rise to Type IIA string theory \cite{Witten:1995ex}. Therefore, whenever a $G_2$ manifold admits a circle-fibration, there should be a relation of the M-theory compactification to a Type IIA compactification. In general the IIA compactification will include D6-branes (which geometrize once uplifted to M-theory). A large class of new such $G_2$ geometries was uncovered in \cite{Foscolo:2018mfs}, motivated by string theory, and their physics was discussed in \cite{Acharya:2020vmg}. 

 Mathematically the most important challenge remains the construction of  compact but singular $G_2$-manifolds, which have singularities both in codimension 4 and 7. Physically, these correspond to realizing gauge degrees of freedom (codimension 4 singularities e.g. of ADE-type) and chiral matter in 4d (codimension 7). 
 Note that the singular limits of twisted connected sum $G_2$-manifolds, which have been studied in the physics literature, have codimension 4 and 6 singularities, which only realized gauge groups and non-chiral matter, and  cannot be deformed within this class of $G_2$-manifolds to realize codimension 7 singularities that correspond to chiral matter \cite{Braun:2018vhk}\footnote{Although one can get deformations that result in codimension 7 singularities, these always appear pair-wise and correspond to non-chiral matter.}.
 
 For  Calabi-Yau manifolds, Yau's proof of the Calabi conjecture provides a simple criterion for $SU(N)$ holonomy spaces. 
 In the realm of $G_2$ and Spin$(7)$ holonomy, no such theorem (or even conjecture) exists.  This makes identifying geometries with exceptional holonomy 
 particularly challenging to construct and identify, requiring highly advanced differential geometric and analytic tools. 
 
 Spin$(7)$ compactifications in M-theory result in 3d $\mathcal{N}=1$ supersymmetry and in heterotic string theory with 2d $\mathcal{N}= (1,0)$ spacetime supersymmetry. The case of one supercharge is a particularly challenging subset of SQFTs, and furthering that study of Spin$(7)$ holonomy can shed light on these QFTs. Most constructions (both in geometry and string theory) are obtained by considering anti-holomorphic involutions on Calabi-Yau four-folds due to Joyce \cite{Joyce:1999nk}, with several recent applications in terms of connected sum constructions in string theory \cite{Braun:2018joh, Cvetic:2021maf}. The main challenge is the resolution of singularities that arise in these constructions, in a  way that is compatible with the Spin$(7)$ structure.

\subsection{Dualities Across Dimensions}
\label{sec:DimRedux}

In lower space-time dimensions, $D\leq 4$, one can construct weakly coupled field theories using standard Lagrangian techniques which describe interacting CFTs. In particular one can start from a free theory deformed by relevant couplings and flow to an interacting CFT in the IR, or, in certain cases in $D=4$, deform the free theory by an exactly marginal coupling and describe an SCFT without any RG flow involved.
One can thus study CFTs, and SCFTs in particular, in $D\leq 4$, using a variety of purely field theoretic techniques. In cases when one has some supersymmetry, {\it e.g.} four supercharges in $D=4$, much non trivial physics in the IR can be directly and indirectly deduced from the Lagrangian descriptions. One very nice example is the  notion of IR duality where two or more different weakly coupled theories flowing to the same theory in the IR \cite{Seiberg:1994pq}. Another example is conformal duality where two or more weakly coupled (or partially weakly coupled) theories are related by tuning continuously an exactly marginal coupling \cite{Leigh:1995ep,Witten:1997sc,Green:2010da,Gaiotto:2009we,Razamat:2020pra,Razamat:2019vfd}. A third example is the emergence of symmetry (flavor symmetry and supersymmetry) in the IR \cite{Razamat:2017wsk,Dimofte:2012pd,Razamat:2018gbu,Sela:2019nqa,Maruyoshi:2016tqk,Maruyoshi:2016aim,Agarwal:2016pjo}. The list goes on. When studied directly in $D\leq4$ these strong coupling phenomena can be hard to understand or predict, but new insight can be obtained from considering higher dimensional theories and employing dimensional reduction. For example using known IR dualities in $D=4$ one can place the relevant theories on a circle, which can be viewed as a supersymmetric relevant deformation, flow to low energies where the theory is effectively three dimensional and deduce IR dualities in $D=3$ and adding another circle one can go to $D=2$ \cite{Niarchos:2012ah,Aharony:2013dha,Aharony:2013kma,Park:2013wta,Dolan:2011rp,Gadde:2011ia,Pasquetti:2019hxf,Hwang:2020wpd,Bottini:2021vms,Nii:2014jsa}. In this manner many previously discovered dualities ({\it e.g.} \cite{Aharony:1997gp,Giveon:2008zn,Hori:2006dk,Gadde:2013lxa,Hori:2000kt,Jafferis:2011ns}) in $D<4$ can be related to $D=4$ dualities \cite{Aganagic:2001uw,Gadde:2015wta,Dedushenko:2017osi,Sacchi:2020pet,Aharony:2017adm}. However, what about the dualities (and other phenomena) in $D=4$? 

An answer lies in connecting the lower dimensional QFTs to higher dimensional ones. One can start in $D=6$, the highest dimension admitting interacting supersymmetric SCFTs \cite{Nahm:1977tg} and turn on geometric deformations. In $D=6$ there are no interesting supersymmetric relevant deformations \cite{Cordova:2016xhm}. However, we can deform a $D=6$ theory in an interesting way by placing it on a compact surface. We already have discussed going to $D=5$ by placing the $D=6$ models on a circle. This procedure can be generalized to placing the $D=6$ SCFTs on higher dimensional surfaces. For example, we can flow to $D=4$ by utilizing a Riemann surface. Such surfaces in general are curved and thus to preserve some supersymmetry we need to perform some twisting procedure. This procedure can be used to predict existence of huge classes of $D=4$ SCFTs (and using three dimensional and four dimensional manifolds one can generalize to predicting existence of $D=3$ and $D=2$ SCFTs). One can read numerous properties of the lower dimensional theories directly from geometry. This includes, symmetries (zero form and higher form), anomalies, number and charged of relevant and marginal supersymmetric deformations, and in many cases duality relations. The canonical example of such a construction is generating $D=4$ ${\cal N}=2$ SCFTs starting with a $(2,0)$ theory \cite{Witten:1997sc,Klemm:1996bj,Gaiotto:2009hg,Gaiotto:2009we}: this is known as class ${\cal S}$. This has a vast generalization that lead to constructions of $D=4$ ${\cal N}=1$ SCFTs in \cite{Benini:2009mz,Bah:2011je,Bah:2011vv,Bah:2012dg}.  The same techniques from $(2,0)$ SCFTs can be applied also to studying compactifications of $(1,0)$ SCFTs to generate even larger classes of ${\cal N}=1$ SCFTs in $D=4$ \cite{Gaiotto:2015usa,Pasquetti:2019hxf,Kim:2017toz,Kim:2018bpg,Razamat:2018gro,Kim:2018lfo,Razamat:2020bix,Nazzal:2021tiu,Hwang:2021xyw}.  One can try to connect  the explicit field theoretic constructions of theories in $D\leq 4$ and compactifications of $D>4$ SCFTs. For example one can seek for $D=4$ field theoretic descriptions of the SCFTs obtained by compactifying $D=6$ SCFTs on a surface. Whether this can always be done and if not what are the obstructions, are important open problems. If one does find such a connection between compactifications and $D=4$ Lagrangians one can speak about an {\it across dimensions} IR duality: we have a $D=6$ SCFT deformed by geometry flowing to a $D=4$ SCFT deformed by a relevant coupling flowing to the same SQFT. Once such a dictionary between $D=4$ QFTs and $D=6$ QFTs is established many of the otherwise sporadic lower dimensional understandings can be explained and predicted using the geometry explicitly present in the construction. A similar program could be pursued starting from $D=5$ SCFTs and flowing to $D=3$ \cite{Sacchi:2021afk,Sacchi:2021wvg}.

More generally we can start from different $D=6$ SCFTs and deform them by different geometries. In general we would flow to different lower dimensional SQFTs. However in some cases we can end up with same lower dimensional models. In this case we can speak of $D=6$ IR dualities. Explaining and understanding such dualities probably should be done by going back to the geometric construction of $D=6$ SCFTs themselves in string/M/F-theory. Some examples of $D=6$ IR dualities can be found in \cite{Ohmori:2015pua,Ohmori:2015pia,Kim:2018bpg,Razamat:2019vfd,Baume:2021qho}. 

This line of research leads to many conceptual questions in SQFTs as well as to a lot of interesting physical mathematics. Let us list some of the interesting questions one can address.

\begin{enumerate}

 \item  {\bf Lagrangians vs no-Lagrangians}: Can any lower dimensional SCFT geometrically constructed starting with higher dimensional one have a lower dimensional IR (or conformal) dual? Many of the theories obtained in compactifications are often called non-Lagragian paying tribute to the fact that a Lagrangian dual is not known at a certain point in time. So this question can be phrased as whether non-Lagrangian theories in lower dimensions truly exist or not? For  example, a canonical instance of a ``non-Lagrangian'' theory, the Minahan-Nemeschansky $E_6$  ${\cal N}=2$ SCFT \cite{Minahan:1996fg},  has several Lagrangian ${\cal N}=1$ constructions \cite{Zafrir:2019hps,Etxebarria:2021lmq,Gadde:2015xta}. When one does find Lagrangian duals often the UV weakly coupled theories exhibit less symmetry, and in some cases supersymmetry, than the IR fixed point (See references above). To find across dimension dualities one thus needs to give up insisting on having all the symmetries of the IR manifest. In some cases a $D=4$ Lagrangian dual is not known, however constructions exist which start from a vanilla Lagrangian in the UV but then tune the parameters (couplings or flow to the IR) so that certain symmetries emerge and are gauged \cite{Gadde:2015xta,Razamat:2019ukg,Razamat:2020bix,Gaiotto:2015usa,Kim:2017toz,Razamat:2016dpl,Agarwal:2018ejn} (See also \cite{Ohmori:2015pia} for a $D=5$ example). Thus one can wonder whether restricting to such constructions one can construct any conceivable lower dimensional SCFT. 
 The same question can be raised regarding the higher dimensional SCFTs. We can try to construct $D=6$ or $D=5$ SQFTs giving up some symmetry and in particular space-time symmetry. For example one can consider $D=4$ quiver gauge theories in the limit the size of the quiver is large and obtain $D=6$ SCFTs in a process called dimensional deconstruction \cite{Arkani-Hamed:2001wsh,Hayling:2018fmv,Hayling:2017cva}.

 \item  {\bf Geometric origin of lower dimensional SCFTs}: Do all lower dimensional SCFTs have a geometric construction starting with higher dimensional supersymmetric ones? The geometric construction starts with a $(1,0)$ SCFT in $D=6$ which has eight supercharges so the question can be phrased as whether any supersymmetric  SCFT (or even non-supersymmetric CFT) in lower dimensions is related to a theory with eight supercharges.

 \item  {\bf Structure of the space of lower dimensional QFTs:} SQFTs in lower dimensions are interconnected by  relevant and exactly marginal deformations. What is the structure of this space? 
 How much of it is determined by higher dimensional geometry? When the $D=4$ SCFTs are obtained by direct dimensional reduction of $D=6$ SCFTs one can generally identify the relevant and the marginal deformations with the KK reduction of the stress-tensor (leading to exactly marginal deformations related to complex structure moduli) and the conserved currents of the $D=6$ SCFT \cite{Benini:2009mz,Razamat:2016dpl,Beem:2012yn,babuip}. An example of an interesting question here which received little attention is of studying and classifying ${\cal N}=1$ conformal dualities in $D=4$, that is weakly coupled cusps of $D=4$ conformal manifolds \cite{Razamat:2020pra}. Similarly, the structure of conformal manifolds in $D=3$ seserves much more intense investigation, see {\it e.g.} \cite{Baggio:2017mas,Bachas:2019jaa,Beratto:2020qyk}.
 
 \item {\bf Structure of moduli spaces}: A related question is the geometry of the moduli spaces of the SCFTs. 
 The moduli spaces of theories with eight supercharges have deep connections with mathematical physics and physical mathematics \cite{Seiberg:1994rs,Nekrasov:2002qd}. One approach to this subject, which has been vigorously pursued in recent years, is to classify   the possible geometries of $D=4$ ${\cal N}=2$ Coulomb branches and relate this to classification of $D=4$ ${\cal N}=2$ SCFTs \cite{Argyres:2015ffa,Argyres:2015gha,Argyres:2017tmj,Argyres:2018zay,Argyres:2019ngz,Argyres:2020wmq,Martone:2021ixp}. One can also study correlation functions of Coulomb branch operators in $D=4$ ${\cal N}=2$ using localization methods leading to beautiful results \cite{Gerchkovitz:2016gxx}.
 Yet another fruitful direction is to try to understand the moduli spaces of the strongly coupled SCFTs in $D=5$ and $D=6$ (again with eight supercharges) using various group theoretical methods (the so called magnetic quivers) \cite{Cabrera:2018jxt,Cabrera:2019izd,Bourget:2020asf,vanBeest:2020kou}, see above. Finally, one can also consider the  interplay between moduli space flows in higher dimensions, compactifications, and lower dimensional moduli spaces \cite{Razamat:2019mdt}. This leads to interesting interconnections between geometric engineering of lower dimensional SCFTs and the deeper structure of moduli spaces. In particular studying the interplay between flows and moduli spaces has already led to derivations of across dimensional dualities \cite{Razamat:2019ukg,Etxebarria:2021lmq} and studying this direction further could lead to many  novel insights.

 \item  {\bf Mirror symmetry in lower dimensions and physics in $D=4$:} Mirror symmetry in $D=3$ and $D=2$ is an extremely rich laboratory for deriving and studying results in physical mathematics. We understand well string theoretic and brane origins of many instances of such dualities: this structure is tightly tied to eight supercharges which for example allows for a clean distinction between Higgs and Coulomb branches. However, one can ask whether these dualities can also be understood purely using field theoretic constructions starting in $D=6$. In particular, we can try to find $D=4$ dualities which lead upon dimensional reduction to mirror symmetry. This necessarily leads to constructions with lower amounts of symmetry and supersymmetry. In fact some of the canonical cases of mirror symmetry in $D=3$  were recently imbedded in ${\cal N}=1$ $D=4$ dynamics \cite{Pasquetti:2019hxf,Hwang:2020wpd,Bottini:2021vms}. What is the mathematical implication of such an embedding?

 \item  {\bf Supersymmetric quantities:} Supersymmetry allows us to compute various quantities exactly with no regard to the values of coupling or the scale of RG flow. Often, if not always, such quantities can be related to counting problems of protected operators: various types of supersymmetric indices and limits thereof. This leads to deep relations between the theory of special functions and supersymmetric physics. 
For example, various dualities lead to non-trivial identities of elliptic-hypergeometric integrals and limits thereof, Painlev{\'e} tau-functions, etc. A set of pairs of concrete examples of the intimate relation between the mathematical works and the supersymmetric physics is \cite{rains,Seiberg:1994pq}, \cite{Ruijsenaars:2020shk,Razamat:2018zel},
\cite{2014arXiv1408.0305R,Pasquetti:2019hxf}, \cite{debultthesis,Benini:2011mf}, and   \cite{MR2787288,Gadde:2009kb}. 
Many times these quantities can be related to topological, conformal, and integrable QFTs in even lower dimensions (not necessarily supersymmetric). The canonical, concrete, and most studied example of those relations, generally called the BPS/CFT correspondence \cite{NN2004:BPSCFT0, NN2004:BPSCFT1}, is the AGT correspondence \cite{Alday:2009aq}  (See also \cite{Alday:2009fs,Drukker:2009tz,Kozcaz:2010af,Dimofte:2010tz}), relating specific four dimensional quantum field theories with ${\cal N}=2$ supersymmetry and specific two dimensional conformal field theories. See \cite{LeFloch:2020uop} for a comprehensive review and list of references. Another example is the association of an integrable quantum mechanical system to supersymmetric theories with two dimensional ${\CalN}=2$ super-Poincare invariance, the Bethe/gauge correspondence \cite{Nekrasov:2009uh}, including those originating as four or six dimensional theories \cite{Gadde:2009kb,Gaiotto:2012xa,Gaiotto:2015usa,Maruyoshi:2016caf,Yagi:2017hmj,Nazzal:2018brc,Nazzal:2021tiu,Razamat:2018zel,Ruijsenaars:2020shk,Chen:2020jla,Chen:2021ivd,Chen:2021rek}. In the geometric program of relating compactifications of six dimensional theories to four dimensional ones the quantum mechanical integrable models appear for example as operators acting on supersymmetric indices \cite{Kinney:2005ej,Romelsberger:2005eg,Dolan:2008qi} of the four dimensional theories introducing surface defects in to the index computation \cite{Gaiotto:2012xa}. For a given six dimensional theory one then obtains one, or more \cite{Nazzal:2021tiu} related, integrable models. The problem of classifying $D=6$ SQFTs  thus can be tied to studying properties of spaces of integrable quantum mechanical models. In addition to the above, the  supersymmetric quantities often have surprising features appearance of which is not always understood, see {\it e.g.} \cite{Buican:2019kba}.

\end{enumerate}

 \subsection{Holography And Classification of QFTs}

 Another fruitful direction of research that complements the geometric engineering methods above is to study the space of $AdS$ solutions in string or M-theory which are related to reduction of higher-$d$ theories on various manifolds similar to class $S$.  For example, the classification of holographic duals of $\mathcal{N}=1$ SCFTs in $D=4$ in M-theory corresponds to characterising the space of supergravity solution in $D=11$ of the form $AdS_5 \times M_6$ \cite{Gauntlett:2004zh}.  In the case of theories that are dual to reductions of 6D SCFTs on Riemann surfaces, the supergravity equation reduces to a single Monge-Amp{\`e}re equation whose solution space governs the allowed choices of $M_6$ and therefore the space of such SCFTs \cite{Bah:2013qya,Bah:2015fwa}. These equations generalize the $SU(\infty)$ Toda system that governs the holographic duals of $\mathcal{N}=2$ SCFTs \cite{Gaiotto:2009gz,Lin:2004nb}.  Similarly the classification problem for the holographic duals of $\mathcal{N}=1$ SCFTs from massive type IIA corresponds to solving a pair of Monge-Amp{\`e}re equations \cite{Bah:2017wxp}.  
 
 Of course, the very Calabi-Yau metric is a solution to a (complex) Monge-Amp{\'e}re equation on its K{\"a}hler potential, and the toric
 Calabi-Yau metrics are similarly described by the solutions of real Monge-Amp{\'e}re equations in three dimensions. 
 
The compactification program of SCFTs can be studied in holography by considering near horizon limits of branes systems wrapped on a compact manifold.  The canonical examples, which have lead to much interesting progress, are obtained from geometric duals of the topological twist, as initiated in \cite{Maldacena:2000mw}. There must exist a wider class of constructions beyond the paradigm of the topological twist where branes wrap non-constant curvature manifolds.  A systematic classification of what is possible subject to the constraint of supersymmetry is lacking. Indeed this is an active arena of research.  Recent important progress in these directions come from branes wrapping "spindles" (manifolds with orbifold fixed points) \cite{Ferrero:2020laf,Faedo:2021nub,Ferrero:2021etw,Ferrero:2021wvk} and discs with interesting holonomies at  their boundary \cite{Bah:2021hei,Bah:2021mzw}.  The latter methods have lead to the construction of holographic duals of Argyres-Douglas theories. For further developments see \cite{Couzens:2021rlk,Suh:2021aik,Couzens:2021cpk,Karndumri:2022wpu,Suh:2021hef}.   

In general the classification of BPS equations that describe the $AdS$ duals of systems of branes and their reductions on various manifolds have been obtained from reducing supergravity with $G$-structure.  This has been the program of Gauntlett, Martelli, Sparks, Waldram, Kim over a large body of literature \cite{Gauntlett:2003di,Gauntlett:2002sc,Gauntlett:2004hs,Gauntlett:2005ww,Gauntlett:2006ux,Gauntlett:2006af,Gauntlett:2006qw,Gauntlett:2006ns,Gabella:2010laf}. (Another approach uses pure spinors. See for example \cite{Grana:2004bg} for Poincar\'e compactification.)   With the classification of possible BPS structures from supergravity,  It is an important question to understand the solution space of the PDEs and their associated $AdS$ geometries.  One of the important successes in this program is the classification of holographic six dimensional SCFTs by the explicit realization of all $AdS_7$ solutions of supergravity in \cite{Apruzzi:2013yva} and their characterization in \cite{Gaiotto:2014lca}.  A similar classification program for five dimensional holographic SCFTs in type II supergravity theories was initiated in \cite{Apruzzi:2014qva} where the basic PDEs that govern all $AdS_6$ solutions are established.  An important objective will be to understand the full solution space of such PDEs.  

Significant steps have been made in understanding and characterizing the $AdS_6$ solutions in type IIB and their field theory duals in \cite{Bergman:2012kr,Bergman:2013aca}. More recently in \cite{ DHoker:2016ysh,DHoker:2016ujz,DHoker:2017mds,DHoker:2017zwj} a systematic construction of a large family of $AdS_6$ solutions is obtained, these are dual to $(p,q)$ brane engineering of five dimensional SCFTs.  
 
Another interesting direction in holography is a systematic characterization of Holographic RG flows \cite{Freedman:1999gp} by the study of geometric flows of metrics.  Flow equations are obtained in supergravity where they interpolate between $AdS\times \Sigma$.  Here different $\Sigma$'s identify different CFT fixed point.  The metric flow can encode various aspects of deformability of CFTs and provide a mathematical framework for exploring the space of CFTs.  In particular important results about uniformatization of metric flows in the context of class $\mathcal{S}$ are described in \cite{Anderson:2011cz}.  

As importantly, using holographic RG flows one can study various aspects of extremalization principles that can fix volumes of cycles in fixed backgrounds, and determine existence conditions of specific metrics in a given K\"{a}hler class \cite{Martelli:2006yb,Couzens:2018wnk,Gauntlett:2019roi}.  Such principles can be used to identify dual counterparts of  field theory phenomena such as a-maximization in four dimensions \cite{Intriligator:2003jj} and c-extremization in two dimensions \cite{Benini:2013cda,Benini:2012cz}.


\section{Some Important Topics In Physical Mathematics Not Covered Here}\label{sec:Omissions}

We are painfully aware that there are numerous exciting developments, and indeed entire sub-fields of physical mathematics which have not been covered in the above review. Some, but not all,  of these omitted  topics were covered in a previous section $10$ of this review. That section has evolved into a separate publication \cite{Commando}. Among the topics to be covered in \cite{Commando} are 

\begin{enumerate} 

\item Relations to the higher algebra of operads, BV algebras, etc. 

\item Further development of topological string theory and string field theory

\item Open-closed string duality

\item Topological Recursion and non-perturbative Dyson-Schwinger

\item Twisted, topological, and hybrid holography

\item Numerous connections to integrability, including 4d CS, Bethe/gauge and ODE/IM correspondences, $QQ$-systems and $qq$-characters.

\item Emergent and hidden hydrodynamics

\end{enumerate}

\section{Acknowledgements}

We would like to thank  David Ben-Zvi, Lakshya Bhardwaj, Daniel Brennan, Nikolay Bobev, Miranda Cheng, Mykola Dedushenko,
Emanuel Diaconescu, Tudor Dimofte, Lance Dixon, Yakov Eliashberg, Mohamed Elmi, Michael Freedman, Arthur Jaffe, Jeff Harvey, Mike Hopkins, Theo Johnson-Freyd, 
Dominic Joyce, Anton Kapustin, Ahsan Khan, Alexei Kitaev, Zohar Komargodski, Maxim Kontsevich, Igor Krichever, Craig Lawrie, Andrei Losev, Jan Manschot, Marcos Marino, Dave Morrison, Andy Neitzke, Alexei Oblomkov, Andrei Okounkov, Natalie Paquette, Sara Pasquetti, Boris Pioline, Peter E. Pushkar, Pavel
Safronov, Vivek Saxena, Christoph Schweigert, Nathan Seiberg, Shu-Heng Shao, Dennis Sullivan, Yuji Tachikawa, Constantin Teleman, Thomas Walpuski,
Maxim Zabzine for useful correspondence and discussions.

The work of IB is supported
in part by NSF grant PHY-2112699 and in part by the ``Simons Collaboration on Global Categorical Symmetry".
DF is partially supported by the
National Science Foundation under Grant Number DMS-2005286 and partially by the ``Simons Collaboration on Global Categorical Symmetries".
The work of GM is supported by US Department of Energy under grant DE-SC0010008.
SSN is supported in part by the European Union's Horizon 2020 Framework: 
ERC grant 682608 and in part by the ``Simons Collaboration on Special Holonomy in Geometry, Analysis and Physics''. The research of SSR was supported in part by Israel Science Foundation under grants no. 2289/18 and 2159/22, by a Grant No. I-1515-303./2019 from the GIF, the German-Israeli Foundation for Scientific Research and Development, and by BSF grant no. 2018204. This work was in part performed at the Aspen Center for Physics, which is supported by National Science Foundation grant PHY-1607611.

\bibliographystyle{ytphys} 
\bibliography{Snowmass}


\end{document}